\documentclass[11pt,letterpaper]{article}

\usepackage{jheppub}
\usepackage{color}
\usepackage[dvipsnames]{xcolor}
\usepackage[table]{xcolor}
\usepackage{colortbl}
\usepackage{graphicx}
\usepackage{subfig}
\usepackage{wrapfig,enumerate,slashed}
\usepackage{stmaryrd}
\usepackage{multirow,enumitem}
\usepackage[noabbrev]{cleveref}

\newcommand{\la}[1]{{\lambda}_{#1}}
\newcommand{\lab}[1]{{\lambda}^{\ast}_{#1}}

\newcommand{\ftr}[1]{\lfloor {#1} \rfloor}
\usepackage[normalem]{ulem}
\usepackage{footmisc}
\usepackage{amsmath}
\usepackage{wasysym} 
\usepackage{graphicx}
\usepackage{color}
\usepackage{comment}
\usepackage{hyperref}
\usepackage{enumitem}
\usepackage{pifont}
\DeclareMathAlphabet{\mathsfit}{\encodingdefault}{\sfdefault}{m}{sl}

\newcommand{\nc}{\newcommand}
\newcommand{\cmark}{\ding{51}} 
\newcommand{\xmark}{\ding{55}} 
\nc{\eps}{\varepsilon}
\nc{\vp}{\varphi}
\nc{\tvp}{\widetilde{\varphi}}
\nc{\D}{\mbox{$\not\!\!D$}}
\nc{\Db}{\mbox{${\raisebox{2mm}{\boldmath ${}^\leftarrow$}\hspace{-4mm} D}$}}
\nc{\Dfb}{\mbox{$\raisebox{2mm}{\boldmath ${}^\leftrightarrow$}\hspace{-4mm} D$}}
\nc{\vpj }{\mbox{${\vp^\dag i \,\raisebox{2mm}{\boldmath ${}^\leftrightarrow$}\hspace{-4mm} D_\mu\,\vp}$}}
\nc{\vpjt}{\mbox{${\vp^\dag i \,\raisebox{2mm}{\boldmath ${}^\leftrightarrow$}\hspace{-4mm} D_\mu^{\,I}\,\vp}$}}

\definecolor{darkgreen}{rgb}{0,0.5,0}

\definecolor{verde}{cmyk}{0.92,0,0.59,0.25}

\newcommand{\Dfbd}{\mathord{\buildrel{\lower3pt\hbox{$\scriptscriptstyle\leftrightarrow$}}\over {D}_{\mu}}}

\usepackage{bbold}
\usepackage{comment}

\begin{document}
	\flushbottom
	\allowdisplaybreaks
	
	\title{Probing Higgs and Top Interactions through the Muon \\ \vspace{1.0mm} Lens at  multi-TeV Muon Colliders}
    
    
	
	\author[a,b,c]{Tisa Biswas}
	\author[d]{, Anindya Datta}
	\author[e]{, and Barbara Mele}
	
	\affiliation[a]{Indian Institute of Technology Kanpur, Kalyanpur, Kanpur 208016, Uttar Pradesh, India}
    \affiliation[b]{The Institute of Mathematical Sciences, Taramani, 600113 Chennai, India}
    \affiliation[c]{Homi Bhabha National Institute, Training School Complex, Anushakti Nagar, Mumbai 400094,
India}
	\affiliation[d]{Department of Physics, University of Calcutta, 92 Acharya Prafulla Chandra Road, Kolkata 700009, India}
	\affiliation[e]{INFN, Sezione di Roma, c/o Dipartimento di Fisica, ``Sapienza'' Universit\`a di Roma, P.le Aldo Moro 2, I-00185 Rome, Italy}
	
	\emailAdd{tisab@imsc.res.in, adphys@caluniv.ac.in, barbara.mele@roma1.infn.it}
	
	\abstract{We investigate the sensitivity of a future 10 TeV muon collider to dimension-6 operators in the Standard Model Effective Field Theory (SMEFT), focusing on Higgs and top quark production processes. The analysis includes two-fermion and four-fermion operators that induce electroweak vector and axial-vector interactions, as well as dipole, scalar, and tensor interactions involving muons. Many of these operators are only weakly constrained or difficult to probe at the LHC due to limited sensitivity and large SM backgrounds. We study the processes $\mu^+\mu^- \to Zh$, $\mu^+\mu^- \to \mu^+\mu^-h$, $\mu^+\mu^- \to t\bar t$, and $\mu^+\mu^- \to t\bar t h$, exploiting the energy-enhanced SMEFT effects  at multi-TeV scales accessible to a muon collider. Using detailed simulations that incorporate differential information and angular distributions, we derive projected bounds on the relevant Wilson coefficients. We find that a 10 TeV muon collider can strengthen existing limits on muon--Higgs--gauge and muon--top interactions by up to an order of magnitude, surpassing even FCC-ee projections. Finally, we interpret these bounds in the context of representative UV scenarios, including models with vector-like lepton and scalar leptoquarks, highlighting the potential of a muon collider to probe new physics at scales well beyond the LHC reach.}

\maketitle
\section{Introduction}
\label{sec:intro}
The efforts to look for experimental evidence of all Standard Model (SM) particles culminated in the discovery of the Higgs boson at the Large Hadron Collider (LHC) by the ATLAS and CMS collaborations~\cite{Aad:2012tfa,Chatrchyan:2012xdj}. As the scalar degree of freedom associated with electroweak symmetry breaking, the Higgs boson provides direct access to the underlying electroweak dynamics and serves as a sensitive probe of possible ultraviolet (UV) completions of the SM. The top quark, owing to its large mass and strong coupling to the Higgs field, also stands out as a promising probe for uncovering new interactions at scales above the electroweak scale. At the same time, despite the remarkable agreement between SM predictions and a broad array of experimental results, several issues remain unresolved. These include the fine tuning of the Higgs mass, the hierarchical structure in fermion masses, and the baryon asymmetry of the Universe, to name a few.  Addressing these issues requires a more fundamental theory underlying the SM, thereby {\em motivating} the Beyond Standard Model (BSM) programme.

In the absence of direct signatures from resonant production of new particles, a robust approach is to probe short-distance interactions through accurate measurements of total rates and kinematic distributions of SM processes. This has marked the beginning of the precision physics era at the LHC, constraining new physics indirectly with sensitivities extending to multi-TeV scales. Within the theoretical framework, two promising avenues emerge. The first considers the possible presence of light new particles below the GeV scale with feeble couplings to SM fields~\cite{Biswas:2017lyg,Biswas:2023ksj,Fabbrichesi:2017zsc}. The second concerns heavy new states well above the TeV scale, whose effects are encapsulated in a set of higher-dimensional effective operators (with mass dimensions $> 4$)~\cite{Biswas:2021qaf,Biswas:2022fsr,Adhikary:2025gdh}. In this work, we focus on the second scenario. We consider interactions of heavy SM particles such as the Higgs boson, top quark, and electroweak gauge bosons ($W/Z$) mediated via higher-dimensional operators and  suppressed by the new physics scale. For certain operator structures, these contributions exhibit characteristic energy growth. We therefore focus on muon collider physics at multi-TeV centre-of-mass (c.o.m.) energies. Muon colliders are ideally suited for both energy and precision frontiers. Being leptonic machines, they maximise energy transfer while reducing backgrounds compared to hadron colliders~\cite{Skrinsky:1981ht,Neuffer:1983xya,Barger:1995hr,Ankenbrandt:1999cta,deBlas:2944678}. In particular, the large muon mass ($m_\mu \approx 207 m_e$) suppresses synchrotron radiation, enabling circular operation at multi-TeV scales, well beyond the reach of $e^+ e^-$ machines~\cite{FCC:2025lpp}. Projected luminosities of $\mathcal{O}(1)$~ab$^{-1}$ or higher~\cite{MuonCollider:2022xlm,Aime:2022flm} make the muon collider a powerful probe of indirect effects of new physics. Effectively functioning as an electroweak gauge-boson collider~\cite{Buttazzo:2018qqp,Han:2020uid,Ruiz:2021tdt,Garosi:2023bvq}, muon colliders offer enhanced sensitivity to colour-neutral extensions of the SM that remain weakly constrained at hadron colliders. Its reach has been explored in the context of precision Higgs and electroweak measurements~\cite{Chiesa:2020awd,Han:2020pif,Buttazzo:2020uzc,Liu:2021jyc,Franceschini:2021aqd,Chiesa:2021qpr,Forslund:2022xjq,Forslund:2023reu,Ruhdorfer:2023uea}, flavour-sensitive processes~\cite{Asadi:2025dii,Chen:2021rnl,Huang:2021biu,Azatov:2022itm,Altmannshofer:2023uci,Han:2023njx,Glioti:2025zpn}, dark-matter searches~\cite{Han:2020uak,Bottaro:2021srh,Asadi:2023csb,Saha:2025npi}, and several other BSM frameworks~\cite{De:2024tbo,Costantini:2020stv,AlAli:2021let}. Recent studies also highlight the potential of muon-beam dumps as powerful probes of long-lived particles~\cite{Cesarotti:2022ttv,Cesarotti:2023sje}.

In the following analysis, we will assume as a theoretical framework the Standard Model Effective Field Theory (SMEFT). The latter parametrises any possible effects beyond the SM by extending the SM Lagrangian with new operators (of mass dimensions $>4$) constructed solely from SM fields and respecting all the symmetries of the SM. 
The physical Higgs field is embedded in the $SU(2)_L \times U(1)_Y$ scalar doublet.  The new operators  can in general  either slightly affect the SM processes, by altering the shapes of kinematic distributions, or give rise to final states  which in the SM have either vanishing or very feeble cross sections. Such indirect signatures in the SMEFT scale at most as $m_W^2/\Lambda^2$ or $s/\Lambda^2$, where $s$ denotes the squared c.o.m.  energy  of the collision. 

In this work, we investigate the sensitivity of a future high-energy muon collider to dimension-6 SMEFT operators involving muons that modify Higgs and top production. We identify the subset of operators that give leading contributions to processes such as $\mu^+\mu^- \to Zh$, $Z$-boson fusion Higgs production, $\mu^+\mu^- \to t\bar t$ and $t\bar t h$, focusing on their energy-growing behaviour at multi-TeV scales. Assuming a common suppression scale, we perform a dedicated high-energy fit to quantify the projected operator constraints. While earlier studies~\cite{Buttazzo:2020ibd,Paradisi:2022vqp,Yin:2020afe} focused mainly on dipole operators relevant to the $(g-2)_\mu$ muon anomaly via $\mu^+\mu^- \to h\gamma$, our analysis shows that Higgs and top-associated production processes offer complementary sensitivity to electroweak and four-fermion interactions involving muons, going beyond dipole-driven probes. This improves the indirect reach for new physics, enabling probes of scales up to several tens of TeV.

Beyond the model-agnostic analysis, it is important to consider the UV-complete frameworks that can generate the corresponding effective operators. Several UV scenarios such as supersymmetric models \cite{Chattopadhyay:2001vx,Heinemeyer:2003dq,Banerjee:2018mnw}, Grand Unification models with scalar or vector leptoquarks \cite{Chakraverty:2001yg,Parashar:2022wrd,Bhaskar:2022vgk}, models with extended scalar~\cite{Babu:2022pdn,Iguro:2018qzf,Li:2018aov} or  fermionic sectors~\cite{Saad:2020ihm,Bobeth:2016llm,Capdevilla:2021rwo} can generate interactions that modify the 
  $\mu \mu Z/\gamma$ or $t\bar{t}\mu^+\mu^-$ couplings. Such deviations from the SM predictions at high energies translate into constraints on the masses and couplings of the underlying heavy states.  
We connect our results to representative UV models, including vector-like leptons and scalar leptoquarks, where the relevant operators arise either at tree- or loop-level.

The salient features of  this work are the following:
\begin{itemize}
\item We identify the set of SMEFT operators \emph{involving muons} that give leading contributions to Higgs and top pair production processes. We begin by assessing the sensitivity of the LHC to small values of the Wilson coefficients of certain two-fermion and four-fermion operators, and find that its reach is limited in this regime. This underscores the muon collider as a more suitable probe of these effects. To the best of our knowledge, this work presents the first comprehensive study of such operator effects in the environment of a high-energy muon collider.
\item We analyse the production channels  $hZ, \mu^+ \mu^- h, t\bar{t} $ and $t\bar{t}h $ at a 10 TeV muon collider and demonstrate that non-standard contributions can be probed at sub-per-mille precision. We employ differential information to further enhance the sensitivity to these effects.
\item We present a comparative study of sensitivities at HL-LHC, FCC\mbox{-}ee and the muon collider. In particular, simple projections for high-energy muon colliders illustrate their potential in  future dedicated analyses. A multi-TeV muon collider can probe four-fermion interactions involving the top quark that are inaccessible at low energies.
\item Our results are useful for deriving new bounds on BSM particle masses. As representative examples, we discuss UV completions that generate these dimension-6 operators, such as the SM extended with vector-like leptons, and singlet and triplet scalar leptoquarks.  The resulting bounds on the BSM parameters after matching to the projected SMEFT constraints 
can surpass those from current direct searches at the LHC.
\end{itemize}

The paper is organised as follows. In section~\ref{section2}, we briefly discuss the SMEFT parametrisation relevant in the present analysis. In section~\ref{section3}, we study the constraints that can be obtained at the hadron colliders, followed in section~\ref{section4} by a detailed collider analysis of Higgs and top production at multi-TeV $\mu^+ \mu^-$ colliders. In section~\ref{section5}, we quantify the constraining potential of the Effective Field Theory (EFT)-driven signals considered in this study across different centre-of-mass energies and integrated luminosity configurations. We also compare the reach of the HL-LHC, FCC\mbox{-}ee, and the muon collider. 
In section~\ref{section7}, we provide the corresponding bounds on BSM parameters with vector-like leptons and scalar leptoquarks. We summarise our conclusions in section~\ref{section8}.
\section{The SMEFT Parametrisation}
\label{section2}
\begin{figure}[b!]
	\centering
			\begin{tabular}{cccc}
	\includegraphics[width=0.19\textwidth]{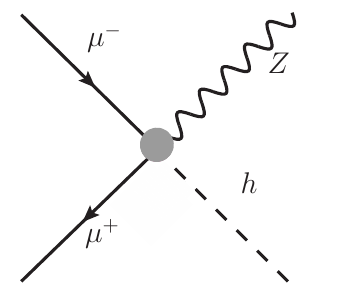}&
	\includegraphics[width=0.19\textwidth]{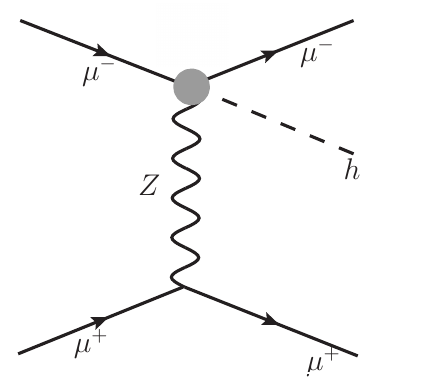}&
	\includegraphics[width=0.21\textwidth]{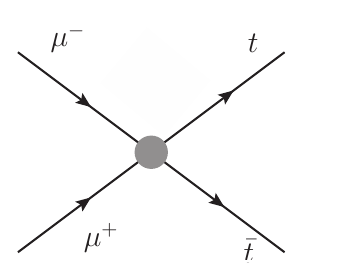}&
	\includegraphics[width=0.21\textwidth]{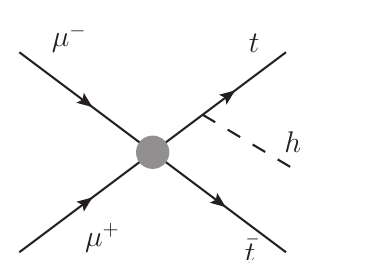}\\
				(a)&(b)&(c)&(d)
		\end{tabular}  \vspace{-2mm}
	\caption{Representative Feynman diagrams for the high-energy processes considered in this study. The grey blobs denote insertions of dimension-6 effective interactions.}
	\label{feyn_diag_coll}
\end{figure}

We now delve into a discussion of  the SMEFT framework and its use in parametrising the effects of BSM dynamics in interactions involving the Higgs boson and top quarks. Within this framework, the effects of heavy new physics above the scale $\mu$ are captured by higher-dimensional operators.  The SMEFT Lagrangian is given by:
\begin{equation}
	\mathcal{L}_{\rm SMEFT}=
	\mathcal{L}_{\rm SM}+\sum_{d=5}^{\infty} \sum_{i=1}^{n}{C_i^{(d)} (\mu) \over\Lambda^{d-4}}\mathcal{O}_i^{(d)}
	\label{eq:smeft}
\end{equation}
The scale $\Lambda (\gg \mu)$ is associated with the masses of heavy degrees of freedom, whose effects enter indirectly through virtual exchange. The operators $\mathcal{O}_i^{(d)}$ form the complete set of dimension-$d$ operators. The Wilson coefficients (WCs) $C_i^{(d)} (\mu)$ encapsulate any non-standard effects. 
We adopt the Warsaw basis parametrisation of these operators as defined in~\cite{Grzadkowski:2010es}. In this work, we focus on the general set of  operators that modify Higgs production via $\mu^+ \mu^- \to Zh, \mu^+ \mu^- \to \mu^+ \mu^- h$ and top pair production in the $\mu^+ \mu^- \to t \bar{t}, \mu^+ \mu^- \to t \bar{t} h$ processes.  The Feynman diagrams relevant for these processes are presented in Fig.~\ref{feyn_diag_coll}. The blobs on the vertices of diagrams in Fig.~\ref{feyn_diag_coll}(a)-(d) signify possible insertions of  operators as discussed in the following. At high energies, Higgs production via the $Z$ boson fusion ($ZBF$) process $\mu^+\mu^- \to \mu^+\mu^- h$ and the $W$ boson fusion process $\mu^+\mu^- \to \nu_\mu \bar{\nu}_\mu h$ probes closely related combinations of SMEFT operators. This follows from the underlying electroweak gauge structure and the dominance of longitudinal gauge boson contributions at high energies. In this work, we focus on the $ZBF$ channel, which provides a clean and representative probe of these energy-enhanced effects.
\begin{itemize}[leftmargin=*] \vspace{0mm}
	\item The dimension-6 operators constructed from the Higgs doublet $\vp$ and its covariant derivatives are given by:
\begin{equation}
	\mathcal{O}_{ \varphi \Box}  =  (\vp^\dag \vp)\raisebox{-.5mm}{$\Box$}(\vp^\dag \vp)~,~~~
	\mathcal{O}_{\varphi D} =  \left(\vp^\dag D^\mu\vp\right)^\ast \left(\vp^\dag D_\mu\vp\right)  
\end{equation}
These operators renormalise the Higgs field and lead to a universal rescaling of Higgs interactions.  The covariant derivative $D_\mu$ carries its usual definition, involving $SU(2)_L$ and $U(1)_Y$ gauge bosons and their respective gauge couplings.
\item The operators inducing lepton-lepton-gauge boson-Higgs ($\ell\ell Vh$) or lepton-lepton-gauge boson ($\ell\ell V$)  interactions are as follows: 
	\begin{equation}
	\mathcal{O}_{\varphi  \ell}^{(1)} = (\vpj)(\bar \ell_p \gamma^\mu \ell_r)\, ,~~
	\mathcal{O}_{\varphi  \ell}^{(3)} = (\vpjt)(\bar \ell_p \tau^I \gamma^\mu \ell_r)\, ,~~
	\mathcal{O}_{\varphi  e} = (\vpj)(\bar e_p \gamma^\mu e_r) 
    \label{llV_llVh}\, ,
	\end{equation}
where we use the notation  $\ell$ for the lepton doublet under $SU(2)_L$ and $e$ for the $SU(2)$ singlet leptons. The indices $p$ and $r$ label lepton generations. We define $\vpj= i \vp^\dagger  D_\mu \vp-i (D_\mu \vp)^\dagger \vp$ and $\vpjt= i \vp^\dagger \tau^I D_\mu \vp-i (D_\mu\vp)^\dagger\tau^I \vp$.  
\item  {\it Dipole} operators can also induce $\ell\ell Vh$ or $\ell\ell V$ interactions. They have a magnetic dipole structure, distinct from the operators discussed earlier. In the massless fermion limit, dipole operators couple left- and right-handed fermions, reflecting their chirality-flipping structure.  Their dependence on the gauge field strength tensors introduces momentum dependent effects that significantly alter the final-state kinematics relevant for this analysis. The operators are:
	\begin{equation}
	\mathcal{O}_{\mu W} = (\bar \ell_p \sigma^{\mu\nu} \mu_r) \tau^I \vp\, W_{\mu\nu}^I \, ,~~~~~~
	\mathcal{O}_{\mu B} = (\bar \ell_p \sigma^{\mu\nu} \mu_r) \vp\, B_{\mu\nu}\, ,
	\end{equation}
    with coefficients $C_{\mu W}$ and $C_{\mu B}$, respectively.
Note that these dipole operators contribute to the muon $(g-2)_\mu$ both at tree level and through loop-effects. We will discuss this issue in the next section.

After electroweak symmetry breaking, the interactions of the muon with the electroweak gauge bosons take the form:%
\begin{eqnarray}
\mathcal{L} &\supset&
- e\, A_\mu \,\bar{\mu}\gamma^\mu \mu
- \frac{g}{\sqrt{2}}\, W_\mu^+\, \bar{\nu}_{\mu L}\gamma^\mu \left(1+\delta g_L^{W\mu}\right)\mu_L + \mathrm{h.c.}
\nonumber \\ &&
- \frac{g}{c_W} Z_\mu
\left[
\bar{\mu}_L \gamma^\mu \left(-\frac{1}{2}+s_W^2+\delta g_L^{Z\mu}\right)\mu_L
+ \bar{\mu}_R \gamma^\mu \left(s_W^2+\delta g_R^{Z\mu}\right)\mu_R
+ \bar{\nu}_{\mu L}\gamma^\mu \left(\frac{1}{2}+\delta g_L^{Z\nu_\mu}\right)\nu_{\mu L}
\right]
\nonumber \\ &&
+ \frac{1+h/v}{v}
\left[
C_{\mu\gamma}\,\bar{\mu}_L \sigma^{\mu\nu}\mu_R\,A_{\mu\nu}
+ C_{\mu Z}\,\bar{\mu}_L \sigma^{\mu\nu}\mu_R\,Z_{\mu\nu}
+ C_{\mu W}\,\bar{\nu}_{\mu L}\sigma^{\mu\nu}\mu_R\,W^+_{\mu\nu}
+ \mathrm{h.c.}
\right] \, ,
\label{eq:La_rephrased}
\end{eqnarray}
where $e$ and $g$ are the electromagnetic and weak gauge couplings, respectively, and $\theta_W$ is the weak mixing angle.  
$\delta g_{L}^{Z\mu}$ and $\delta g_{R}^{Z\mu}$ parameterise the corrections arising from dimension-6 SMEFT operators. These shifts are expressed in terms of the Wilson coefficients as follows:
\begin{equation}
\delta g_{L}^{Z\mu} = \frac{v^2}{2 \Lambda^2}\left(C_{\varphi \ell}^{(1)} + C_{\varphi \ell}^{(3)}\right) \ , 
\hspace{1cm}
\delta g_{R}^{Z\mu} = \frac{v^2}{2 \Lambda^2} C_{\varphi  e}  \,, 
\hspace{1cm}
\delta g_L^{W \mu} = \frac{v^2}{\Lambda^2} C_{\varphi \ell}^{(3)}  \,,
\end{equation}
where $\{C_{\varphi \ell}^{(1)} ,C_{\varphi \ell}^{(3)}, C_{\varphi e} \}$ correspond to the coefficients of operators $\{\mathcal{O}_{\varphi \ell}^{(1)} ,\mathcal{O}_{\varphi \ell}^{(3)}, \mathcal{O}_{\varphi e} \}$ (in Eq.~\ref{llV_llVh}), respectively. The coefficients, $C_{\mu Z}$ and $C_{\mu \gamma}$ are associated to dipole operators and are given by:
\begin{equation}
C_{\mu Z} =  \cos\theta_W \, C_{\mu W} + \sin\theta_W \, C_{\mu B}  \ , 
\hspace{1cm}
C_{\mu \gamma} = \cos\theta_W \, C_{\mu B} - \sin\theta_W \, C_{\mu W} .
\label{eq:cmuZA}
\end{equation}
\item Finally, we consider the two-quark--two-lepton operators in the Warsaw basis that contribute to interactions involving muons and top quarks. These operators are classified according to their Lorentz structures:
\end{itemize}
\begin{equation}
\begin{array}{rl@{\,}cc}
	\mathcal{O}_{\ell q}^{(1)}
		&=
		&(\bar{\ell}_p\gamma^\mu \ell_r)&(\bar{q}_s\gamma_\mu q_t )
	\, ,\\
	\mathcal{O}_{\ell q}^{(3)}
		&=
		&(\bar{\ell}_p\tau^I\gamma^\mu \ell_r)&(\bar{q}_s\tau^I\gamma_\mu q_t)
	\, ,\\
	\mathcal{O}_{\ell u}
		&=
		&(\bar{\ell}_p\gamma^\mu  \ell_r)&(\bar{u}_s\gamma_\mu u_t)
	\, ,\\
	\mathcal{O}_{eq}
		&=
		&(\bar{e}_p\gamma^\mu e_r)&(\bar{q}_s\gamma_\mu q_t)
	\, ,\\
	\mathcal{O}_{eu}
		&=
		&(\bar{e}_p\gamma^\mu e_r)&(\bar{u}_s\gamma_\mu u_t)
	\, ,
\end{array}
\quad
\begin{array}{rlccc}
	\mathcal{O}_{\ell equ}^T
		=
		 (\bar{\ell}^a_p\sigma_{\mu\nu} e_r) &\epsilon_{ab} &(\bar{q}^b_s\sigma^{\mu\nu} u_t)
	\, ,
\end{array}
\quad
\begin{array}{rlcc}
	\mathcal{O}_{\ell equ}^S
		=
		(\bar{\ell}^a_p e_r)&\epsilon_{ab}(\bar{q}^b_s u_t)
	.
\end{array}
\label{eq:op_2q2l}
\end{equation}
\noindent
We denote the quark doublet under $SU(2)_L$ by $q$, and the corresponding up- and down-type singlet quarks by $u$ and $d$, respectively. These operators have flavour-dependence, with $p,r$ denoting the lepton flavour indices and $s,t$ representing the quark flavour indices.
Due to their different transformation properties under parity and distinct sensitivities to quark currents,  it is useful to define the following vector and axial-vector combinations:
\begin{equation}
	\begin{aligned}
		C_{\ell q}^V	&\equiv C_{\ell u} + (C_{\ell q}^{(1)}-C_{\ell q}^{(3)})\equiv C_{\ell u} +C_{\ell q}^{(-)}~,	\\
		C_{\ell q}^A	&\equiv C_{\ell u} - (C_{\ell q}^{(1)}-C_{\ell q}^{(3)})\equiv C_{\ell u} -C_{\ell q}^{(-)}~,
	\end{aligned}
	\qquad\quad
	\begin{aligned}
		C_{eq}^V	&\equiv C_{eu} + C_{eq}	~~,	\\
		C_{eq}^A	&\equiv C_{eu} - C_{eq}	~~.
	\end{aligned}
\end{equation}
\noindent
with $\{C_{\ell q}^{(1,3)}\!, C_{\ell u}, C_{eu}, C_{eq}\}$ being the WCs of $\{\mathcal{O}_{\ell q}^{(1,3)}\!, \mathcal{O}_{\ell u}, \mathcal{O}_{eu},  \mathcal{O}_{eq} \}$ in Eq.~\ref{eq:op_2q2l}, respectively. Owing to their chirality structure, the tensor and scalar operators\footnote{In this study, we will focus on two-top--two-muon interactions, and we will denote these tensor and scalar operators as $\mathcal{O}_{\ell equ}^T=\mathcal{O}^T_{\mu t}$ and $\mathcal{O}_{\ell equ}^S=\mathcal{O}^S_{\mu t}$ with coefficients $C_{\mu t}^T$ and $C_{\mu t}^S$, respectively.} do not interfere with the SM amplitudes in the limit of vanishing lepton masses. We now proceed to discuss the experimental constraints on these operator coefficients.
\vspace{-2mm}
\section{Constraints from electroweak precision data, LHC and flavour observables}
\label{section3}
In this section, we examine the current experimental constraints on the effective couplings introduced in Section~\ref{section2}, using electroweak precision observables, LHC measurements, and flavour data. The resulting bounds are summarised in Tables~\ref{other_constraints_flavour_universal_1} and~\ref{other_constraints_flavour_universal_2}. We express the available collider and flavour measurements in terms of the SMEFT Wilson coefficients defined earlier. We also include the weak-boson-fusion process $pp \to \mu^+ \mu^- jj$ as an additional novel probe of dipole interactions.

\begin{itemize}[leftmargin=*, labelsep=0.5em]
\item Constraints on lepton-gauge and dipole operators from $Z$-pole and LHC data
\end{itemize}
\noindent
\underline{$Z \to \mu^+\mu^-$ and $h \to \mu^+\mu^- Z$ decays}:
The partial width $\Gamma(Z \to \mu^+\mu^-)$ provides a stringent probe of operators that modify the $\mu\mu Z$ vertex at tree level. 
Using the measured $Z$-pole observables reported by the PDG~\cite{ParticleDataGroup:2024cfk}, we extract the corresponding 95\% C.L. bounds on $C_{\varphi \ell}^{(1)}$, $C_{\varphi \ell}^{(3)}$, and $C_{\varphi e}$, as summarised in Table~\ref{other_constraints_flavour_universal_1}. The lepton dipoles $C_{\mu B}$ and $C_{\mu W}$ enter the $Z$ interaction through the tensor form factor (cf.  Eq.~\ref{eq:cmuZA}). The resulting constraint on the combination $C_{\mu Z}$ is weak, $C_{\mu Z}\in [-37.98,\,37.98]~\text{TeV}^{-2}$, and is far less stringent than the bounds derived from $(g-2)_\mu$, which is of order $10^{-4}~\text{TeV}^{-2}$.
The same operators contribute to the decay $h\to\mu^+\mu^- Z$ through the modified $\mu\mu Z$ couplings and the contact interaction $\mu\mu hZ$~\cite{Dawson:2024pft}. The bounds on the aforementioned coefficients are given in Table~\ref{other_constraints_flavour_universal_1}.
\begin{figure}[b]
	\centering
	\includegraphics[width=4.cm,height=3.cm]{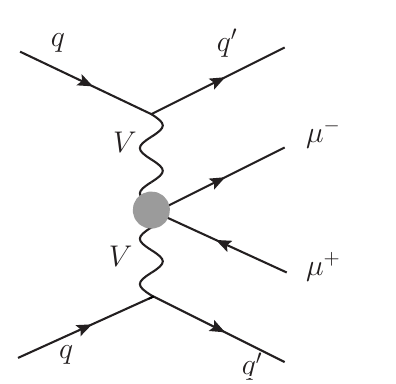} 
	\caption{Representative VBF topology in the SMEFT for $pp \to \mu^- \mu^+ +2j$. The grey blob denotes the induced $\mu^+ \mu^- VV$ interaction generated by leptonic dipole operators.}
	\label{FD}
\end{figure}
\vspace{3mm}

\noindent
\underline{$pp \to \mu^- \mu^+ +2j$ via W fusion}:  This process proceeds via the fusion of two electroweak gauge bosons radiated from the initial-state quarks. Unlike the standard Drell–Yan mechanism, where muon pairs originate from $s$-channel $Z/\gamma^*$ exchange, this process occurs via weak boson fusion in the $t$ channel and is directly sensitive to electroweak interactions beyond the SM (see Fig.~\ref{FD}).
This dipole-operator–induced process has not yet been explored at the LHC. We compute the corresponding sensitivity to the $\mathcal{O}{\mu W}$ operator within our framework, obtaining $C{\mu W} \in [-18.88,18.88]$~TeV$^{-2}$.
\begin{itemize}[leftmargin=*, labelsep=0.5em]
\item Constraints on $C_{\mu \gamma}$ from radiative Higgs decays, $(g-2)_\mu$ and LFV transitions
\end{itemize}
\noindent
\underline{$h \to \mu \mu \gamma$ decay}: We examine the impact of the dipole coefficient $C_{\mu \gamma}$ on the decay $h \to \mu^+ \mu^- \gamma$, where the Higgs is produced through SM processes and decays via this non-standard interaction. The decay width is given by~\cite{Buttazzo:2020ibd}
\begin{equation}
\Gamma(h \to \mu^+ \mu^- \gamma) = \Gamma_{\rm SM} + \frac{e ~m_h^3~ m_\mu}{64 \pi^3 v}\frac{C_{\mu \gamma}}{\Lambda^2} + \frac{m_h^5}{768 \pi^3}\frac{C_{\mu \gamma}^2}{\Lambda^4}.
\label{htomumugamma_decay}
\end{equation}
\noindent
Experimentally, the dominant background arises from non-resonant $\mu^+\mu^-\gamma$ production, while the SM Higgs contribution to this final state is comparatively small. The kinematic configuration of this decay is characterised by relatively soft and collinear final-state particles, which affects the experimental sensitivity and motivates dedicated analysis strategies. The CMS Collaboration has performed a search for such rare decays at $\sqrt{s}=13$~TeV~\cite{CMS:2018myz}, analysing both $h\to Z\gamma$ and $h\to \gamma^*\gamma$ channels. No significant excess was observed, and upper limits between 1.4 and 4.0 times the SM prediction were reported. Using the current CMS limit~\cite{CMS:2018myz} on this decay, we derive bounds on $C_{\mu \gamma}$ as follows:
\begin{equation}\label{eq:cmugamma_h2llgamma}
C_{\mu \gamma} \in [-0.0730,-0.0599] \cup [0.0097,0.0228]~\text{TeV}^{-2}.
\end{equation}

\vspace{2mm}

\noindent
\underline{$(g-2)$ of muon}: 
The anomalous magnetic moment of the muon provides a stringent constraint on dipole operators. The Fermilab experiment of Muon $g-2$ has reported its final value of the anomalous magnetic moment, $a_\mu = 1165920705(148) \times 10^{-12}$~\cite{Muong-2:2025xyk}, giving a new world average of $a_\mu^{\rm exp} = 1165920715(145)\times10^{-12}$. The updated SM prediction~\cite{Aliberti:2025beg} is $a_\mu^{\rm SM}=116592033(62) \times 10^{-11}$. The difference,
\begin{equation}
\Delta a_\mu = (38 \pm 63) \times 10^{-11}
\end{equation}
which shows no statistically significant deviation from the SM prediction, replacing the earlier difference $(249\pm 48)\times10^{-11}$ which was a $5.1\sigma$ discrepancy~\cite{Muong-2:2023cdq,Aoyama:2020ynm}. This substantially weakens the earlier motivation for new-physics explanations of the $(g-2)_\mu$ anomaly.
In SMEFT, the leading contribution to $(g-2)_\mu$ arises from $C_{\mu\gamma}$, with additional one-loop running effects from two-fermion $C_{\mu Z}$ and four-fermion tensor operators~\cite{Buttazzo:2020ibd}:
\begin{align}
\vspace{-3mm}
\Delta a_\mu &\simeq \frac{4m_\mu v}{e\Lambda^2} 
\bigg(
C_{\mu \gamma} - \frac{e^2}{16 \pi^2} 12~{{\rm{cot}}~ \theta_{2W}} ~C_{\mu Z} \log \tfrac{\Lambda}{m_Z}
\bigg)
- \sum_{q=c,t} \frac{4m_\mu m_q}{\pi^2} \frac{C^T_{\mu q}}{\Lambda^2}
\log \tfrac{\Lambda}{m_q},
\label{eq:Delta_a_ell}
\end{align}
or equivalently,
\begin{align}
\vspace{-3mm}
\frac{\Delta a_\mu}{3 \times 10^{-9}} \approx  \left( \frac{250 ~ {\rm TeV}}{\Lambda} \right)^{2} 
\left(C_{\mu \gamma} - 0.2 C_{\mu t}^T - 0.001 C_{\mu c}^T - 0.05 C_{\mu Z}\right).
\label{delta_a}
\end{align}
Using the updated world average and Eq.~\eqref{eq:Delta_a_ell}, we extract the corresponding bounds on $C_{\mu \gamma}$, $C_{\mu Z}$ and $C_{\mu t}^T$, summarised in Tables~\ref{other_constraints_flavour_universal_1} and~\ref{other_constraints_flavour_universal_2}. In general, several operators can contribute simultaneously to $(g-2)_\mu$, potentially with opposite signs. Although the current measurement agrees with the SM at the $1\sigma$ level~\cite{Muong-2:2025xyk}, sizeable individual contributions from these operators remain allowed, since cancellations can reproduce the observed value $\Delta a_\mu$. Hence, we aim to study independent  bounds on these operators at future multi-TeV muon colliders.

\noindent
\underline{$\mu \to e \gamma$ transition}: The lepton-flavour-violating decay $\mu \to e \gamma$ provides the most stringent test of the dipole operator. Using the upper limit on the branching ratio 
$\mathcal{B}(\mu^+ \to e^+ \gamma)$ from the MEG experiment~\cite{MEG:2016leq}, we have
\begin{align}
\begin{split}
\mathcal{B}(\mu^+ \to e^+ \gamma) &= \frac{m_\mu^3 v^2}{8\pi \Gamma_\mu}
\frac{(|[C_{\mu \gamma}]_{12}|^2 + |[C_{\mu \gamma}]_{21}|^2)}{\Lambda^4}
< 4.2 \times 10^{-13} \quad \text{(90\% C.L.)},
\end{split}
\end{align}
which implies\footnote{The subscript `12' denotes the flavour indices $e,\mu$.}
\begin{align}
\left|\frac{[C_{\mu \gamma}]_{12(21)}}{\Lambda^2}\right| &\lesssim 2.1 \times 10^{-10}~\mathrm{TeV}^{-2}.
\label{Cegamma12-constraint}
\end{align}
Hence, $\mu \to e \gamma$ sets a tight bound on the off-diagonal elements of WC of the operator $O_{\mu \gamma}$, and therefore does not restrict the flavour-diagonal muon dipole coefficient relevant for our collider study. The resulting limit from such flavour-violating interactions is shown in the grey-shaded row of Table~\ref{other_constraints_flavour_universal_1}.
\begin{itemize}[leftmargin=*, labelsep=0.5em]
\item Higgs field operators
\end{itemize}
The operators involving only the Higgs field and its derivatives, $\mathcal{O}_{\varphi D}$ and $\mathcal{O}_{\varphi \Box}$, modify the Higgs kinetic term and induce an overall rescaling of Higgs couplings. The corresponding Wilson coefficients are constrained by LHC data to be $C_{\varphi D} \in [-0.023,\,0.0027]~\text{TeV}^{-2}$ and $C_{\varphi \Box} \in [-1.00,\,0.47]~\text{TeV}^{-2}$~\cite{Biswas:2022fsr}.
\begin{table}[t]
	{\renewcommand{\arraystretch}{1.2}%
		\centering
		\scalebox{0.8}{
\resizebox{\textwidth}{!}{%
			\begin{tabular}{ |c|c|c| } 
				\hline
				Wilson coefficient & Related processes/Observables & Constraints ($\rm{TeV^{-2}}$)\\ 
				\hline\hline
				$C_{\varphi \ell}^{(1)}/\Lambda^2$&$Z\to \mu^+ \mu^-$  & $[-0.44,1.30]\times10^{-2}$~\cite{ParticleDataGroup:2024cfk} \\
				&$h\to \mu^+ \mu^- Z$ & $[-8.60,8.60]\times10^{-3}$~\cite{CMS:2021ugl}\\
				\hline
				$C_{\varphi \ell}^{(3)}/\Lambda^2$&$Z\to \mu^+ \mu^-$  & $[-1.00,0.30] \times10^{-2}$~\cite{ParticleDataGroup:2024cfk}\\
				&$h\to \mu^+ \mu^- Z$ &  $[-1.11,1.11]\times10^{-1}$~\cite{CMS:2021ugl}\\
				\hline
				$C_{\varphi e}/\Lambda^2$&$Z\to \mu^+ \mu^-$  &$[-1.50,0.71]\times10^{-2}$~\cite{ParticleDataGroup:2024cfk} \\
				&$h\to \mu^+ \mu^- Z$ & $[-6.90,6.90]\times10^{-3}$~\cite{CMS:2021ugl}\\
				\hline
					\cline{2-2} 
				$C_{\mu \gamma}/\Lambda^2$&$h\to \mu^+ \mu^- \gamma$ & $[-7.30,-5.99]\times10^{-2} ~\bigcup ~[0.97,2.28]\times10^{-2}$~\cite{CMS:2018myz}  \\
                 &$(g-2)_\mu$ &  $\left[-1.33\times 10^{-6}, 5.39\times 10^{-6}\right]$~\cite{Aliberti:2025beg,Muong-2:2025xyk}\\
    \rowcolor{gray!20} $[C_{\mu \gamma}]_{12}/\Lambda^2$&$\mu^+ \to e^+ \gamma$ & $[-2.10,2.10] \times 10^{-10}$ ~\cite{Falkowski:2023hsg} \\
				\cline{2-2}
				\hline
				$C_{\mu Z}/\Lambda^2$&$Z\to \mu^+ \mu^-$  &$[-37.98,37.98]$~\cite{ParticleDataGroup:2024cfk} \\
				&$h\to \mu^+ \mu^- Z$ &  $[-15.97, 15.97]$\\
				&$pp \to \mu^+ \mu^- jj$ & $[-18.88,18.88]$ \\
                 &$(g-2)_\mu$ &  $\left[-1.08, 0.27\right]\times 10^{-4}$~\cite{Aliberti:2025beg,Muong-2:2025xyk}\\
				\cline{2-2}
                \hline 
                $[C_{\ell q}^{(1)}]_{22,33}/\Lambda^2$ &  $pp \to t\bar{t}\ell^+\ell^-$  & $[-4.49,4.91]$~\cite{CMS:2023xyc}\\\rowcolor{gray!20}
				$[C^{(1)}_{\ell q}]_{32}/\Lambda^2$& $b\to s \mu \mu$ &$[-2.78,-5.18] \times 10^{-4}$\cite{Ray:2023xjn} \\ \rowcolor{gray!20}
                & ${\cal B}(B_s^0\to \mu^+ \mu^-)$ &$[-6.94,-7.79] \times 10^{-4}$ \\
				\hline
                $[C_{\ell q}^{(3)}]_{22,33}/\Lambda^2$ &  $pp \to t\bar{t}\ell^+\ell^-$  & $[-2.69,2.58]$~\cite{CMS:2023xyc}\\
             & ${\cal{B}}(t \to b \mu \nu_\mu)$  & $[6.76,7.50] \times 10^{-1}$  \\
             & ${\cal{B}}(D_s^\pm \to \mu \nu_\mu)$  & $[-3.92,15.96] \times 10^{-3}$ \\\rowcolor{gray!20}
            $[C^{(3)}_{\ell q}]_{32}/\Lambda^2$& $b\to s \mu \mu$ &$[-2.78,-5.18] \times 10^{-4}$\\\rowcolor{gray!20}
                & $ B \to D^{(*)}\mu \nu$ &$[0.50,2.30] \times 10^{-2}$ \\
            \hline
                $[C_{\ell u}]_{22,33}/\Lambda^2$ & $pp \to t\bar{t}\ell^+\ell^-$  & $[-2.02,2.20]$~\cite{CMS:2023xyc} \\
                $[C_{e q}]_{22,33}/\Lambda^2$ &  $pp \to t\bar{t}\ell^+\ell^-$ & $[-2.04,2.12]$ \\\rowcolor{gray!20}
				$[C_{eq}]_{31}/\Lambda^2$& $b\to d \mu \mu$ &$[-2.11,3.40] \times 10^{-4}$\\ \rowcolor{gray!20}
				$[C_{eq}]_{32}/\Lambda^2$& $b\to s \mu \mu$ &$[-0.98,-1.50] \times 10^{-3}$\\ \rowcolor{gray!20}
                & ${\cal B}(B_s^0\to \mu^+ \mu^-)$ &$[-6.94,-7.80] \times 10^{-4}$ \\
                \hline
		\end{tabular}}}
		\caption{  Summary of constraints on dimension-6 operator Wilson coefficients from various flavour conserving processes and observables. Constraints from flavour-violating processes are shown in grey-shaded entries. }
		\label{other_constraints_flavour_universal_1}}
\end{table} 

\begin{table}[t]
	{\renewcommand{\arraystretch}{1.2}%
		\centering
		\scalebox{0.75}{
\resizebox{\textwidth}{!}{%
			\begin{tabular}{ |c|c|c| } 
				\hline
				Wilson coefficient & Related processes/Observables & Constraints ($\rm{TeV^{-2}}$)\\ 
				\hline \hline               
                $[C_{e u}]_{22,33}/\Lambda^2$ &  $pp \to t\bar{t}\ell^+\ell^-$  & $[-1.91,2.39] $\\
                $[C^{S}_{\ell e qu}]_{22,33}/\Lambda^2$ & $pp \to t\bar{t}\ell^+\ell^-$ & $[-2.80,2.80]$~\cite{CMS:2023xyc} \\\rowcolor{gray!20}
            $[C^{S}_{\ell equ}]_{11}/\Lambda^2$ & ${\cal{B}}(\pi^\pm \to \mu \nu_\mu)$  & $-[5.51,6.13] \times 10^{-4}$ \\\rowcolor{gray!20}
            $[C^{S}_{\ell equ}]_{22}/\Lambda^2$ &${\cal{B}}(D_s^\pm \to \mu \nu_\mu)$  & $-[1.19,7.53] \times 10^{-4}$ \\
            \hline
                $[C^{T}_{\ell e qu}]_{22,33}/\Lambda^2$ & $pp \to t\bar{t}\ell^+\ell^-$ & $[-0.40,0.40]$ \\
		$[C^{T}_{\ell  e q u}]_{22,33}/\Lambda^2$ & $t \to b \mu \nu_\mu$ & $[-68.84,68.61]$~\cite{ParticleDataGroup:2024cfk}   \\
           $C^{T}_{\mu t}/\Lambda^2 $ &$(g-2)_\mu$ &   $\left[-4.45, 2.37 \right] \times 10^{-5}$\\\rowcolor{gray!20}
		$[C^{T}_{\ell  e q u}]_{32}/\Lambda^2$& $B\to D^{(*)} \mu \nu_{\mu}$ &$[-6.94,6.28] \times 10^{-4}$\cite{Ray:2023xjn} \\
		\hline
		\end{tabular}}}
		\caption{Summary of constraints on  dimension-6 operator Wilson coefficients from various flavour-conserving processes and observables.  Constraints from flavour-violating processes are shown in grey-shaded entries. }
		\label{other_constraints_flavour_universal_2}}

\end{table} 
\begin{itemize}[leftmargin=*, labelsep=0.5em]
\item Constraints on two-muon--two-quark operators
\end{itemize}    
\noindent
\underline{Associated top-quark production with a dimuon pair at the LHC:}
The process $pp \to t\bar{t}\mu^+\mu^-$ directly probes four-fermion operators involving two muons and two top quarks. However, current LHC measurements~\cite{CMS:2023xyc} remain statistically limited and thus, yield only weak constraints on these operators, as summarised in Tables~\ref{other_constraints_flavour_universal_1} and~\ref{other_constraints_flavour_universal_2}.

\vspace{2mm}

\noindent
\underline{$t \to b \mu \nu_\mu$ decay}:  Semileptonic top-quark decays provide an important probe of new physics. In the SM, $t \to b \mu \nu_{\mu}$ proceeds via a $W$-mediated charged-current interaction, whereas beyond-SM effects can arise from four-fermion operators $\mathcal{O}^{(3)}_{\ell q}$, $\mathcal{O}^{S}_{\ell equ}$, and $\mathcal{O}^{T}_{\ell equ}$. Using the current $2\sigma$ uncertainty on the measured top-quark width~\cite{ParticleDataGroup:2024cfk}, $C^{(3)}_{\ell q}$, $C^{S}_{\ell equ}$, and $C^{T}_{\ell equ}$ are loosely constrained, as summarised in Tables~\ref{other_constraints_flavour_universal_1} and~\ref{other_constraints_flavour_universal_2}.

\vspace{2mm}

\noindent
\underline{Flavour-changing charged-current (FCCC)  processes $b \to q \ell \nu_\ell$:}
FCCC processes such as $b \to q \ell \nu$ have been  extensively studied~\cite{Bhattacharya:2023beo,Huang:2018nnq,Ray:2023xjn}. We incorporate the constraints derived in~\cite{Ray:2023xjn}, which use experimental inputs from $\bar{B} \to D(D^\ast)\ell^-\bar{\nu}_{\ell}$, $R(D^{(\ast)})$, and $\bar{B} \to \pi \ell^- \bar{\nu}_{\ell}$ decays for $\ell=\mu,~e$\footnote{For the FCCC constraints shown in grey–shaded entries in Table~\ref{other_constraints_flavour_universal_1}, experimental inputs correspond to heavy-meson branching ratios. The corresponding bounds on SMEFT WCs are obtained by first extracting the low-energy effective coefficients using decay constants and form factors, and then matching them onto the SMEFT framework~\cite{Ray:2023xjn}.}. From analyses of $b \to c \mu \nu_{\mu}$ and $b \to u \mu \nu_{\mu}$ channels, we obtain the constraint $[C^{T}_{\ell e q u}]_{32}/\Lambda^2 \in [-6.94,6.28] \times 10^{-4}$~TeV$^{-2}$.

\noindent
Constraints from the above observables impose only mild bounds on the relevant dimension-6 operators, particularly for flavour-conserving interactions while flavour-violating observables place very strong constraints on off-diagonal flavour structures. Irreducible SM backgrounds further limit the achievable precision. Even with conservative assumptions, a future $e^+e^-$ collider up to $3$ TeV would have limited sensitivity, particularly to operators that do not interfere with the SM~\cite{deBlas:2016ojx}. This leaves a gap between existing indirect constraints and the energy-enhanced sensitivity required to probe muon–top contact interactions. Motivated by these challenges, we investigate the potential of a high-energy muon collider~\cite{Black:2022cth} to probe such operators.
\section{Muon Collider Analysis}
\label{section4}
So far, our discussion has focused on constraining the Wilson coefficients of operators involving muons, the Higgs boson, and top quarks using existing experimental data. In this section, we explore the possibility of directly probing these couplings at a high-energy muon collider. We consider final states involving a Higgs boson, top quarks, or both. For single-Higgs production, we analyse the $Zh$, $Z$ boson fusion, and $t\bar{t}h$ channels, with the latter being particularly sensitive to dimension-6 operators associated with top-quark interactions. In addition, we investigate top-quark pair production as an independent probe. The effect of tensor-type operators (two-fermionic dipole and four-fermionic tensor operators) on $Zh$ and $t\bar{t}$ processes, has been discussed briefly in Refs.~\cite{Buttazzo:2020ibd,Paradisi:2022vqp,Yin:2020afe}. We will consider the {\it{ZBF}} and the associated $t\bar{t} h$ production along with the former two processes, as shown in Fig.~\ref{feyn_diag_coll} (a)-(d). These channels are analysed in presence of a subset of additional higher-dimensional operators involving vector currents and other four-fermion operators, which remain unexplored. Operators involving the top quark are essentially inaccessible to low-energy flavour experiments. A high-energy muon collider uniquely overcomes this limitation, as the $t\bar{t}$ and $t\bar{t}h$ channels provide direct sensitivity to muon-top interactions. The $\mu\mu tt$ contact terms induced by the operators are most directly probed in such collider environments. 
The relevant operators fall mainly into two classes. The $\psi^2Vh$-type operators modify the rates of  $\mu^+\mu^- \to Zh$ and $\mu Z \to \mu h$ processes while $\mu^+\mu^- \to t\bar t$ and $\mu^+\mu^- \to t \bar th$ processes are altered by four-fermion operators.

\begin{table}[b]
	\centering
    \resizebox{\textwidth}{!}{
	\renewcommand{\arraystretch}{1.2}
	\begin{tabular}{|c|c|c|c|c|}
		\hline
		Signal &$\mathcal{O}$& Interference with SM &Scaling & Coupling (TeV$^{-2}$)  \\ \hline \hline
		BP1 &  $\psi^2 \varphi^2 D$ &\cmark & $\mathcal{O}(v^2 /\Lambda^2)$ &\rm $C_{\varphi \ell}^{(1)}/\Lambda^{2}=C_{\varphi \ell}^{(3)}/\Lambda^{2} =  1 \times 10^{-3}, C_{\varphi  e}/\Lambda^{2} = 2 \times 10^{-3}$\\ \hline
		BP2 & $\psi^2 \varphi^2 D$&\cmark &$\mathcal{O}(v^2 /\Lambda^2)$ & \rm $C_{\varphi \ell}^{(1)}/\Lambda^{2}=0, C_{\varphi \ell}^{(3)}/\Lambda^{2} =  1 \times 10^{-3}, C_{\varphi  e}/\Lambda^{2} = 1 \times 10^{-3}$  \\ \hline
		BP3 & $\psi^2 X \varphi$ & \xmark &$\mathcal{O}(v^2 s/\Lambda^4)$ & \rm $C_{\mu B}/\Lambda^{2} = 5.3 \times 10^{-4},C_{\mu W}/\Lambda^{2} = 1 \times 10^{-3}$\\ \hline
		BP4 & $\psi^4$&\cmark&$\mathcal{O}(s /\Lambda^2)$ & \rm $C_{\ell q}^{(1)}/\Lambda^{2} = 1 \times 10^{-3} , C_{\ell q}^{(3)}/\Lambda^{2} =2\times 10^{-3}  $ \\ \hline
		BP5 & $\psi^4$ & \cmark&$\mathcal{O}(s /\Lambda^2)$ & \rm $C_{eu}/\Lambda^{2} = 2 \times 10^{-3}, C_{\ell u}/\Lambda^{2}= -2 \times 10^{-3} $\\ \hline
		BP6 & $\psi^4$ &\xmark&$\mathcal{O}(s^2 /\Lambda^4)$ & \rm $C_{\mu t}^S/\Lambda^{2} = -5 \times 10^{-4}$ \\ \hline
		BP7 & $\psi^4$ & \xmark&$\mathcal{O}(s^2 /\Lambda^4)$ & \rm $C_{\mu t}^T/\Lambda^{2} = 5 \times 10^{-4}$ \\ \hline
	\end{tabular}}
	\caption{Summary of the benchmark points considered in this study, their operator classes and scaling behaviour with the corresponding values of Wilson coefficients. The third column indicates whether the respective operator-type  interferes with the SM amplitude (\cmark) or not (\xmark).}
	\label{BM_points}
\end{table}
To illustrate the impact of these operators on both total and differential rates, we select benchmark values for the WCs, listed in Table~\ref{BM_points}. These benchmark points (BPs) satisfy the constraints in Tables~\ref{other_constraints_flavour_universal_1} and~\ref{other_constraints_flavour_universal_2}. BP1–BP3 are used for the $Zh$ and $\mu^+\mu^-h$ channels, while BP4–BP7 correspond to $t\bar{t}$ and $t\bar{t}h$ production. Together, these benchmark scenarios span different interference patterns, symmetry structures, and energy scalings, allowing a comprehensive exploration of SMEFT effects.
\begin{itemize}[leftmargin=*, labelsep=0.5em]
\item BP1 and BP2 involve leptonic current operators $C_{\varphi \ell}^{(1)}, C_{\varphi \ell}^{(3)}$, and $C_{\varphi e}$, which modify muon couplings to electroweak bosons. BP1 alters $W$ and $Z$ couplings uniformly, whereas BP2 induces non-universal modifications of the electroweak couplings.
\item BP3 lies on the photophobic line $C_{\mu B} = {\rm tan}{\theta_W}~ C_{\mu W}$, where $g_{\mu\mu\gamma}=0$, $g_{\mu\mu Z}\neq0$.
\item BP4 and BP5 test vectorial four-fermion contact interactions.
\item BP6 and BP7 probe scalar and tensor four-fermion operators, which do not interfere with the SM but have energy-growing contributions.
\end{itemize}
We use {\tt FeynRules}~\cite{Alloul:2013bka} to implement the effective Lagrangian and generate the UFO model file. We interface this with {\tt Madgraph5\_aMC@NLO}~\cite{Alwall:2014hca} to generate signal and background events at leading order. We use {\tt Pythia8}~\cite{Sjostrand:2014zea} for parton showering and simulate detector effects with {\tt Delphes-v3.5.0}~\cite{deFavereau:2013fsa}, using a muon-collider detector card\footnote{The performances encoded in the {\tt{delphes\_card\_MuonColliderDet.tcl}} card represent a benchmark parametrisation rather than a finalised detector design. The dependence of our results on the dominant detector parameters is discussed in section~\ref{section5}.}. We reconstruct jets using the anti-$k_t$ algorithm implemented in {\tt FastJet}-v3.3.2~\cite{Cacciari:2011ma}. In this study, we consider muon-collider setups with c.o.m energies $\sqrt{s} = [3, 10, 14, 30]$~TeV. The corresponding integrated luminosities are $[1, 10, 20, 90]$~ab$^{-1}$~\cite{InternationalMuonCollider:2025sys}. The increase in luminosity with energy compensates for the suppression of $s$-channel processes, ensuring sufficient event yields at high energies~\cite{Delahaye:2019omf}. We are now all set to analyse the kinematic features that distinguish SMEFT signals from SM backgrounds in the aforementioned processes.
\subsection{\textit{Zh} Production channel}
\label{subsec:_Zhprod}
\begin{figure}[t]
	\centering
	\includegraphics[width=7.7cm,height=5.2cm]{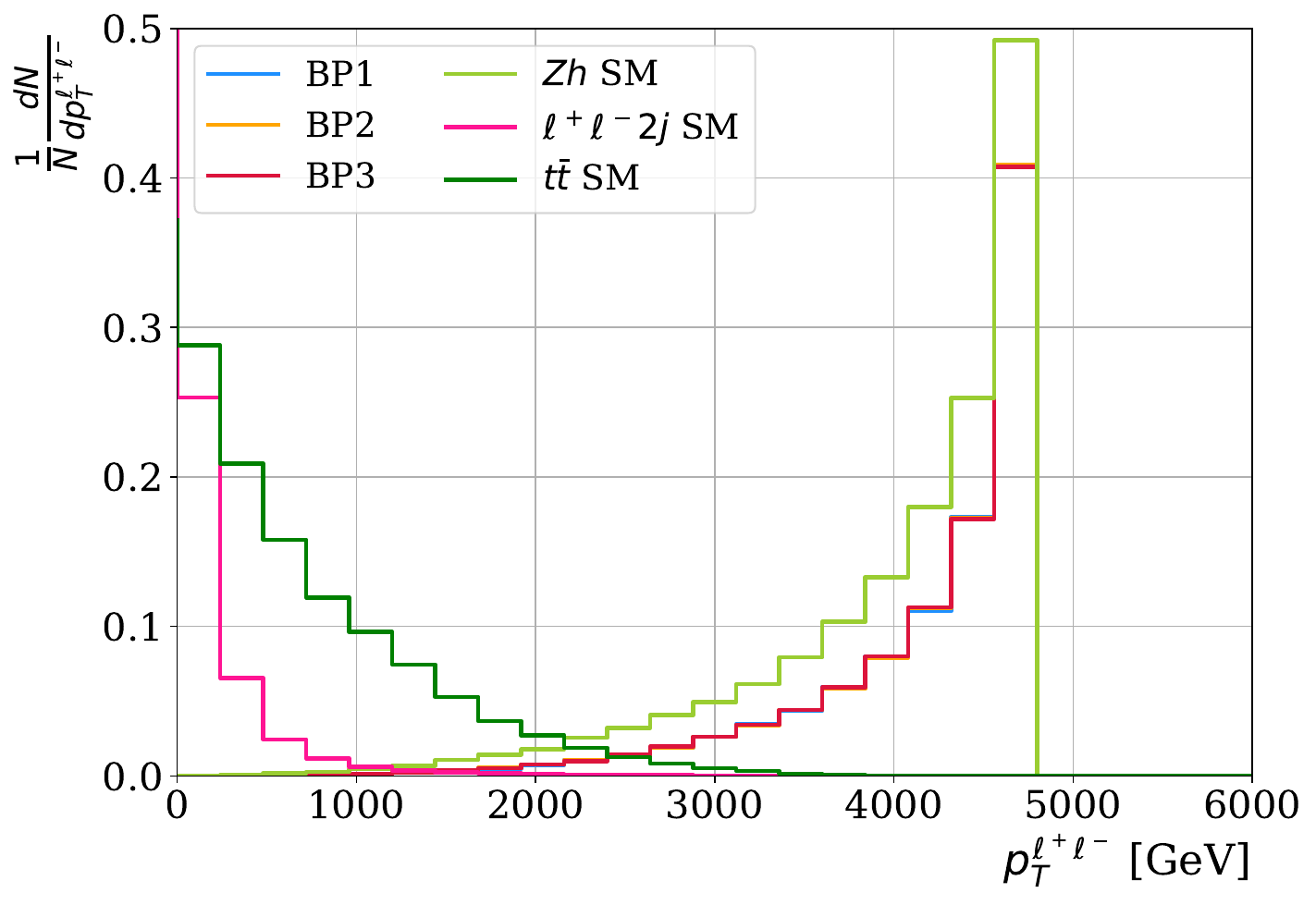}~
	\includegraphics[width=7.7cm,height=5.2cm]{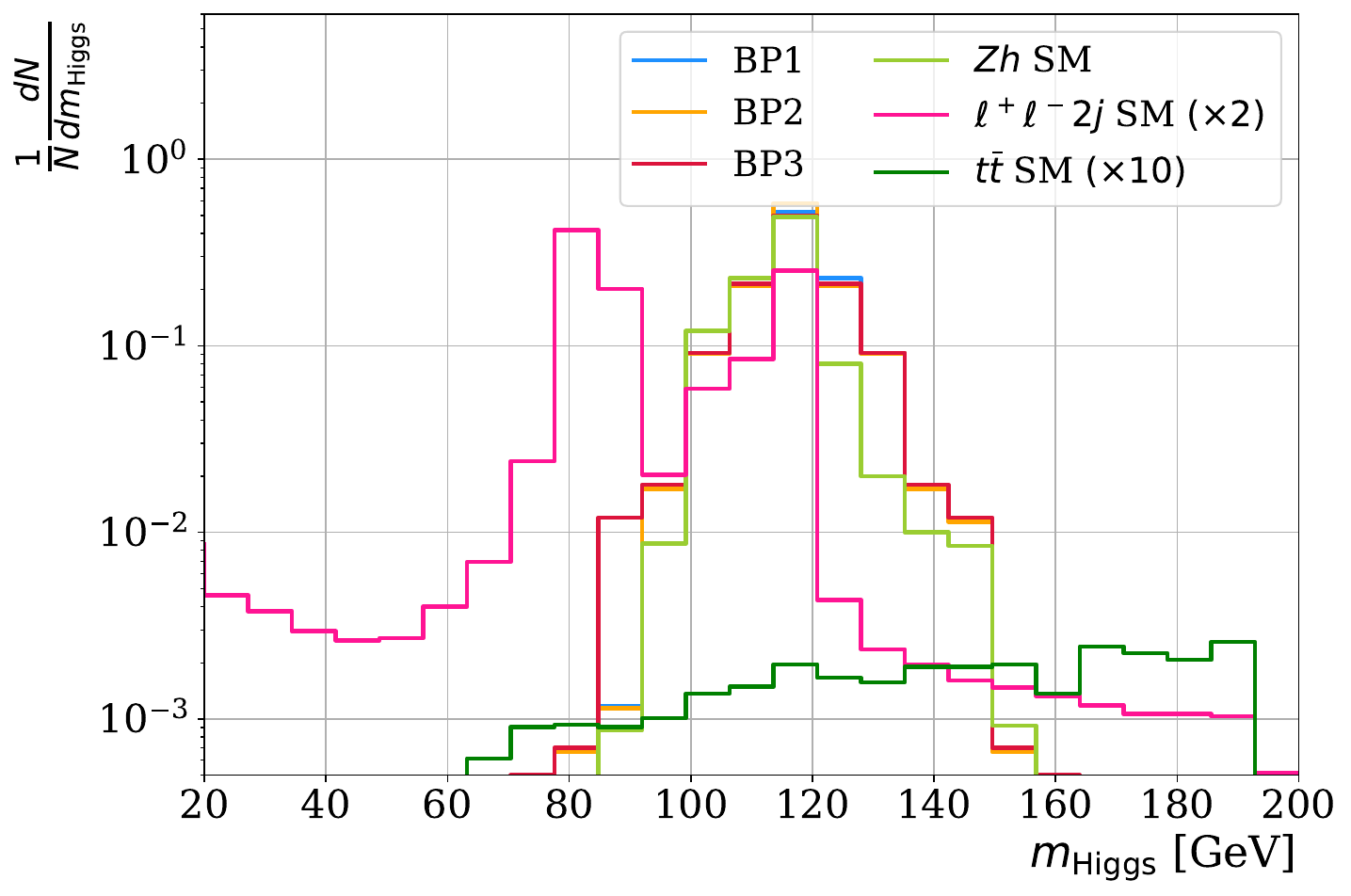}
	\caption{Normalised distribution of $p_T$ of  dilepton (left) and reconstructed Higgs mass  for $\mu^+\mu^- \to Zh$ in $\ell^+\ell^- b\bar{b}$ final state  for signal and background processes at a muon collider with $\sqrt{s}=10$ TeV. The signal is driven by SMEFT couplings  BP1,  BP2 and BP3 respectively as summarised in Table~\ref{BM_points}.}
		\label{fig:Zh_selection}
\end{figure}

We perform a cut-and-count analysis for the process $\mu^+\mu^- \to Zh$ to isolate the impact of  two-muon operators in the high-energy regime. In the SM, Higgs-strahlung  dominates at low energies. We decay the Higgs in the $b\bar{b}$ final state and the Z boson leptonically. We focus on leptonic $Z$-boson decays, which provide a clean signature through an opposite-sign same-flavour lepton pair. The hadronic $Z \to q\bar{q}$ mode has a larger branching fraction and could improve the statistical reach, but it would require a separate optimisation of jet reconstruction, flavour tagging, and detector-level effects. Here, we focus on the leptonic channel.  The selection criteria, varying with $\sqrt{s}$, are listed in the first two columns of Table~\ref{Higgsproduction_cuts}. 
The Higgs decaying to two b-jets is reconstructed using a fat jet of radius $R = 1$. A stringent mass window around Higgs peak is imposed to suppress the backgrounds.
Fig.~\ref{fig:Zh_selection} presents the transverse momentum distribution $p_{T}^{\ell^+\ell^-}$ (left) of the dilepton system and the reconstructed Higgs mass distribution (right) at a 10 TeV muon collider. The $p_{T}^{\ell^+ \ell^-}$ distribution exhibits a similar shape for the signal for the benchmarks BP1, BP2 and BP3 and the SM $Zh$ background, but with an overall enhancement in the presence of SMEFT interactions. This is due to the contact four-point interaction and presence of momentum dependence in the vertex. The cross sections scale as: 
\begin{equation}
    |\mathcal{A}_{\rm SM}|^2 \sim g_{\rm SM}^4 \frac{v^2}{s} ~, \quad {\rm Re}(\mathcal{A}_{\rm SM} \mathcal{A}_{\rm D6}^\ast) \sim g_{\rm SM}^2 c_{\rm D6}^{i} \frac{v^2}{\Lambda^2} ~, \quad |\mathcal{A}_{\rm D6}|^2 \sim (c_{\rm D6}^{i})^2 \frac{v^2 s}{\Lambda^4} ~.
    \label{zh_amp_scaling}
\end{equation} 
\begin{table}[b]
\centering
\scalebox{0.84}{
\resizebox{\textwidth}{!}{%
\begin{tabular}{|l|l||l|l|}
			\hline
			\multicolumn{2}{|c||}{(a) Zh channel }&\multicolumn{2}{|c|}{(b) {\it{ZBF}} channel} \\
			\hline
			Observable & Selection &Observable & Selection \\
			\hline \hline
			Exactly one pair of OSSF leptons & &$y_{\mu^+} y_{\mu^-}$ & $<0$ \\
			$m_{\ell^+ \ell^-}$  & $\in [86,96]$ GeV &	$|\Delta y_{\mu^+ \mu^-}|$  & $\in [3.5,6]$\\
			$p_{T}^{\ell^+ \ell^-}$& $\in [\frac{4}{10},\frac{1}{2}] \sqrt{s}$&$p_{T}^{\mu^+ \mu^-}$& $\in [\frac{3}{5},\frac{14}{15}] \sqrt{s}$ \\
			Reconstructed Higgs mass & $\in [115,135]$ GeV &$m_{\mu^+ \mu^-}$  & $\in [\frac{1}{20},\frac{4}{15}] \sqrt{s}$ \\
            &&$\Delta \phi_{\mu^+ \mu^-}$ & $\in[0.6,2.8]$\\
			&&Reconstructed Higgs mass& $\in [115,135]$ GeV\\
			\hline
\end{tabular}}}
\caption{Selection criteria for $Zh$ production in $\ell^+ \ell^- b\bar{b}$ final state and {\it{ZBF}} Higgs production channel in $\mu^+ \mu^-  b\bar{b} $ final state.}
\label{Higgsproduction_cuts}
\end{table}
Fig.~\ref{fig:Zh_selection} (right) shows the reconstructed Higgs mass distribution.
The signal processes exhibit a clear peak in the range $110\text{-}135~\mathrm{GeV}$, corresponding to the Higgs boson. In contrast, most background events peak around $90~\mathrm{GeV}$ due to jets reconstructing a $Z$ boson, or around $165\text{-}185~\mathrm{GeV}$ due to top quarks. Applying the selection $m_J \in [115,135]~\mathrm{GeV}$ significantly reduces the $t\bar{t}$ background. Table~\ref{Zhrates} summarises the effect of the selection criteria on the background cross sections at different c.o.m energies. Since the signal yield grows more rapidly than the background, high-energy muon colliders with $\sqrt{s} \ge 10$ TeV and high luminosity can substantially improve the sensitivity to dimension-6 operators in the $Zh$ channel.
\begin{table}[t]\scriptsize
	\centering
	\renewcommand{\arraystretch}{1.2}
	\setlength{\arrayrulewidth}{.3mm}
	\setlength{\tabcolsep}{0.1 em}
\scalebox{0.9}{
\resizebox{\textwidth}{!}{%
	\begin{tabular}{|c|c||c|c||c|c||c|c||c|c|}
		\hline
		&{Process} & 
		\multicolumn{2}{c||}{$3 ~{\rm TeV}~/~ 1~{\rm ab}^{-1}$} &\multicolumn{2}{c||}{$10~{\rm TeV}~/~ 10~{\rm ab}^{-1}$} &\multicolumn{2}{c||}{$14~{\rm TeV}~/~ 20~{\rm ab}^{-1}$} & 	\multicolumn{2}{c|}{$30~{\rm TeV}~/~ 90~{\rm ab}^{-1}$} \\ \cline{3-10}
		&at [$\sqrt{s} ~/~\mathcal{L}~$]& $\sigma~(ab)$ &  N& $\sigma~(ab)$ & N& $\sigma~(ab)$ &N& $\sigma~(ab)$ &N \\ \hline \hline  
		BP1 &$h Z; \ h\rightarrow b\bar{b}, \ Z \rightarrow l^+ l^- $ & $1.23 \times 10^{2}$  & $101$ & $1.04 \times 10^{2}$   &$875$ &$1.67 \times 10^{2}$ & $1151$  & $6.52 \times 10^{2}$  & $49158$  \\
        BP2 &$h Z; \ h\rightarrow b\bar{b}, \ Z \rightarrow l^+ l^- $ &  $1.19 \times 10^{2}$ & $97$ &  $5.21 \times 10^{1}$ & $434$ &$6.48\times 10^{1}$&$1087$ &$1.83 \times 10^{2}$  & $14043$ \\
        BP3 &$h Z; \ h\rightarrow b\bar{b}, \ Z \rightarrow l^+ l^- $ &  $1.18 \times 10^{2}$ & $97$ & $4.47 \times 10^{1}$  & $368$ & $5.01 \times 10^{1}$& $830$ & $1.16 \times 10^{2}$ & $8768$ \\ \hline
		Bkgds &$h Z; \ h\rightarrow b\bar{b}, \ Z \rightarrow l^+ l^- $  & $1.17\times 10^{2}$& $96$ &$3.49 \times 10^{1}$&$286$&$3.09 \times 10^{1}$&$508$& $2.77 \times 10^{1}$ &$2041$ \\
		&$ l^+ l^- b \bar{b}$ &$6.3\times 10^{4}$ & $3$ & $4.97\times 10^{4}$ &$3$  &$3.41\times 10^{4}$  &  $3$&$1.08 \times 10^{4}$ &$5$ \\
		&$ l^+ l^- j j$  & $7.28 \times 10^{4}$ & $0$  &$5.10 \times 10^{4}$  & $0$ &  $3.89 \times 10^{4}$& $0$  &$1.82 \times 10^{4}$& $0$ \\
		&$ t\bar{t} \to  l^+ l^- 2b+MET$  & $1.84\times 10^{4}$ &  $0$ &$1.66 \times 10^{3}$  & $0$ &  $8.48 \times 10^{2}$&  $0$&$1.85\times 10^{2}$& $0$ \\
		\hline
	\end{tabular}}}
	\caption{Cross sections ($\sigma$) for signal and backgrounds for $Zh$ production channel before applying the event selection cuts and $N$ denotes the number of events after the cuts. The signal is driven by SMEFT couplings  BP1,  BP2 and BP3 respectively as summarised in Table~\ref{BM_points}.}
	\label{Zhrates}
\end{table}
\subsection{$Z$ boson fusion to Higgs channel}
\label{subsec:_ZBFprod}
We next consider Higgs production via the \textit{ZBF} process, $\mu^+\mu^- \to \mu^+\mu^-h$, which proceeds through $t$-channel topology and exhibits a cross section that grows with energy. At low energies, it interferes with the $\mu^+\mu^- \to \ell^+\ell^-h$ via \textit{Zh} channel, making their separation difficult because the available phase space is small. The leptons cannot be produced with large boosts, so the final-state kinematics of {\it{ZBF}} closely resemble those of Higgs-strahlung. While \textit{Zh} involves an on-shell $Z$ boson, \textit{ZBF} is instead characterized by a highly energetic $\mu^+\mu^-$ pair in the forward and backward directions. The \textit{ZBF} cross section eventually dominates over \textit{Zh} production beyond $\sqrt{s}\sim 850$ GeV~\cite{Forslund:2022xjq,InternationalMuonCollider:2024jyv}. 
We find for the {\it{ZBF}} process, the energy-scaling of the subprocess $\mu Z \to \mu h$ goes as:
\begin{equation}
    |\mathcal{A}_{\rm SM}|^2 \sim g_{\rm SM}^6 \frac{v^2}{t} ~, \quad {\rm Re}(\mathcal{A}_{\rm SM} \mathcal{A}_{\rm D6}^\ast) \sim g_{\rm SM}^3 c_{\rm D6}^{i} \frac{v^2 }{ \Lambda^2} ~, \quad |\mathcal{A}_{\rm D6}|^2 \sim (c_{\rm D6}^{i})^2 \frac{v^2t}{\Lambda^4} ~.
    \label{mumuh_amp_scaling}
\end{equation}
We take the Higgs decaying to two b-jets and perform a cut-based analysis on final-state objects after detector simulation. 
\begin{figure}[b]
	\centering
	\includegraphics[width=7.7cm,height=5.2cm]{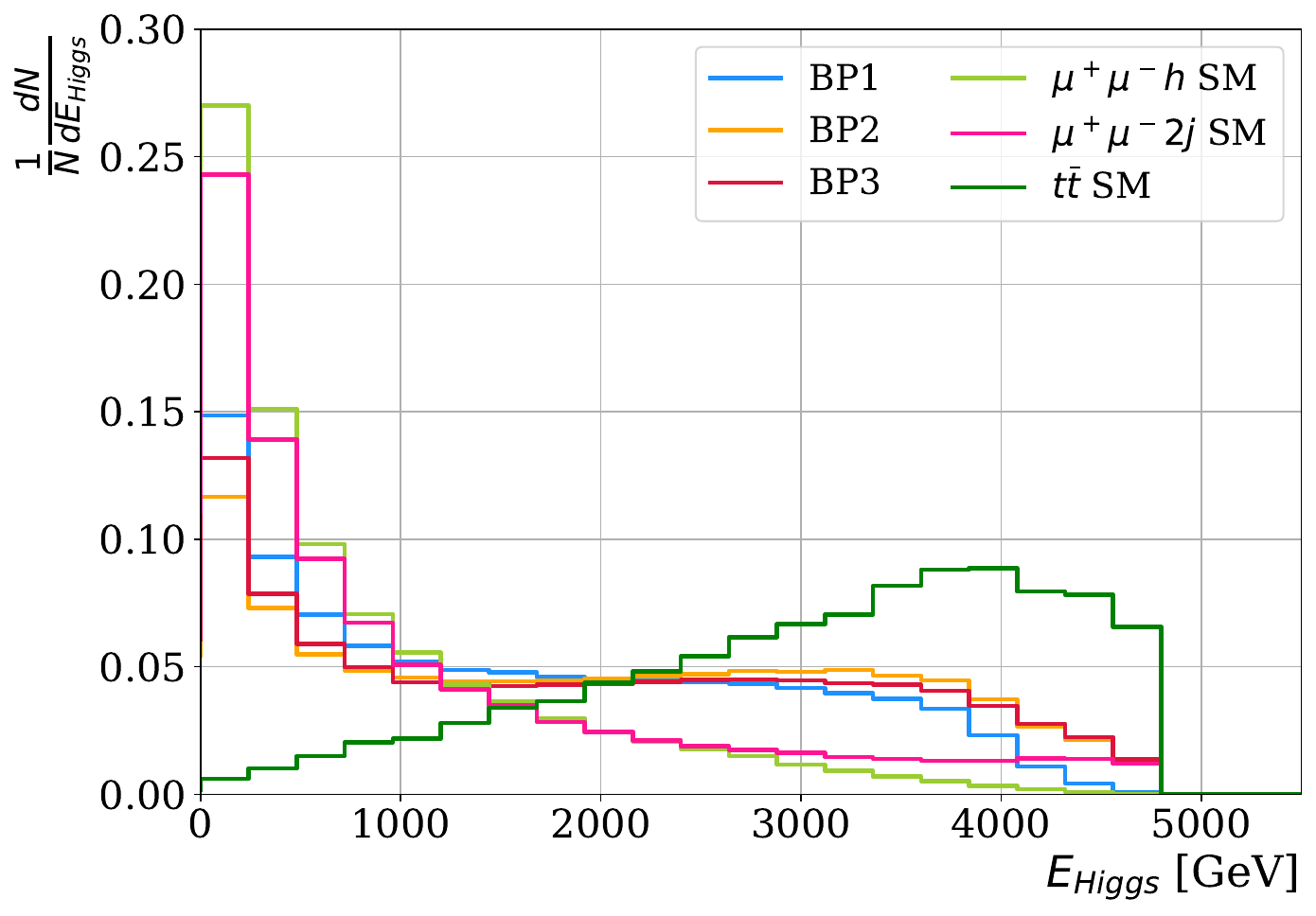}~
	\includegraphics[width=7.7cm,height=5.2cm]{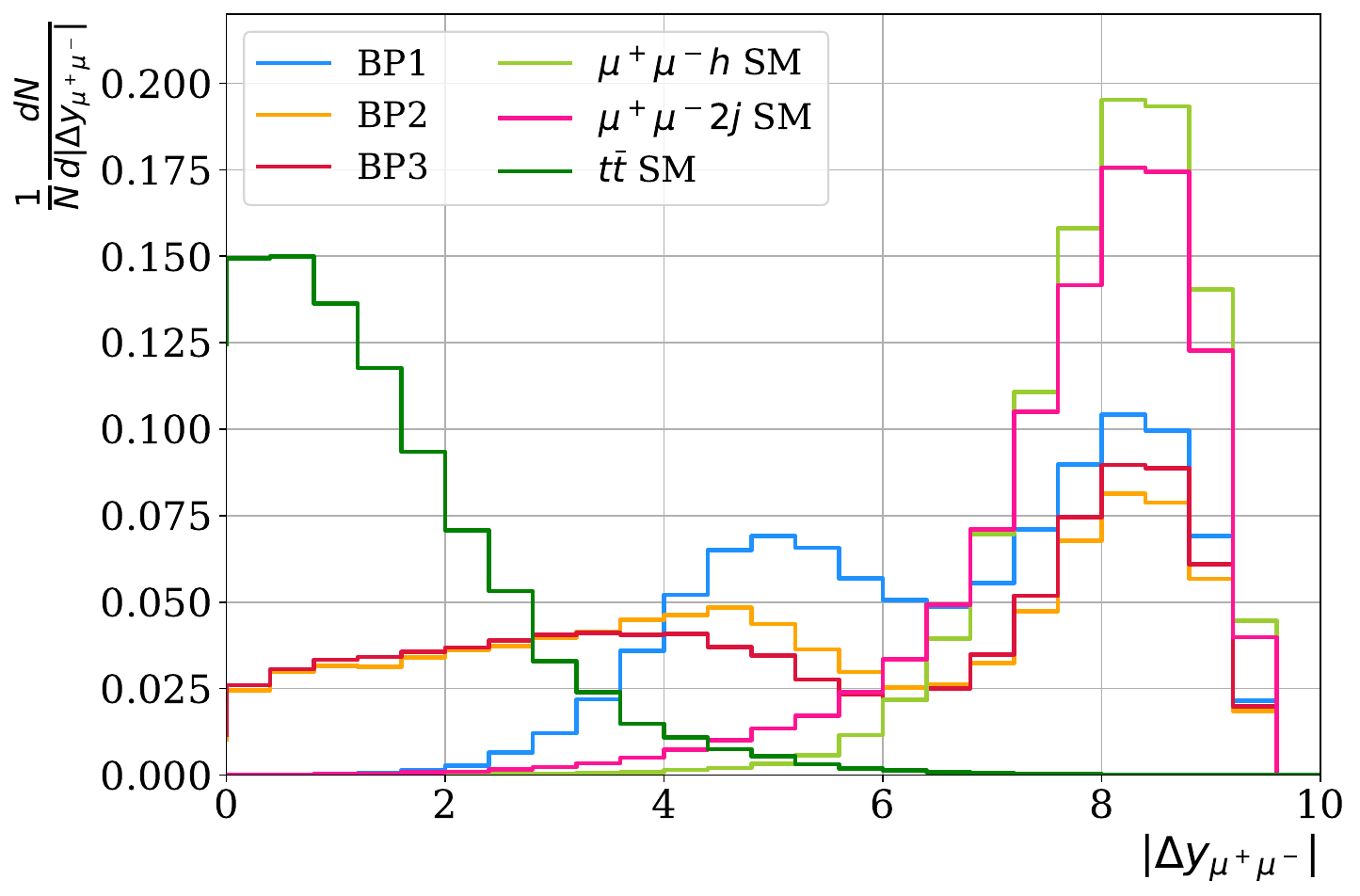}
	\caption{  Normalised event distribution for $E_{\rm Higgs}$ and $|\Delta y_{\mu^+\mu^-}|$ containing signal and background processes at a muon collider with $\sqrt{s}=10$ TeV , $\mathcal{L}_{\rm int}=$ 10 $\rm{ab}^{-1}$. The signal is driven by SMEFT couplings  BP1,  BP2 and BP3 respectively as summarised in Table~\ref{BM_points}.
		\label{fig:ZBF_selection}}
\end{figure}
\begin{table}[t]\scriptsize
	\centering
	\renewcommand{\arraystretch}{1.2}
	\setlength{\arrayrulewidth}{.3mm}
	\setlength{\tabcolsep}{0.1 em}
\scalebox{0.9}{
\resizebox{\textwidth}{!}{%
	\begin{tabular}{|c|c||c|c||c|c||c|c||c|c|}
		\hline
		&{Process} & 
		\multicolumn{2}{c||}{$3 ~{\rm TeV}~/~ 1~{\rm ab}^{-1}$} &\multicolumn{2}{c||}{$10~{\rm TeV}~/~ 10~{\rm ab}^{-1}$} &\multicolumn{2}{c||}{$14~{\rm TeV}~/~ 20~{\rm ab}^{-1}$} & 	\multicolumn{2}{c|}{$30~{\rm TeV}~/~ 90~{\rm ab}^{-1}$} \\ \cline{3-10}
		&at [$\sqrt{s} ~/~\mathcal{L}~$]& $\sigma~(ab)$ &  N& $\sigma~(ab)$ & N& $\sigma~(ab)$ &N& $\sigma~(ab)$ &N \\ \hline \hline  
		BP1 &$\mu^+\mu^- h; \ h\rightarrow b\bar{b}$ &  $3.89 \times 10^4$  & $106$   & $3.72 \times 10^4$ & $555$  & $2.36 \times 10^4$ & $599$  & $4.14 \times 10^3$  & $4313$  \\
		BP2 &$\mu^+\mu^- h; \ h\rightarrow b\bar{b}$ &  $3.89 \times 10^4$  & $109$   & $3.73 \times 10^4$ & $600$  & $2.35 \times 10^4$ & $663$  & $4.14 \times 10^3$  & $4322$  \\
		BP3 &$\mu^+\mu^- h; \ h\rightarrow b\bar{b}$ & $3.90 \times 10^4$  & $122$   & $3.72 \times 10^4$ & $537$  & $2.36 \times 10^4$ & $797$  & $4.30 \times 10^3$  & $7926$  \\ \hline
		Bkgds &$\mu^+\mu^- h; \ h\rightarrow b\bar{b}$  &  $3.90 \times 10^{4}$& $94$ &$3.73 \times 10^{4}$& $458$ & $2.35 \times 10^{4}$ &  $522$& $4.14 \times 10^{3}$ &$3830$ \\
		&$\mu^+\mu^- b \bar{b}$  & $6.18 \times 10^{4}$ & $4$ & $5.01\times 10^{4}$ & $32$  &$3.43 \times 10^{4}$  &  $43$&$1.08 \times 10^{4}$ &$61$ \\
		&$\mu^+\mu^- j j$  & $7.28 \times 10^{4}$ &$0$  &$7.17\times 10^{4}$  & $0$ &  $5.10 \times 10^{4}$&$1$  &$1.82 \times 10^{4}$& $1$ \\
            &$ t\bar{t} \to \mu^+ \mu^- 2j 2b$  & $1.84\times 10^{4}$ & $0$  &$4.61\times 10^{3}$  & $0$ &  $1.66 \times 10^{3}$& $0$ &$1.85\times 10^{2}$& $0$   \\
		\hline
	\end{tabular}}}
	\caption{Cross sections  $(\sigma)$ for signal and backgrounds for $\mu^+ \mu^- h$  production channel before applying the event selection cuts and $N$ denotes the number of events after the cuts. The signal is driven by SMEFT couplings  BP1,  BP2 and BP3 respectively as summarised in Table~\ref{BM_points}.}
	\label{tab:mumuhbackgrounds}
\end{table}
The reconstructed Higgs energy ($E_{\text{Higgs}}$) and the rapidity gap between the forward muons ($|\Delta y_{\mu^+\mu^-}|$) at $\sqrt{s}=10$ TeV are shown in Fig.~\ref{fig:ZBF_selection}. The $E_{\text{Higgs}}$ distribution in Fig.~\ref{fig:ZBF_selection} (left) extends up to about 5 TeV, with EFT-induced events populating the high-energy tail. Note that the $t\bar{t}$ background can be reduced significantly by the reconstruction of Higgs mass selection cut. The rapidity gap in Fig.~\ref{fig:ZBF_selection} (right) serves as another discriminating observable. With increasing $\sqrt{s}$, final-state muons become more forward, resulting in larger $|y^\mu|$. While the SM-like {\it{ZBF}} process dominantly yields events with larger rapidity gap, for the signal (BP1, BP2 and BP3) driven by the electroweak  operators, events are populated with relatively smaller rapidity gaps ($\Delta y_{\mu \mu} \le 4$). 
A set of selection cuts applied to suppress background events in this channel is listed in the last two columns of Table~\ref{Higgsproduction_cuts}. The invariant masses $m_{\mu^+\mu^-}$ and $m_{\text{Higgs}}$ are used to reduce the $Zh$ and non-Higgs backgrounds, respectively, with the Higgs resonance $\in [115,135]$ GeV.  To mitigate the effects of ISR, FSR, and beamstrahlung on the shape of the $m_{\text{Higgs}}$ distribution, a cut on transverse momentum of the dimuon system is imposed, $p_T^{\mu^+\mu^-} \in [3/5,14/15]\sqrt{s}$. The effect of these cuts on the dominant SM processes at different energies, $\sqrt{s}\in[3,10,14,30]$  TeV is given in Table~\ref{tab:mumuhbackgrounds}.
\subsection{Top pair production channel}
\label{subsec:_ttbarprod}
\begin{figure}[t]
	\centering
	\includegraphics[width=7.7cm,height=5.2cm]{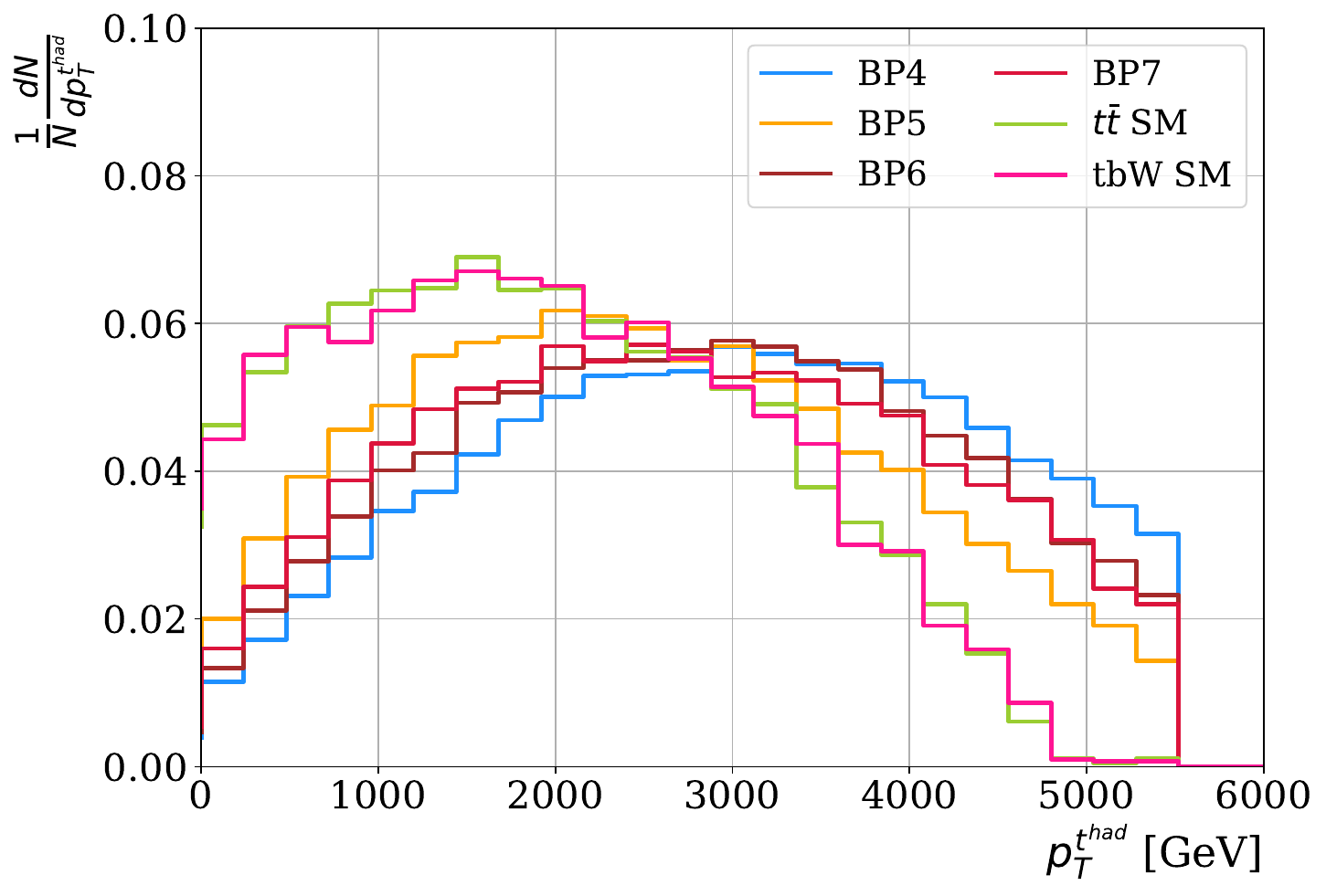}
	\includegraphics[width=7.7cm,height=5.2cm]{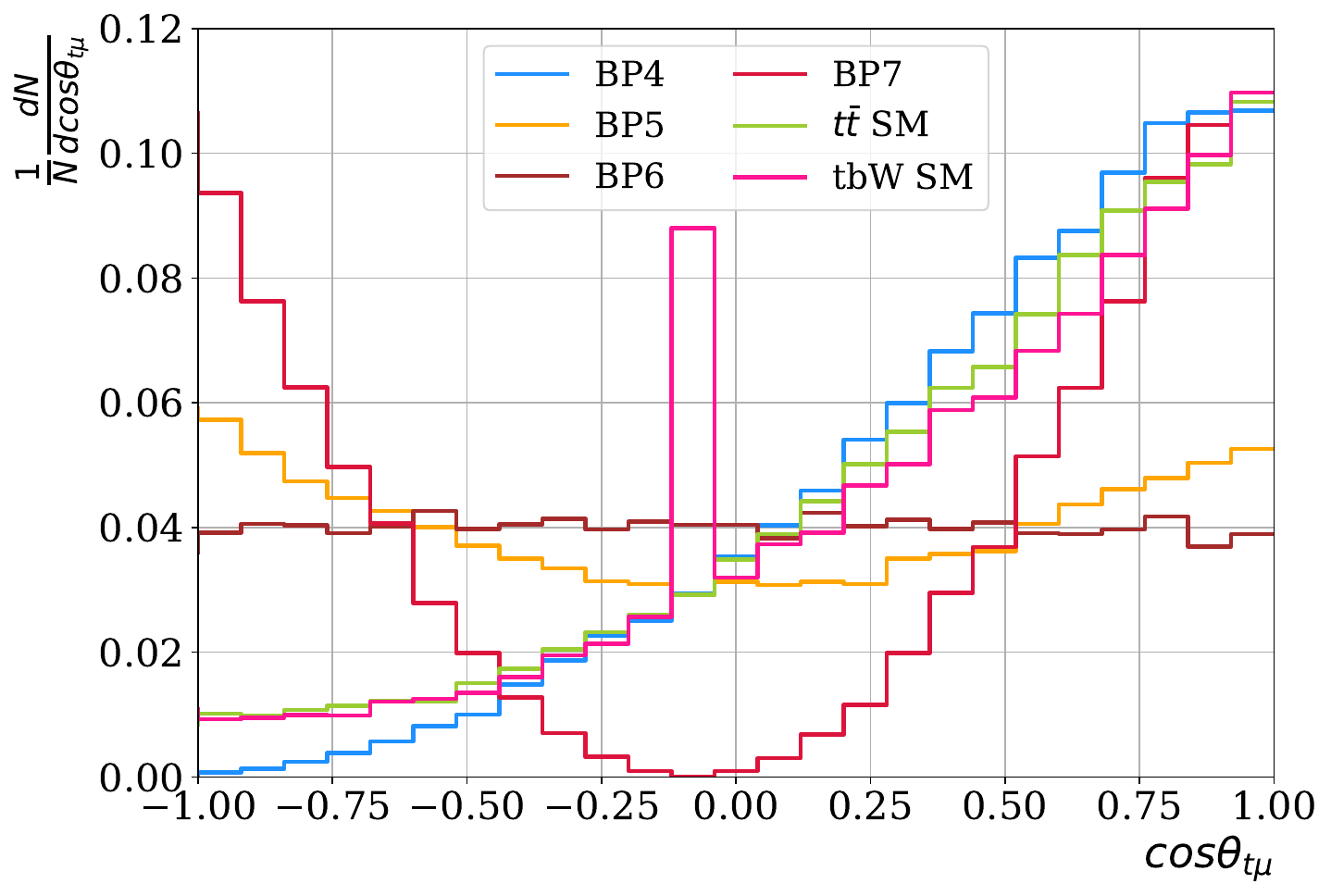}
	\caption{Normalised distributions of $p_{T}^{t^{\rm had}}$ and scattering angle of top with respect to muon beam, cos $\theta_{t \mu}$ for signals and  SM backgrounds for $\mu^+\mu^- \to t \bar{t}$ at $\sqrt{s}=10$ TeV. The signal is driven by SMEFT couplings  BP4,  BP5, BP6 and BP7 respectively as summarised in Table~\ref{BM_points}.}
	\label{ttbar_distributions}
\end{figure}

The three sets of BPs : $\{ \rm (BP4, BP5), BP6, BP7 \}$ (see Table~\ref{BM_points}) correspond to four-fermion interactions with \emph{distinct} Lorentz structures. As discussed earlier, operators with vector currents of fermions interfere with the SM amplitudes. However,  in the limit of vanishing lepton masses, those with scalar or tensor structures do not. The energy scaling for $\mu^+\mu^- \to t\bar{t}$ process due to interfering vector-current interactions follows:
\begin{equation}
    |\mathcal{A}_{\rm SM}|^2 \sim g_{\rm SM}^4 \frac{t^2+u^2}{s^2} ~, \quad {\rm Re}(\mathcal{A}_{\rm SM} \mathcal{A}_{\rm D6}^\ast) \sim g_{\rm SM}^2 c_{\rm D6}^{i} \frac{t^2+u^2}{s\Lambda^2} ~, \quad |\mathcal{A}_{\rm D6}|^2 \sim (c_{\rm D6}^{i})^2 \frac{t^2+u^2}{\Lambda^4} ~.
    \label{tt_amp_scaling}
\end{equation}
\begin{table}[t]
\centering
\scalebox{0.84}{
\resizebox{\textwidth}{!}{%
\begin{tabular}{|l|l||l|l|}
			\hline
			\multicolumn{2}{|c||}{(a) 10 TeV $t\bar{t}$ selection} & \multicolumn{2}{c|}{(b) 10 TeV $t\bar{t}h$ (3 TeV) selection} \\
			\hline
			Observable & Selection & Observable & Selection \\
			\hline \hline
			Lepton isolation & 1 isolated lepton & Lepton isolation & 3 (1) isolated leptons \\
			Exclusive jet clustering & $R=1.5$ & Exclusive jet clustering & $R=1.5$ \\
			Number of light jets & $> 1$ & No. of light jets & At least 2 \\
			$p_T$ cut on jets &  $> 20$ GeV  & $b$-tagging & At least 4  \\
			$b$-tagging  & At least 2  &  $p_T$ cut on jets & $> 20$ GeV  \\
			$m_{\nu\nu}$ & $> 150$ GeV &  $m_{h},m_{t,\bar{t}}$ & $\in [110,140]$ GeV, $\in [140,200]$ GeV   \\
			$m_{t,\bar{t}}$ & $\in [140,200]$ GeV &  $m_{th},m_{tt},m_{\bar{t}h}$ &  $\in [1,8.2]$ TeV ($\in [0.5,2.5]$ TeV)\\
			 $p_{T}^{t,\bar{t}}$  &  $>2$ TeV   &  $|{\rm cos}~\theta_{tt}|$ &  $<0.95$ \\
			$m_{t\bar{t}}$  &  $\in [1.4,9.5]$ TeV  &  $E_{W_{1,2}}$ &  $> 180$ GeV\\
            &&$E_h$  &  $> 900$ GeV  ($> 500$ GeV) \\
            && $m_{\text{tot}}$  &   $\in [0.9,9.0]$ TeV ($\in [0.5,2.5]$ TeV) \\
			\hline
\end{tabular}}}
\caption{Phase space cuts defining the selection criteria for (a) $t\bar{t}$ production at 10 TeV and (b) $t\bar{t}h$ signal regions at 10 TeV (3 TeV) in the $h \to b\bar{b}$ decay channel. These selection cuts help isolate signal events while reducing background contamination.}
\label{tab:tt_ttH_selection}
\end{table}
We consider the semileptonic decay channel for the top pair. The main backgrounds consist of the SM $t\bar{t}$ and the non-resonant $tbW$ production processes.
Fig.~\ref{ttbar_distributions} shows the distribution of the transverse momentum of the hadronically decaying top,  $p_{T}^{t^{\rm had}}$ (left) and the angular variable cos $\theta_{t\mu}$ (right), defined with respect to the muon beam,  at a $10$ TeV muon collider. These observables distinctly characterise signal features over the background.  Signal events exhibit a harder $p_T$ spectrum extending to higher values than the SM backgrounds, making $p_T^{t^{\rm had}} > 2$ TeV an efficient discriminator of new interactions.
	\begin{table}[b]\scriptsize
		\centering
		\renewcommand{\arraystretch}{1.2}
		\setlength{\arrayrulewidth}{.3mm}
		\setlength{\tabcolsep}{0.1 em}
\scalebox{0.9}{
\resizebox{\textwidth}{!}{%
		\begin{tabular}{|c|c||c|c||c|c||c|c||c|c|}
			\hline
		&{Process} & 
		\multicolumn{2}{c||}{$3 ~{\rm TeV}~/~ 1~{\rm ab}^{-1}$} &\multicolumn{2}{c||}{$10~{\rm TeV}~/~ 10~{\rm ab}^{-1}$} &\multicolumn{2}{c||}{$14~{\rm TeV}~/~ 20~{\rm ab}^{-1}$} & 	\multicolumn{2}{c|}{$30~{\rm TeV}~/~ 90~{\rm ab}^{-1}$} \\ \cline{3-10}
		&at [$\sqrt{s} ~/~\mathcal{L}~$]& $\sigma~(ab)$ &  N& $\sigma~(ab)$ & N& $\sigma~(ab)$ &N& $\sigma~(ab)$ &N \\ \hline \hline  
			BP4 &$t\bar{t} $&  $5.37 \times 10^3$ & $1139$ &  $1.08 \times 10^3$ & $2965$ &$1.05 \times 10^3$ & $6816$  &$2.23 \times 10^3$ & $60330$ \\
			BP5 &$t\bar{t} $&  $4.73 \times 10^3$ & $1009$ &  $1.65 \times 10^3$ & $4613$ &$2.91 \times 10^3$ & $20213$  &$1.36 \times 10^4$ & $395336$  \\
			BP6 &$t\bar{t} $&  $4.91 \times 10^3$ & $1042$ &  $5.14 \times 10^2$ & $1406$ &$3.66 \times 10^2$ & $2435$  &$6.98 \times 10^2$ & $18035$  \\
			BP7 &$t\bar{t} $&  $4.94 \times 10^3$ & $1043$ &  $8.29 \times 10^2$ & $2525$ &$9.81 \times 10^2$ & $6361$  &$3.52 \times 10^2$ & $11932$  \\ \hline
			Bkgds & $t\bar{t} $ & $4.91 \times 10^{3}$ &$527$ &   $4.43\times 10^{2}$ & $738$ & $2.26 \times 10^{2}$& $806$& $4.92 \times 10^{1}$ & $775$\\
			 &$t\bar{b}W$& $4.22 \times 10^{4}$ &$1716$ &   $4.88\times 10^{3}$ & $1986$ &$2.68 \times 10^{3}$& $2197$&  $6.92\times 10^{2}$ &$2533$ \\
			&$W^+ W^- Z $ &$5.58 \times 10^{3}$ & $0$&   $1.59\times 10^{3}$ &$0$  & $1.06 \times 10^3$& $0$&  $3.90 \times 10^{2}$ & $0$ \\
			&$W^+ W^- + $ jets  & $7.47 \times 10^3$ & $0$& $1.63 \times 10^3$ & $0$ &$2.09 \times 10^3$&$1$& $2.32 \times 10^3$& $0$ \\
			\hline
		\end{tabular}}}
		\caption{Cross sections $(\sigma)$ for signal and backgrounds for $t \bar{t}$  production channel before applying the event selection cuts and $N$ denotes the number of events after the cuts. The signal is driven by SMEFT couplings  BP4,  BP5, BP6 and BP7 respectively as summarised in Table~\ref{BM_points}.} 
	\label{tab:ttbarbackgrounds}
	\end{table}
For cos $\theta_{t \mu}$ distribution  in Fig.~\ref{ttbar_distributions} (right), the signal events show forward-backward symmetry while the background $t\bar{t}$ process (which proceeds through an $s$–channel $\gamma/Z$ exchange) tends to have a forward-biased production. The non-resonant $tbW$ background follows a topology similar to single-top production, dominated by $t$-channel exchange. In this configuration, the lighter final-state particles ($b$ and $W$) are preferentially emitted in the forward direction along the beam axis. As a result, the top quark recoils against these forward objects and is produced more centrally, giving rise to a distribution that peaks around $\cos\theta_{t\mu} \simeq 0$. The selection criteria for $t\bar{t}$ channel are given in Table~\ref{tab:tt_ttH_selection}. The signal and background events at the different collider setups are given in Table~\ref{tab:ttbarbackgrounds}. It clearly demonstrates that $t\bar{t}$ production is an excellent probe of the considered SMEFT operators at high-energy muon colliders.
\subsection{Associated production of a Higgs boson and a top pair}
\label{subsec:_ttbarhprod}
We now analyse $t\bar{t}h$ production at a muon collider. In the SM, this process receives contributions from both $s$-channel and $VBF$ topologies. The  $s$-channel cross section for $t\bar{t}h$ production decreases with energy, becoming subdominant to VBF $t\bar{t}h$ production above $\sim7$ TeV. Consequently, the kinematic features differ significantly between low and high energies, requiring separate analyses. The selection criteria for the process at 10 TeV (3 TeV) collisions are presented in the last two columns of Table~\ref{tab:tt_ttH_selection} to suppress background contributions. As already mentioned, the SMEFT effects preferentially populate the high invariant mass and high transverse momentum tails.
For both topologies, we consider the semileptonic decay mode of the top pair and Higgs decaying to $b\bar{b}$. This leads to a final state of one isolated lepton, six jets, four of which are $b$-jets and missing energy. 
The SM background contributions arise from the following processes: $\mu^+ \mu^- \to t\bar{t}h$, $t\bar{t}Z$, and $t\bar{t}b\bar{b}$.
	\begin{table}[b]\scriptsize
		\centering
		\renewcommand{\arraystretch}{1.2}
		\setlength{\arrayrulewidth}{.3mm}
		\setlength{\tabcolsep}{0.1 em}
\scalebox{0.9}{
\resizebox{\textwidth}{!}{%
		\begin{tabular}{|c|c||c|c||c|c||c|c||c|c|}
			\hline
		&{Process} & 
		\multicolumn{2}{c||}{$3 ~{\rm TeV}~/~ 1~{\rm ab}^{-1}$} &\multicolumn{2}{c||}{$10~{\rm TeV}~/~ 10~{\rm ab}^{-1}$} &\multicolumn{2}{c||}{$14~{\rm TeV}~/~ 20~{\rm ab}^{-1}$} & 	\multicolumn{2}{c|}{$30~{\rm TeV}~/~ 90~{\rm ab}^{-1}$} \\ \cline{3-10}
		&at [$\sqrt{s} ~/~\mathcal{L}~$]& $\sigma~(ab)$ &  N& $\sigma~(ab)$ & N& $\sigma~(ab)$ &N& $\sigma~(ab)$ &N \\ \hline \hline  
			BP4 &$t\bar{t}h $& $1.16 \times 10^2$  &  $20$& $3.02 \times 10^1$  & $129$ & $3.00 \times 10^1$ &  $216$ &  $6.99 \times 10^1$ &  $2258$ \\
			BP5 &$t\bar{t}h $& $1.06 \times 10^2$  &  $18$& $4.40 \times 10^1$  & $186$ & $7.87 \times 10^1$ &  $553$ &  $4.20 \times 10^2$ &  $13288$ \\
			BP6 &$t\bar{t}h $& $1.08 \times 10^2$  &  $18$& $1.58 \times 10^1$  & $72$ & $1.15 \times 10^1$ &  $82$ &  $2.17 \times 10^1$ &  $696$ \\ 
			BP7 &$t\bar{t}h $& $1.08 \times 10^2$  &  $18$& $2.41\times 10^1$  & $105$ & $2.88 \times 10^1$ &  $205$ &  $1.14 \times 10^1$ &  $3663$  \\\hline
			Bkgds & $t\bar{t} h$ & $6.49 \times 10^{2}$ &$24$ &   $8.59 \times 10^{1}$ &$37$ &$4.79 \times 10^{1}$&$41$ &  $2.06 $ & $48$\\
			&$t\bar{t}Z $ & $4.68 \times 10^{2}$ &$1$  &$8.70 \times 10^{1}$ & $1$ &$5.21 \times 10^{1}$&$1$& $1.56 \times 10^{1}$& $1$\\
			&$t\bar{t} b\bar{b}$  & $1.39 \times 10^{2}$ & $10$ &$2.33\times 10^{4}$ &  $11$&$1.38 \times 10^{1}$&$13$&$3.68 \times 10^{4}$ & $17$ \\
			\hline
		\end{tabular}}}
		\caption{Cross sections $(\sigma)$ for signal and backgrounds for $t \bar{t} h$  production channel before applying the event selection cuts and $N$ denotes the number of events after the cuts. The signal is driven by SMEFT couplings  BP4,  BP5, BP6 and BP7 respectively as summarised in Table~\ref{BM_points}.} 
		\label{tab:ttbarhbackgrounds}
	\end{table}
\begin{figure}[t]
	\centering
	\includegraphics[width=7.7cm,height=5.2cm]{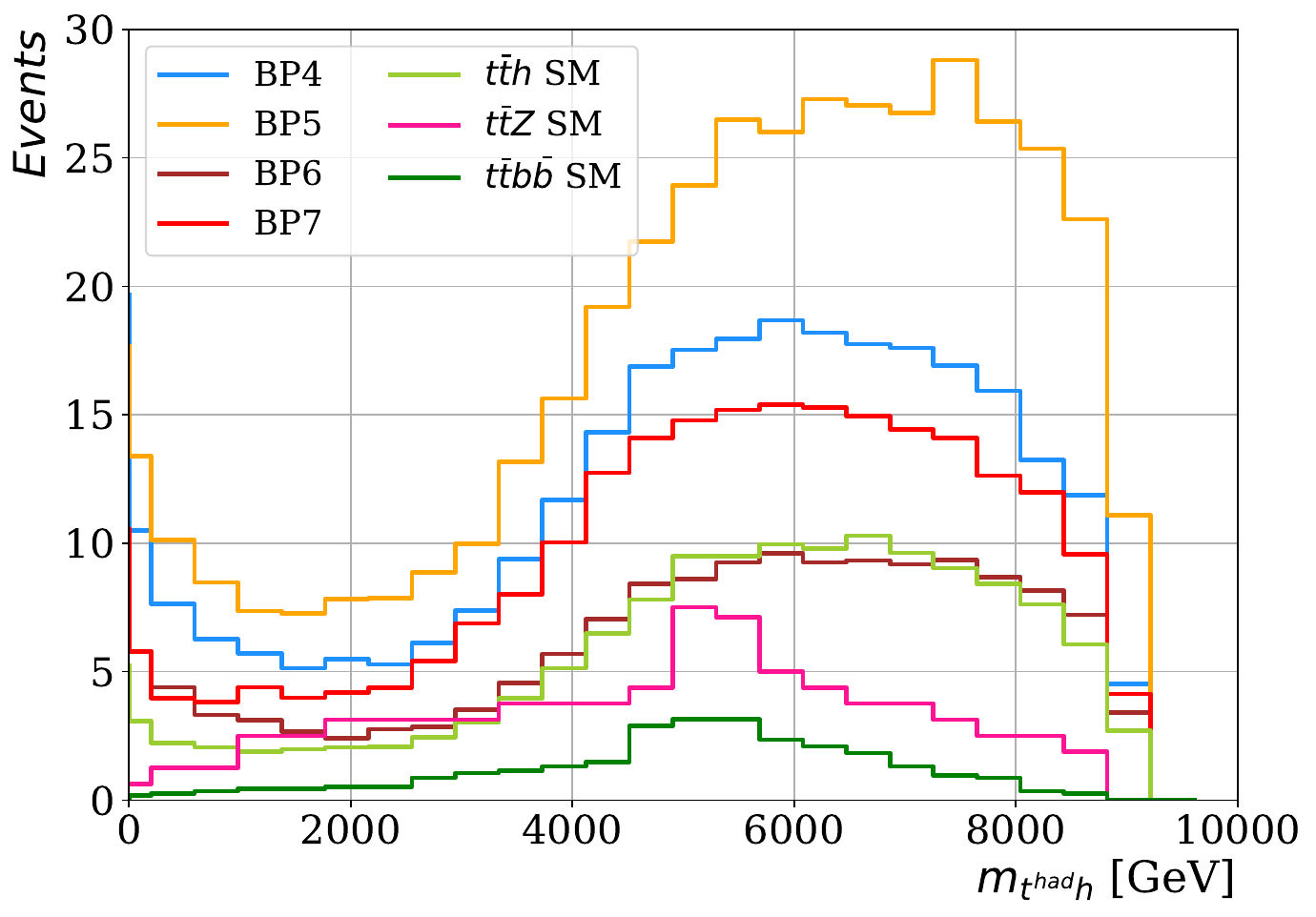}
	\includegraphics[width=7.7cm,height=5.2cm]{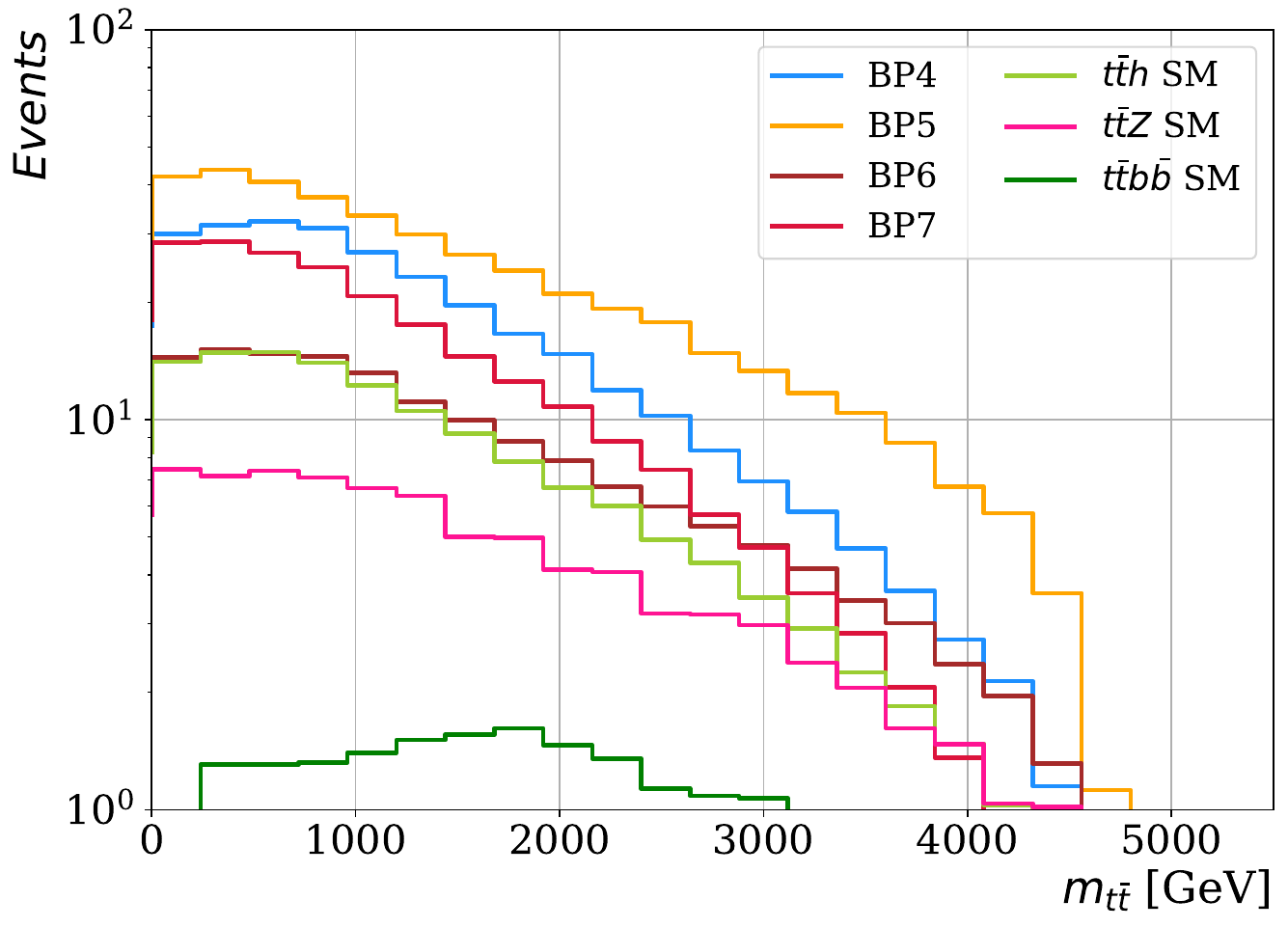}
	\caption{Event distributions of $m_{t^{\rm had}h}$ and $m_{t\bar{t}}$ for signal and  SM backgrounds for $\mu^+\mu^- \to t \bar{t} h$ at $\sqrt{s}=10$ TeV, $\mathcal{L}_{\rm int}=10$ ab$^{-1}$. The signal is driven by SMEFT couplings  BP4,  BP5, BP6 and BP7 respectively as summarised in Table~\ref{BM_points}.}
	\label{ttbarh_distributions}
\end{figure}
In Fig.~\ref{ttbarh_distributions}, we show the event distributions for the invariant masses of the hadronically decaying top–Higgs system, $m_{t^{\rm had}h}$ (left) and the top-pair system, $m_{t\bar{t}}$ (right) in $\mu^+\mu^-\to t\bar{t}h$ at $\sqrt{s}=10$ TeV with $\mathcal{L}_{\rm int}=10$ ab$^{-1}$. The SM backgrounds populate the lower invariant-mass region and fall rapidly with increasing mass, whereas the signal peaks at higher values. Although the $t\bar{t}h$ cross section remains small at these energies, it can still yield measurements competitive with $pp$ or $e^+e^-$ colliders above the $t\bar{t}h$ threshold. A summary of signal and background yields based on the selection criteria of Table~\ref{tab:tt_ttH_selection} for $\sqrt{s} = [3,10,14,30]$ TeV is presented in Table~\ref{tab:ttbarhbackgrounds}.

\subsection{Angular Observables}
In addition to total rates and differential distributions, angular asymmetries provide a complementary handle on the effective interactions. Such observables are often more sensitive to interference effects that are diminished in cross-section measurements and help resolve degeneracies among operator coefficients that affect the overall rate in similar directions. Asymmetry variables that originate from the angular structure of the final state, rather than the rate of the process can also be a useful probe to distinguish the signal from the background in the channels $\mu^{+}\mu^{-}h$ and $ t\bar t h$.

\begin{figure}[t]
		\centering
		{\begin{tabular}{cc}
		\includegraphics[width=7.5cm,height=5.2cm]{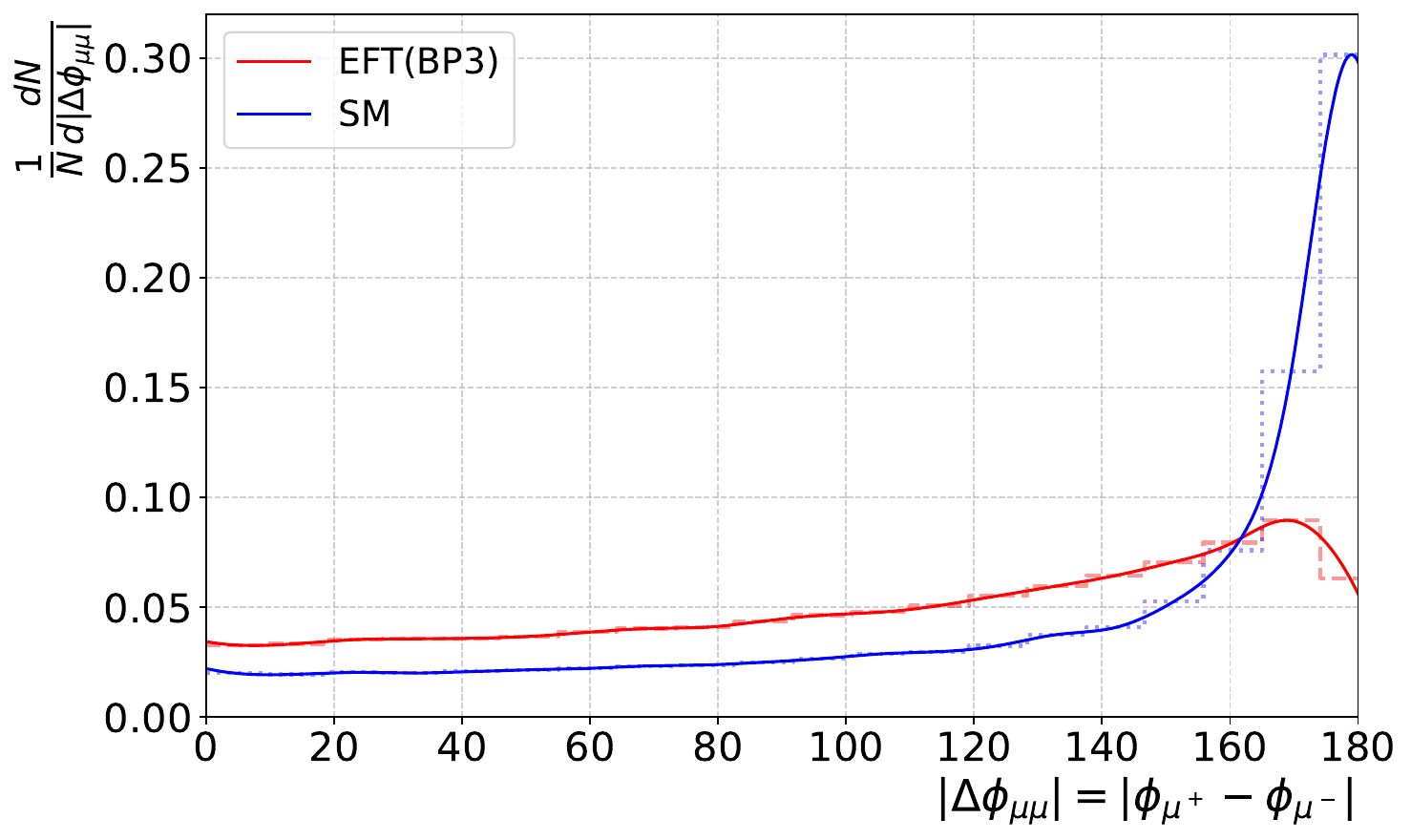}&
		\includegraphics[width=7.5cm,height=5.2cm]{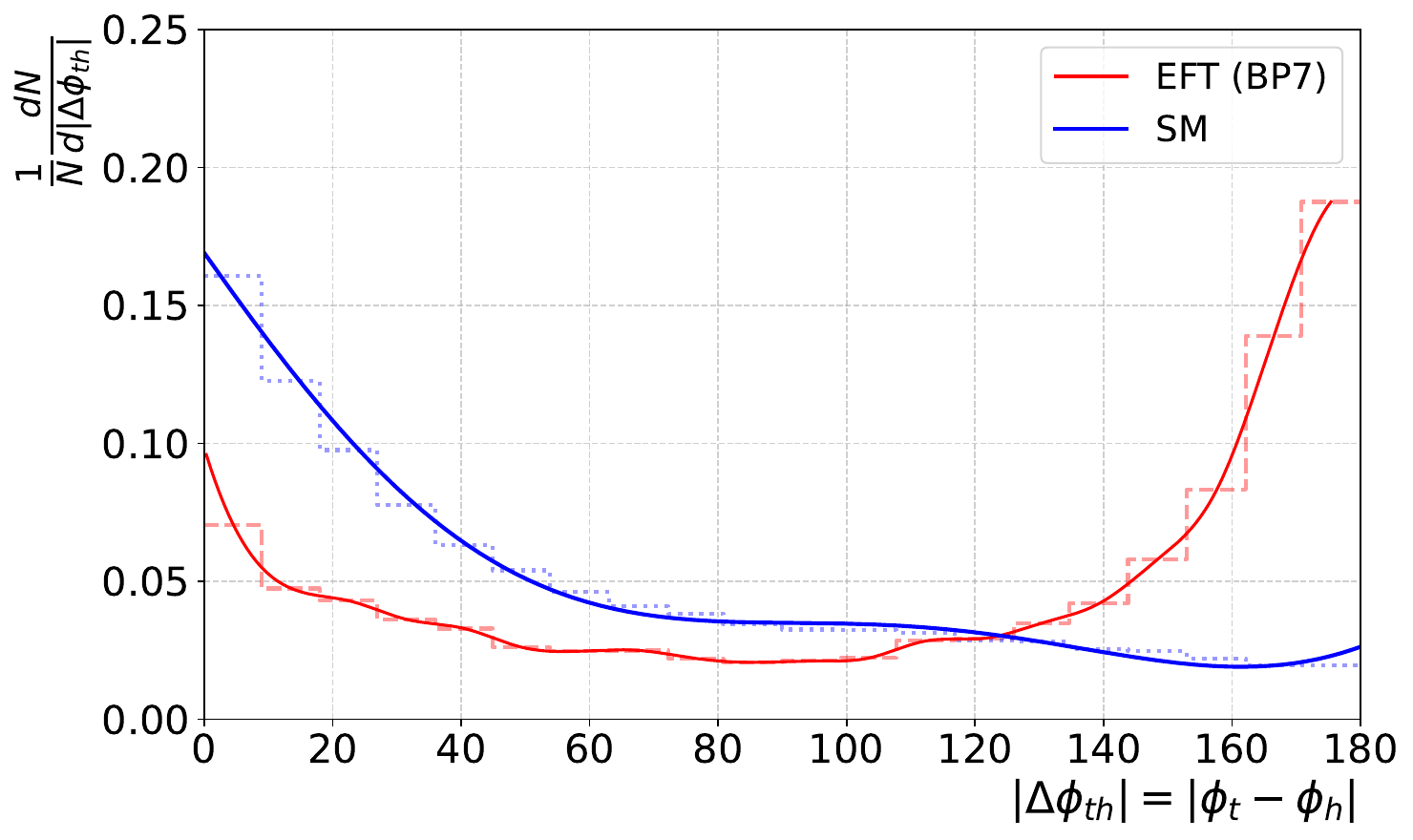}\\
			(a)&(b)
	\end{tabular}}
		\caption{(a) Normalised $\Delta\phi_{\mu\mu}$ distribution for $\mu^+ \mu^- \to \mu^+ \mu^- h \to \mu^+ \mu^- b\bar b$ at $\sqrt{s}=10$ TeV, comparing SM background and SMEFT benchmark BP3. (b) Normalised  $\Delta\phi_{th}$ distribution between the outgoing top and the Higgs in $\mu^+ \mu^- \to t \bar{t} h$ at $\sqrt{s}=10$ TeV, comparing SM background and SMEFT benchmark BP7. }
		\label{fig:phimumu_asy_SM}
	\end{figure}

For $\mu^+ \mu^- \to \mu^+ \mu^- h$, we study the azimuthal separation between the final-state muons. Fig.~\ref{fig:phimumu_asy_SM}(a) shows the normalised distribution of $\Delta\phi_{\mu\mu} = |\phi_{\mu^+} - \phi_{\mu^-}|$, comparing the SM background $\mu^+ \mu^- b \bar{b}$ with the SMEFT benchmark point BP3 involving the $\mathcal{O}_{\mu Z}$ interaction. The SM distribution peaks sharply near $\Delta\phi_{\mu\mu} \sim 180^{\circ}$, indicating that muons are mostly emitted back-to-back while the dipole-type EFT contribution produces a flatter and  more isotropic distribution. Based on this, the following angular asymmetry variable can serve as a complementary probe of the new coupling:
\begin{equation}
A_{\Delta \phi_{\mu \mu}} = \frac{N(|\Delta \phi_{\mu \mu}|<\pi/2) - N(|\Delta
\phi_{\mu \mu}|>\pi/2)}{N(|\Delta \phi_{\mu \mu}|<\pi/2) + N(|\Delta
\phi_{\mu \mu}|>\pi/2)}~.
\label{asymm_mumuh}
\end{equation}
For the $\mu^+ \mu^- \to t \bar{t} h$ process, we consider the azimuthal separation $|\Delta \phi_{th}|$ between the top and the Higgs. As shown in Fig.~\ref{fig:phimumu_asy_SM}(b), the SM distribution disfavours intermediate separations (around $90^\circ$), while the SMEFT benchmark BP7, driven by the tensorial four-fermion operator $\mathcal{O}^{T}_{\mu t}$ exhibits a more uniform angular spread. As the statistics increase with increasing luminosity, an asymmetry variable such as the following:
\begin{equation}
A_{\Delta \phi_{t h}} = \frac{N(|\Delta \phi_{th}|<\pi/2) - N(|\Delta
\phi_{th}|>\pi/2)}{N(|\Delta \phi_{th}|<\pi/2) + N(|\Delta
\phi_{th}|>\pi/2)}~,
\label{asymm_tth}
\end{equation}
 can serve as an independent observable to improve the separation between signal and background. At $\sqrt{s}=10$ TeV, we present the sensitivity obtained from a simple cut based analysis using the selection criterion on the angular variables (Fig.~\ref{fig:phimumu_asy_SM}) as summarised in Table~\ref{tab:cut_vs_asym}. These observables serve to illustrate the additional discriminating power that could be exploited in a more dedicated or global analysis, where their inclusion further enhances the sensitivity to the underlying SMEFT operators.
\begin{table}[t]
\centering
\begin{tabular}{|lccc|}
\hline\hline
Process & Observable & Criteria & $2\sigma$ sensitivity \\ \hline
$\mu^+\mu^-\to\mu^+\mu^-h$ 
& $\Delta\phi_{\mu\mu}$ 
& Cut: $|\Delta\phi_{\mu\mu}|>7\pi/8$ 
&  $[-1.93,1.93]\times10^{-3}$ \\\hline

$\mu^+\mu^-\to t\bar t h$ 
& $\Delta\phi_{th}$ 
& Cut: $|\Delta\phi_{th}|>11\pi/16$ 
& $[-1.51,1.51]\times10^{-3}$ \\\hline\hline
\end{tabular}
\caption{Sensitivity at $\sqrt{s}=10$ TeV obtained from a cut based selection on the angular variables. The reach in the last column corresponds to the minimum WC value that can be distinguished from the SM at the $2\sigma$ level, assuming luminosity $\mathcal{L}_{\rm int} = 10$ ab$^{-1}$.}
\label{tab:cut_vs_asym}
\end{table}

Before closing this section, we comment on the possible impact of higher-order electroweak corrections. At multi-TeV lepton colliders, electroweak Sudakov logarithms and real electroweak radiation can give sizeable corrections to high-energy scattering processes~\cite{Chen:2022msz}. These effects can modify both the overall normalisation and the shapes of differential distributions, and a precision extraction of SMEFT coefficients would require their inclusion, possibly with an appropriate resummation of the leading logarithmic effects. In the present work we restrict to a leading-order analysis of the signal and background processes, with the aim of estimating the relative sensitivity of different Higgs and top channels to the energy-enhanced dimension-six interactions, while a complete treatment of electroweak radiative effects is left for future work.
\section{Projections on SMEFT couplings}
\label{section5}
Armed with the signal and background rates of the processes of our interest, we are now ready to present the projected reach on the WCs involved 
in such processes. 
At a $10$ TeV muon collider with $10$ ab$^{-1}$ integrated luminosity, about 8 million Higgs bosons and 100 million top quarks are expected to be produced, enabling us to determine small deviations in Higgs and top couplings from their SM values.  The estimated number of signal $(S)$ and background $(B)$ events are defined as:
\begin{align}
\label{eq:S}
S &= N_{\textrm{total}} - N_{\textrm{SM}} = \mathcal{L}_{\rm int} \cdot (\sigma_{\textrm{total}}\cdot \varepsilon_{\text{total}} - \sigma_{\textrm{SM}}\cdot \varepsilon_{\text{SM}})  , \\ 
\label{eq:B}
B &= N_{\textrm{SM}} = \mathcal{L}_{\rm int} \cdot \sigma_{\textrm{SM}} \cdot \varepsilon_{\text{SM}} \nonumber
\end{align}
where $\mathcal{L}_{\rm int}$ is the integrated luminosity. $\sigma_{\textrm{SM}}$ and $\sigma_{\textrm{total}}$ denote the SM and total cross sections, respectively, with $\varepsilon_{\rm{SM}}$ and $\varepsilon_{\rm{total}}$ representing the corresponding efficiency factors that account for the event selection criteria and detector acceptance. The signal significance of the signal is defined as ~\cite{Cowan:2010js}:
\begin{equation}
    \mathcal{Z}=\sqrt{2 \times \bigg[(S+B) \ln \bigg(1+\frac{S}{B} \bigg)-S \bigg]}
    \label{signal_signi}
\end{equation}
\vspace{-1mm}
\begin{figure}[t!]
		\begin{center}\vspace{-1mm}
			{\begin{tabular}{cc}
					\includegraphics[width=7.2cm,height=5.cm]{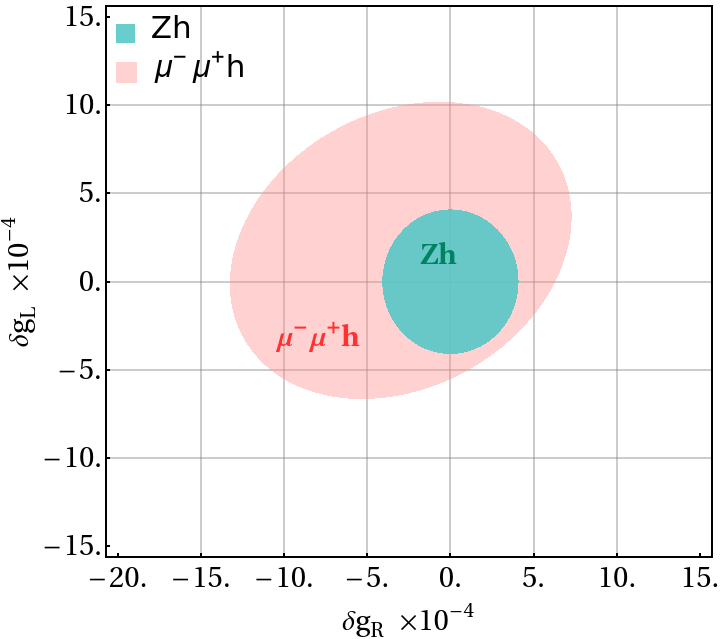}&
					\includegraphics[width=7.2cm,height=5.cm]{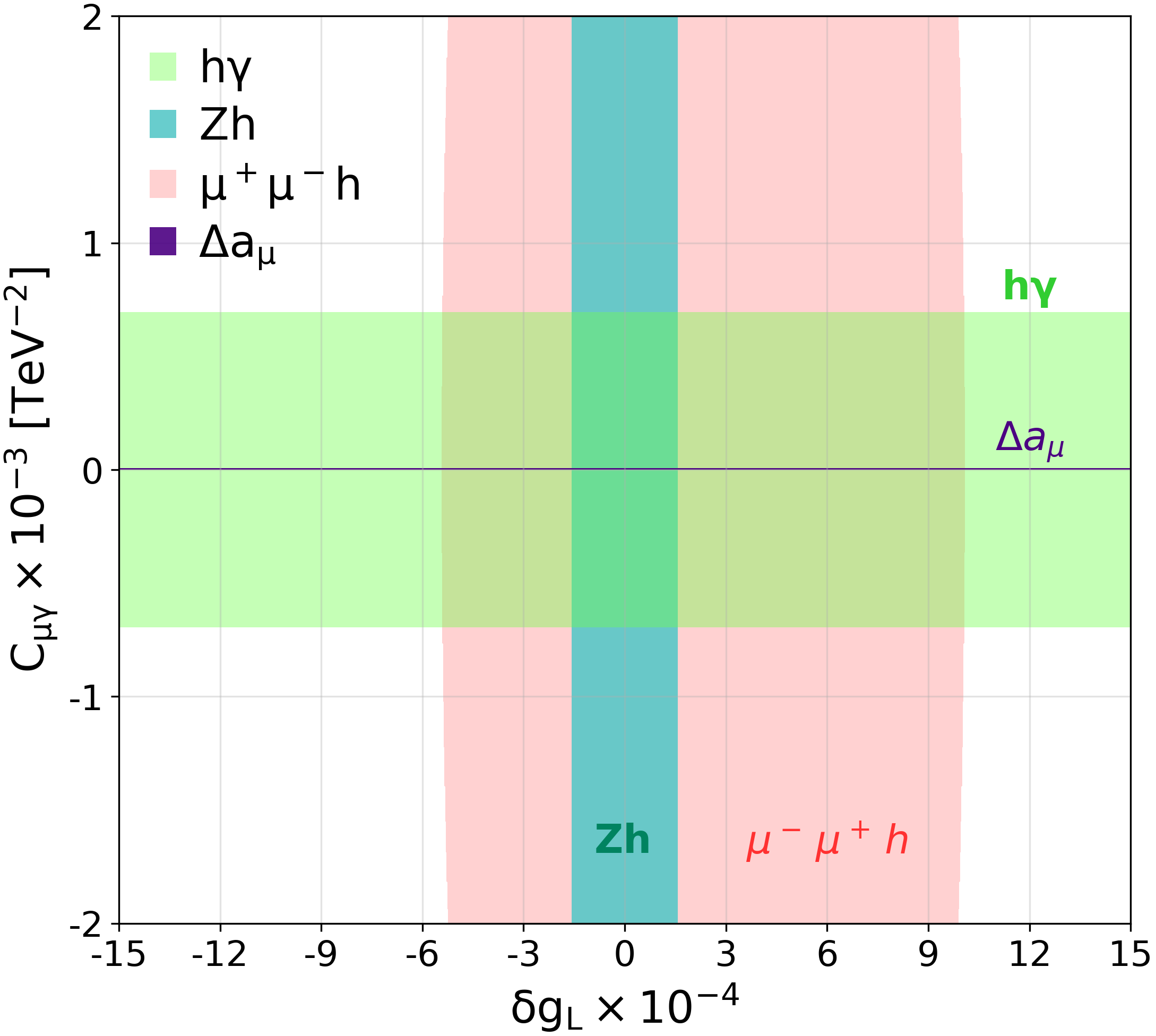}\\
                    (a)&(b)\\
					\includegraphics[width=7.2cm,height=5.cm]{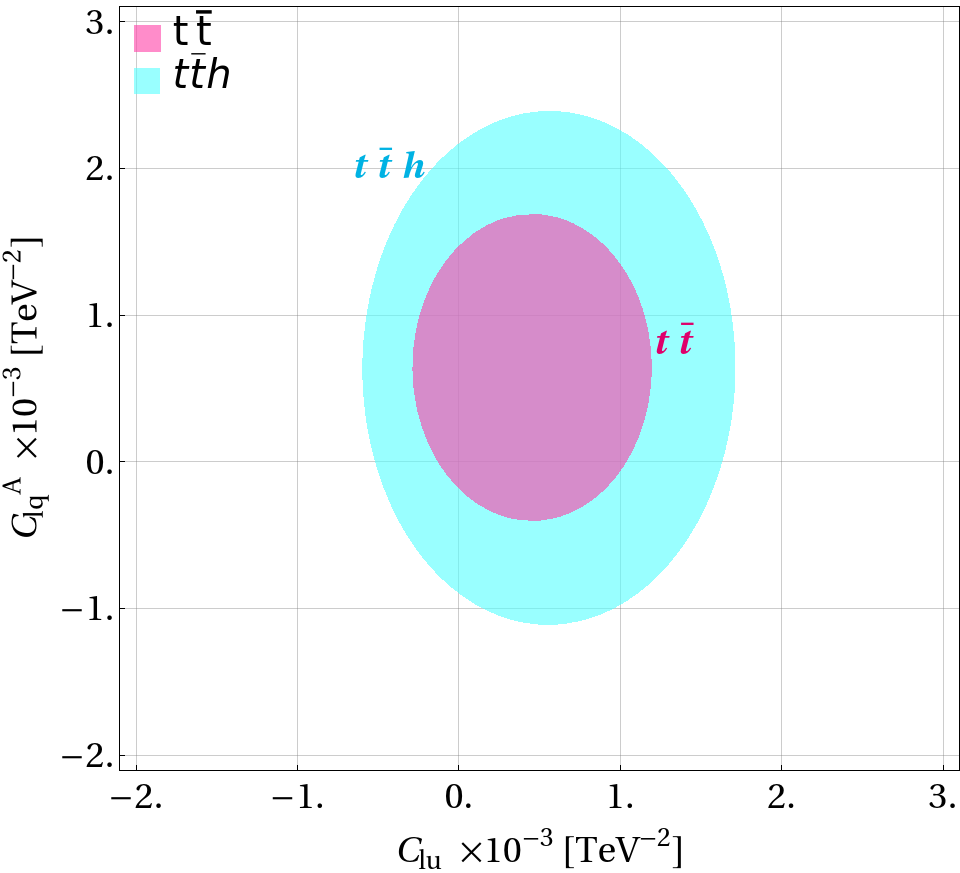}&
					\includegraphics[width=7.2cm,height=5.cm]{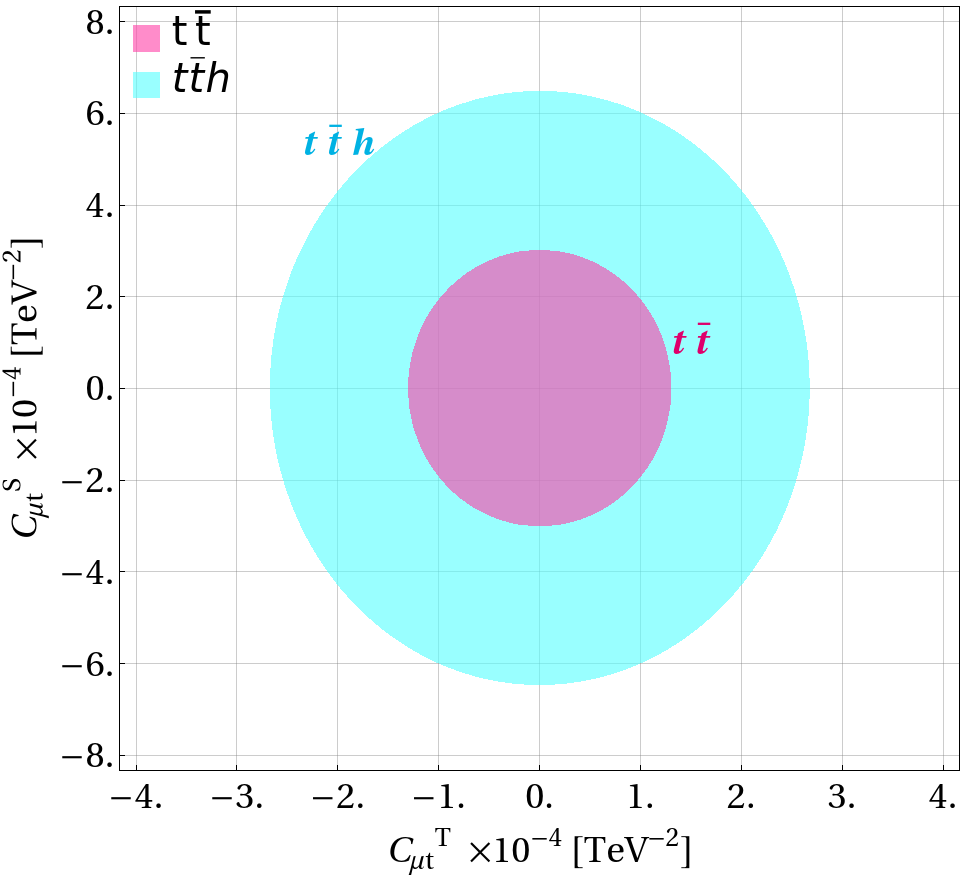}\\
                    (c)&(d)\\
					\includegraphics[width=7.2cm,height=5.cm]{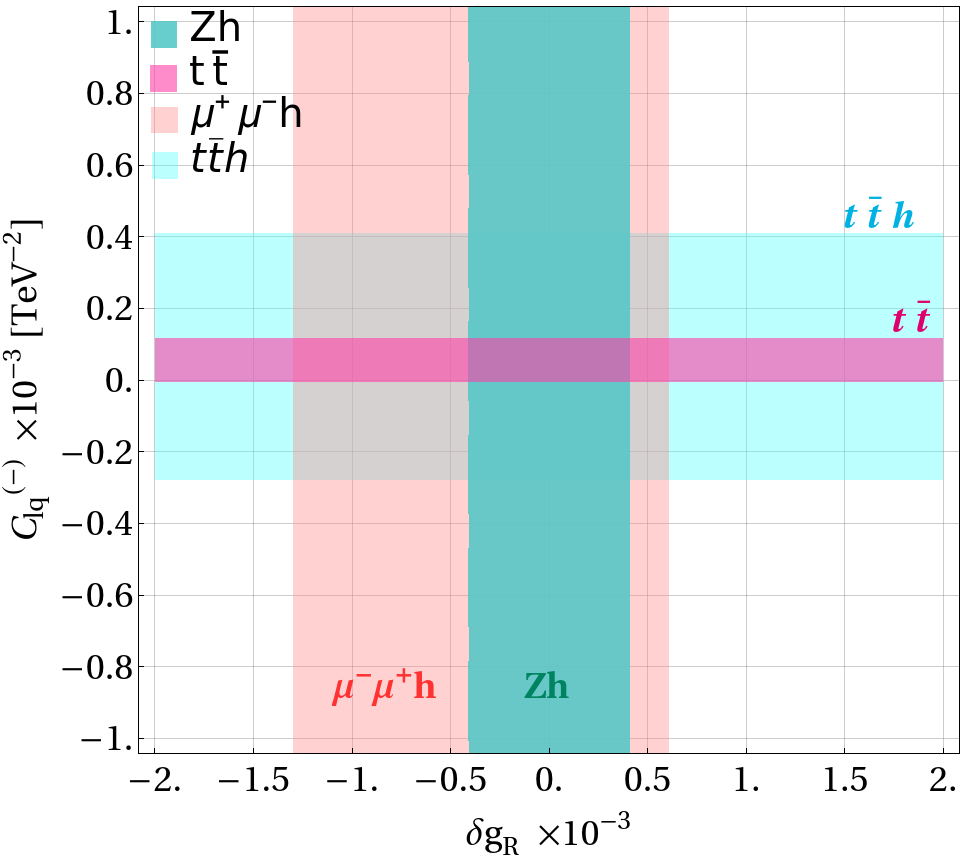}&
					\includegraphics[width=7.2cm,height=5.cm]{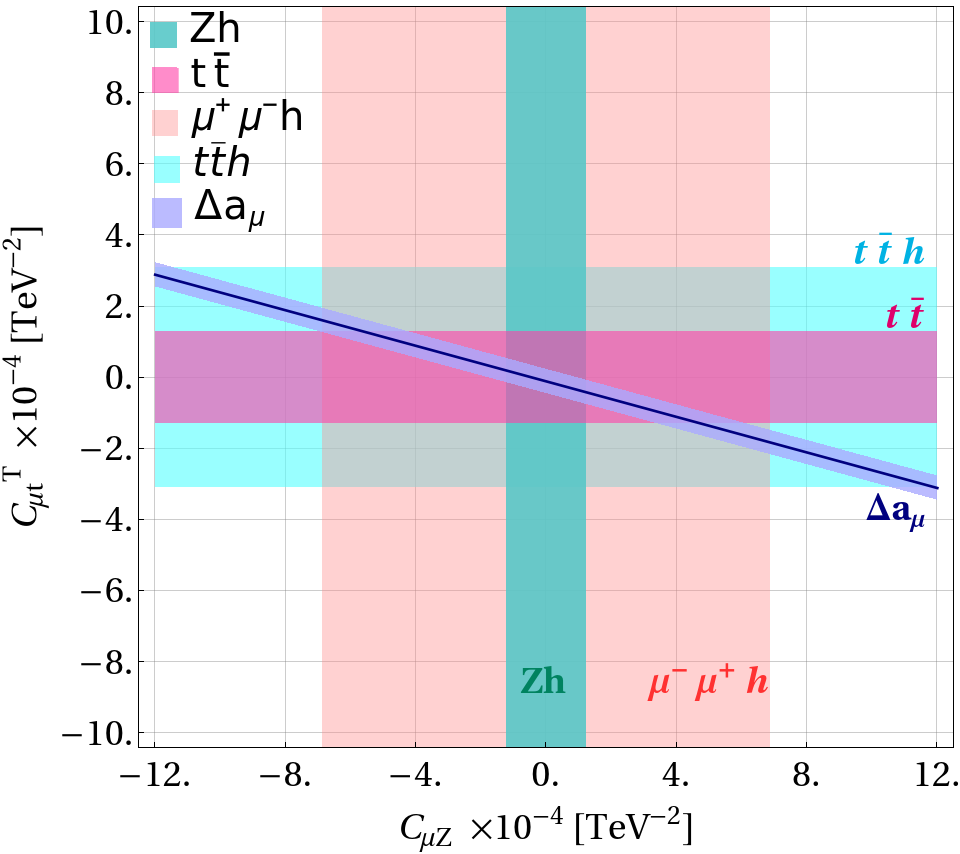}\\
                    (e)&(f)\vspace{-1mm}
			\end{tabular}}\vspace{-3mm}
\caption{Projected sensitivities on the effective couplings for $\mu^+\mu^- \to h Z$ (dark green), $\mu^+\mu^- \to h \mu^+\mu^-$ (light red),   $\mu^+\mu^- \to t \bar{t}$ (pink) ,   $\mu^+\mu^- \to t \bar{t} h $ (blue) and $\mu^+\mu^- \to h \gamma$ (green) at a 10 TeV collider with an integrated luminosity of $10$ ab$^{-1}$ in the plane of (a) $\delta g_L-\delta g_R$, (b) $\delta g_L-C_{\mu \gamma}$, (c) $C_{\ell u}-C_{\ell q}^A$ , (d) $C_{\mu t}^T-C_{\mu t}^S$, (e)  $\delta g_R-C_{\ell q}^{(-)} $ and  (f)  $C_{\mu Z} - C_{\mu t}^T$. The shaded regions mark the projected bounds at $2\sigma$ level. 
The $95\%$ C.L. constraints on the muon anomalous magnetic moment is shown in purple band in (b) and (f).}
\label{fig:collider_constraints_bottomup}
\end{center}
\end{figure}
In Fig.~\ref{fig:collider_constraints_bottomup}, we summarise the projected $2\sigma$ regions ($\mathcal{Z} \geq 2$ outside these shaded regions) for chosen pairs of SMEFT coefficients in six cases, combining sensitivities from $hZ$ (dark green), $h\mu^+\mu^-$ (light red), $h\gamma$ (green), $t\bar{t}$ (pink), and $t\bar{t}h$ (blue) production channels at a $10$ TeV muon collider with $\mathcal{L}_{\rm int}=10$ ab$^{-1}$. Some key features that emerge from these plots are the following:
\begin{itemize}[leftmargin=*, labelsep=0.5em]
\item The shaded regions in the $\delta g_L$–$\delta g_R$ plane (Fig.~\ref{fig:collider_constraints_bottomup} (a)) represent the projected sensitivities from $Zh$ and {\it{ZBF}} production processes. The $Zh$ production rate is equally sensitive to $\delta g_L$ and $\delta g_R$, allowing one to probe these couplings down to the level of $\mathcal{O}(10^{-4})$. On the other hand, the $\mu\mu h$ rate is slightly more sensitive to $\delta g_R$, but with an overall weaker sensitivity, reaching the level of $\mathcal{O}(10^{-3})$. 

\item Fig.~\ref{fig:collider_constraints_bottomup} (b) shows the $\delta g_L$–$C_{\mu\gamma}$ projection from $\mu^+\mu^- \to h\gamma$, $hZ$, and $\mu^+\mu^-h$\footnote{Here in Fig.~\ref{fig:collider_constraints_bottomup}, $\delta g_L$ and $\delta g_R$ are dimensionless couplings while $C_i \equiv C_i/\Lambda^2$ carry units of TeV$^{-2}$.}. 
The $h\gamma$ process depends only on $C_{\mu\gamma}$ (and hence the region is a band parallel to $\delta g_L$), while $\mu\mu h$ constrains both couplings with stronger dependence on $\delta g_L$. The $Zh$ and $h\gamma$ channels provide complementary sensitivity in the $\delta g_L $–$C_{\mu\gamma}$ plane. The current $\Delta a_\mu$ constraint from $(g-2)_\mu$is shown as a purple band along the $C_{\mu\gamma}$ axis. The $h\gamma$ process constrains $C_{\mu\gamma}$ at the level of $\mathcal{O}(10^{-4})$, whereas the $\Delta a_\mu$ measurement, which receives a tree level contribution from $C_{\mu\gamma}$, constrains it at the level of $\mathcal{O}(10^{-6})$. 
\item The shaded regions in $C_{\ell u}$–$C^{A}_{\ell q}$ plane (Fig.~\ref{fig:collider_constraints_bottomup} (c)) represent the projected bounds from $t\bar{t}$ and $t\bar{t}h$ production processes. In $t\bar{t}$ channel, interference between these two operators is negligible and energy growth due to quadratic terms becomes evident, resulting in a reach of the WCs at $\mathcal{O}(10^{-4})$. The $t\bar{t}h$ channel yields a weaker reach of the corresponding WCs at the level of $\mathcal{O}(10^{-3})$ but still provides complementary information because it probes the same operator combinations in a different kinematic regime and with a distinct event topology. 
In Fig.~\ref{fig:collider_constraints_bottomup} (d), the projected reach in the $C^{T}_{\mu t}$–$C^{S}_{\mu t}$ plane resulting from $t\bar t$ and $t \bar th$ channels have been presented. In both cases,  interference with SM is small due to the chirality structure of these operators.  Overall, a higher $t\bar{t}$ cross-section yields a better reach for the WCs. 
\item In the $\delta g_R$–$C_{\ell q}^{(-)}$ plane (Fig.~\ref{fig:collider_constraints_bottomup} (e)), the $Zh$ channel sets a much tighter bound on $\delta g_R$, with $|\delta g_R|\le 4.2\times 10^{-4}$, while $\mu\mu h$ channel allows $-1.3\times 10^{-3}\le \delta g_R \le 6.1\times 10^{-4}$. The $t\bar t$ and $t\bar t h$  channels are sensitive only to $C_{\ell q}^{(-)}$, appearing as horizontal bands, with limits $-5.7\times 10^{-5}\le C_{\ell q}^{(-)} \le 1.7\times 10^{-4}$ and $-2.7\times 10^{-4}\le C_{\ell q}^{(-)} \le 4.0\times 10^{-4}$, respectively. 
\item In the $C_{\mu Z}$–$C_{\mu t}^{T}$ plane (Fig.~\ref{fig:collider_constraints_bottomup}(f)), the constraints from $t\bar{t}$ and $t\bar{t}h$ production lie along similar directions because both processes depend on the  muon–top tensor interaction. In contrast, the $Zh$ and $\mu\mu h$ channels probe combinations dominated by $C_{\mu Z}$, leading to contours that are orthogonal to the top-involving ones. The $\Delta a_\mu$ band (in purple, with $C_{\mu\gamma}=0$) is mainly sensitive to $C_{\mu t}^T$, leaving only a small overlapping region with the collider probe, which further shrinks with increasing luminosity. At $10$~ab$^{-1}$ ($40$~ab$^{-1}$), Higgs-strahlung constrains $g_{hZ}^{\mu\mu}\lesssim2.4(1.7)\times10^{-4}$, and $t\bar{t}$ ($t\bar{t}h$) processes bound $g_{tt}^{\mu\mu}\lesssim1.3(0.8)\times10^{-4}$.\footnote{Here, $g^{\mu \mu}_{hZ}$ denotes the four-point contact interaction of two muons and the Higgs and the Z while $g^{\mu \mu}_{tt}$ refers to the four-point contact interaction of two muons and two tops.} Hence, a $10$~ab$^{-1}$ run would probe deviations at the sub-per-mille level, providing a powerful collider complement to low-energy measurements. 
\end{itemize}
\noindent
At $\sqrt{s}=10~\mathrm{TeV}$, we impose kinematic restrictions to ensure a consistent SMEFT interpretation. In particular, we restrict the analysis to regions where the $\mathcal{O}(1/\Lambda^{4})$ effects are dominated by the squared dimension-6 contributions, while the interference of dimension-8 operators with the SM remains subleading~\cite{Contino:2016jqw}. This is implemented by requiring that the characteristic momentum transfer $Q$ in each channel satisfies $Q^2/\Lambda^2 \ll 1$. We retain events with $p_T^{\ell^+\ell^-}<5~\mathrm{TeV}$ for $\mu^{+}\mu^{-}\to Zh$, $m_{\mu^+\mu^-}<2.7~\mathrm{TeV}$ for $\mu^{+}\mu^{-}\to\mu^{+}\mu^{-}h$ ({\it ZBF}), and $m_{t\bar t}<9.5~\mathrm{TeV}$ and $m_{t\bar t h}<8.2~\mathrm{TeV}$ for $\mu^{+}\mu^{-}\to t\bar t$ and $\mu^{+}\mu^{-}\to t\bar t h$, respectively.  These choices follow the method of Refs.~\cite{Busoni:2013lha,Bhattacharya:2015vja} and retain at least $90\%$ of the signal events for representative values of the Wilson coefficients at the projected sensitivities, thereby keeping the analysis within the EFT validity regime. The projected bounds on the WCs can be translated into the corresponding cutoff scale $\Lambda$. For the $Z\mu\mu$ coupling shifts from $C_{\varphi\ell}^{(1,3)}$ and $C_{\varphi e}$, with $\delta g_L=\tfrac{v^2}{2\Lambda^2}(C_{\varphi\ell}^{(1)}+C_{\varphi\ell}^{(3)})$ and $\delta g_R=\tfrac{v^2}{2\Lambda^2}C_{\varphi e}$, the bounds on $\delta g_{L,R}$ are mapped to $\Lambda\simeq \tfrac{v}{\sqrt{2|\delta g_{L,R}|}}$, leading to a representative lower bound of $\Lambda>14.4$ TeV. For the semileptonic four-fermion operators contributing to $\mu^+\mu^-\to t\bar t$ and $\mu^+\mu^-\to t\bar t h$, the contact coupling scales $\Lambda\ (\simeq1/\sqrt{|C|})$ are obtained by interpreting the bounds on $C/\Lambda^2$ under the assumption $C=1$ when putting bounds on $\Lambda$. \footnote{When interpreting the SMEFT bounds in terms of UV models with heavy states, we further restrict to the decoupling regime where the heavy mass becomes greater than the maximum momentum transfer entering the analysis, ensuring that  SMEFT expansion remains valid and that on-shell production effects are not present.} For this sector, the projected sensitivities correspond to $\Lambda > 26~\mathrm{TeV}$ for the operator combination $(C_{\ell u}, C_{\ell q}^{A})$, representing the vector–axial structures, and $\Lambda > 45~\mathrm{TeV}$ for the scalar–tensor pair $(C_{\mu t}^{T}, C_{\mu t}^{S})$.
\begin{table}[t]
\centering
\renewcommand{\arraystretch}{1.4}
\resizebox{\textwidth}{!}{%
\begin{tabular}{|c|c|c||c|c|c|}
\hline
\multirow{1.5}{*}{WCs} & \multicolumn{2}{c||}{MuC} &
\multirow{1.5}{*}{WCs} & \multicolumn{2}{c|}{MuC} \\
\cline{2-3}\cline{5-6}
(TeV$^{-2}$)& Single parameter & Marginalised & (TeV$^{-2}$) & Single parameter & Marginalised \\
\hline
$C_{\varphi \ell}^{(1)}/\Lambda^{2}$ & $[-4.2,4.2]\times10^{-4}$& $[-2.8,3.7] \times10^{-3}$&$C_{\ell q}^{(-)}/\Lambda^{2}$&$[-0.6,1.7] \times 10^{-4}$ & $[-0.4,2.7] \times 10^{-3}$\\
$C_{\varphi \ell}^{(3)}/\Lambda^{2}$ & $[-4.2,4.2]\times10^{-4}$& $[-8.1,9.8]\times 10^{-3}$&$C_{\ell u}/\Lambda^{2}$& $[-0.2,1.4] \times 10^{-3}$&$[-1.1,2.0] \times 10^{-3}$ \\
$C_{\varphi  e}/\Lambda^{2}$ & $[-4.2,4.2]\times10^{-4}$& $[-0.9,1.4] \times 10^{-2}$&$C_{eu}/\Lambda^{2}$ & $[-2.2,2.2]\times10^{-5}$& $[-7.2,2.4] \times 10^{-3}$\\
$C_{\mu Z}/\Lambda^{2}$ & $[-2.4,2.4]\times10^{-4}$ &$[-2.2,2.2]\times10^{-4}$ &$C_{eq}/\Lambda^{2}$ &$[-7.8,0.2]\times 10^{-5}$ & $[-1.4,1.8] \times 10^{-3}$\\
$C_{\mu \gamma}/\Lambda^{2}$ &$[-4.8,4.8]\times10^{-4}$ & $[-4.3,4.3]\times10^{-4}$&$C_{\mu t}^{S}/\Lambda^{2}$ &$[-3.0,3.0]\times 10^{-4}$ & $[-3.0,3.0] \times 10^{-4}$ \\
 & & &$C_{\mu t}^{T}/\Lambda^{2}$ & $[-1.3,1.3]\times10^{-4}$&$[-1.3,1.3]\times10^{-4}$ \\
\hline
\end{tabular}
}
\caption{Projected limits at 95\% C.L. on various Wilson coefficients under 10 TeV muon collider scenario, including marginalised and single-parameter fits.}
\label{tab:future_limits_1}
\end{table}
\begin{table}[b]
\centering
\renewcommand{\arraystretch}{1.4}
\resizebox{\textwidth}{!}{%
\begin{tabular}{|c|c|c||c|c|}
\hline
\multirow{1.5}{*}{{WCs}} & 
\multicolumn{2}{c||}{{HL-LHC}}&
\multicolumn{2}{c|}{{HL-LHC+FCC\mbox{-}ee}} \\
\cline{2-5}
(TeV$^{-2}$)& Single parameter & Marginalised & Single parameter & Marginalised \\
\hline
$C_{\varphi \ell}^{(1)}/\Lambda^{2}$ & $[-1.6,1.6]\times10^{-1}$& $[-8.9,8.9] \times10^{-1}$& $[-1.0,1.0]\times10^{-2}$&$[-2.6,2.6]\times10^{-2}$ \\
$C_{\varphi \ell}^{(3)}/\Lambda^{2}$ & $[-1.8,1.8]\times10^{-1}$& $[-1.0,1.0]$&$[-9.7,9.7]\times10^{-4}$& $[-7.8,7.8]\times10^{-3}$ \\
$C_{\varphi  e}/\Lambda^{2}$ & $[-1.8,1.8]\times10^{-1}$& $[-9.4,9.4] \times 10^{-1}$&$[-2.5,2.5]\times10^{-3}$&$[-4.6,4.6]\times10^{-3}$ \\
$C_{\mu Z}/\Lambda^{2}$ & $[-8.3,8.3]\times10^{-2}$ &$[-8.5,8.5]\times10^{-2}$ &$[-5.3,5.3]\times10^{-2}$ & $[-5.3,5.3]\times10^{-2}$\\
$C_{\mu \gamma}/\Lambda^{2}$ &$[-1.5,1.5]\times10^{-1}$ & $[-1.6,1.6]\times10^{-1}$&$[-1.6,1.6] \times10^{-2}$ & $[-1.7,1.7]\times10^{-2}$ \\
$C_{\ell q}^{(-)}/\Lambda^{2}$ &$[-5.6,5.6]\times 10^{-1}$ &$-$ & $[-3.0,3.0]\times10^{-3}$& $[-2.8,2.8] \times 10^{-2}$ \\
$C_{\ell u}/\Lambda^{2}$ &$[-1.8,1.8]$ &$-$ & $[-2.9,2.9] \times 10^{-3}$& $[-4.1,4.1]\times 10^{-2}$\\
$C_{eu}/\Lambda^{2}$ &$[-8.3,8.3] \times 10^{-1}$ & $-$& $[-4.2,4.2]\times 10^{-3}$& $[-5.1,5.1] \times 10^{-2}$\\
$C_{eq}/\Lambda^{2}$ & $[-3.6,3.6]$&$-$ & $[-1.9,1.9] \times 10^{-3}$& $[-3.9,3.9]\times10^{-2}$ \\
\hline
\end{tabular}
}
\caption{Projected limits at 95 \% C.L. on various Wilson coefficients under different scenarios, including marginalised and single-parameter fits. The HL-LHC and HL-LHC+FCC\mbox{-}ee bounds are taken from the global fits of refs.~\cite{Celada:2024mcf,Cornet-Gomez:2025jot}, performed at order $\mathcal{O}(\Lambda^{-4})$. Here, FCC\mbox{-}ee refers to the full running scenario with $\sqrt{s}=91$, $161$, $240$, and $365$ GeV added to the HL-LHC dataset.}
\label{tab:future_limits}
\end{table}
\begin{figure}[t]
	  \begin{tabular}{cc}
	\centering%
	\includegraphics[width=7.5cm,height=5.7cm]{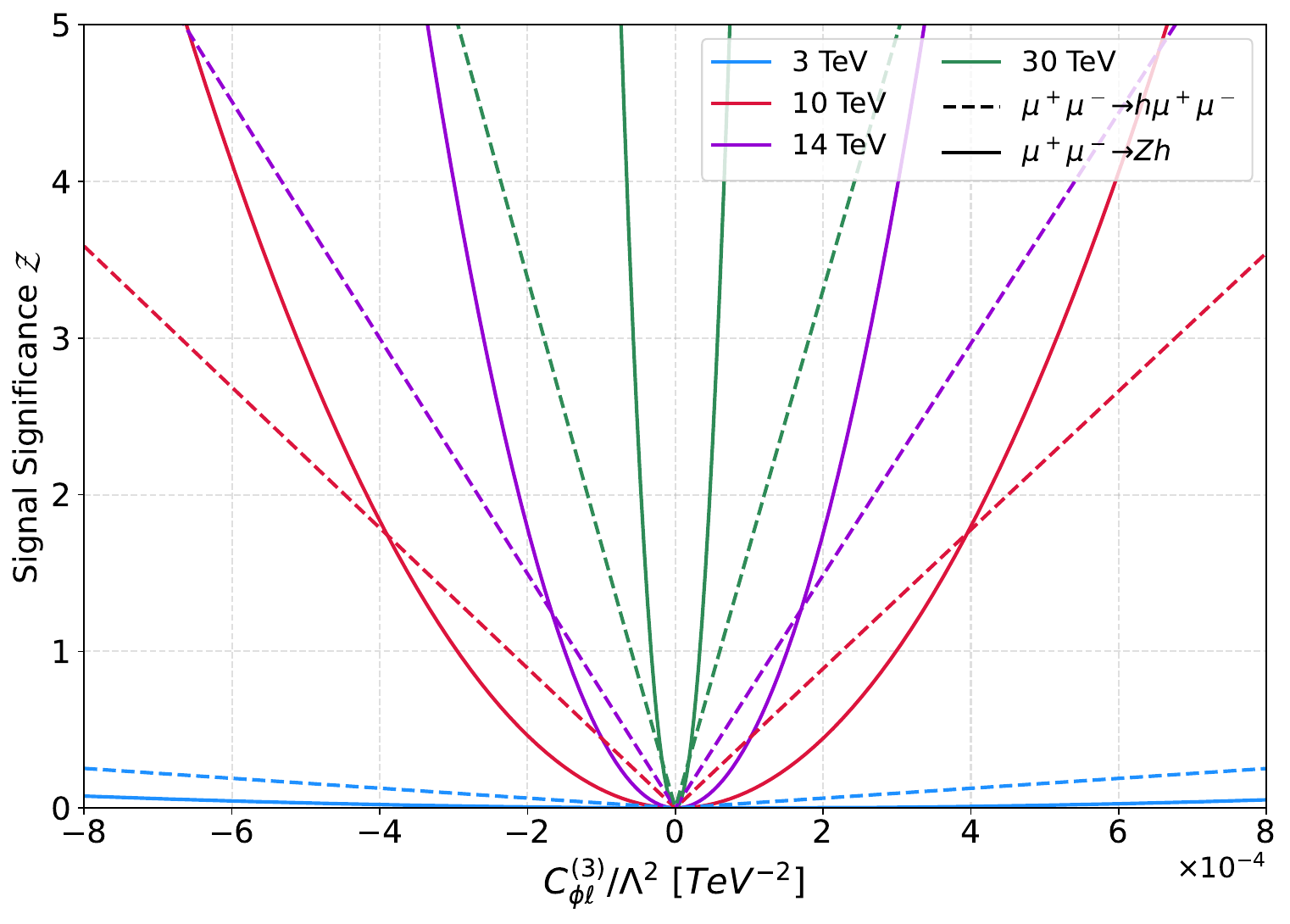}&
    \includegraphics[width=7.5cm,height=5.7cm]{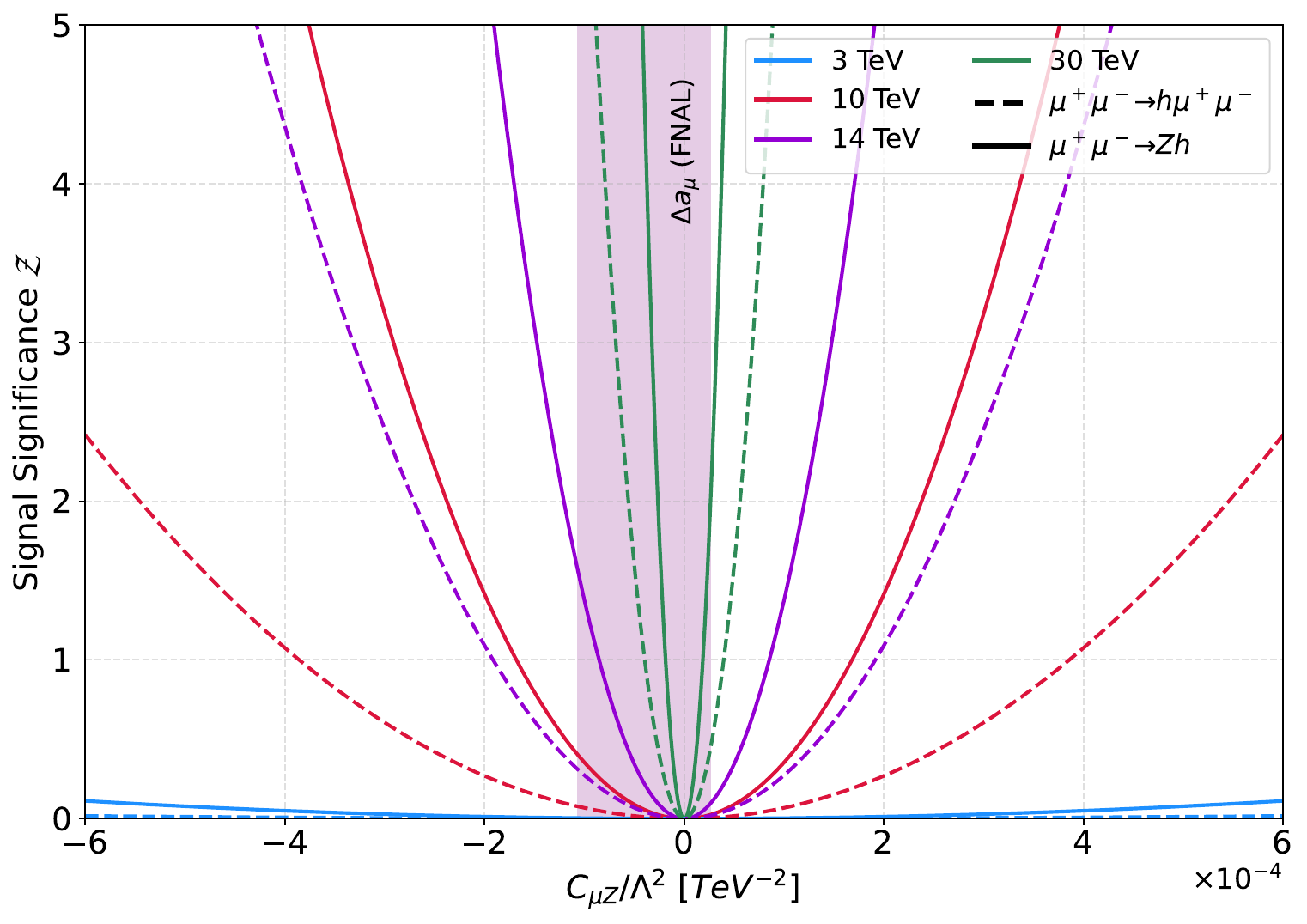}\\
    (a)&(b)
	  \end{tabular}
	\caption{ Sensitivity reach for Wilson coefficients (a) $\frac{C_{\varphi \ell}^{(3)}}{\Lambda^2}$ and  (b) $\frac{C_{\mu Z}}{\Lambda^2}$ with  $\mu^+\mu^- \to h Z$ (solid) and $\mu^+\mu^- \to h \mu^+\mu^-$ (dashed) 
    at different collision energies and corresponding luminosities. The $95\%$ C.L. constraints on anomalous magnetic moment of muon is shown in purple band.}
	\label{fig:significance_vs_C}
\end{figure}
\begin{figure}[b]
	\centering%
	  \begin{tabular}{cc}
	\includegraphics[width=7.5cm,height=5.7cm]{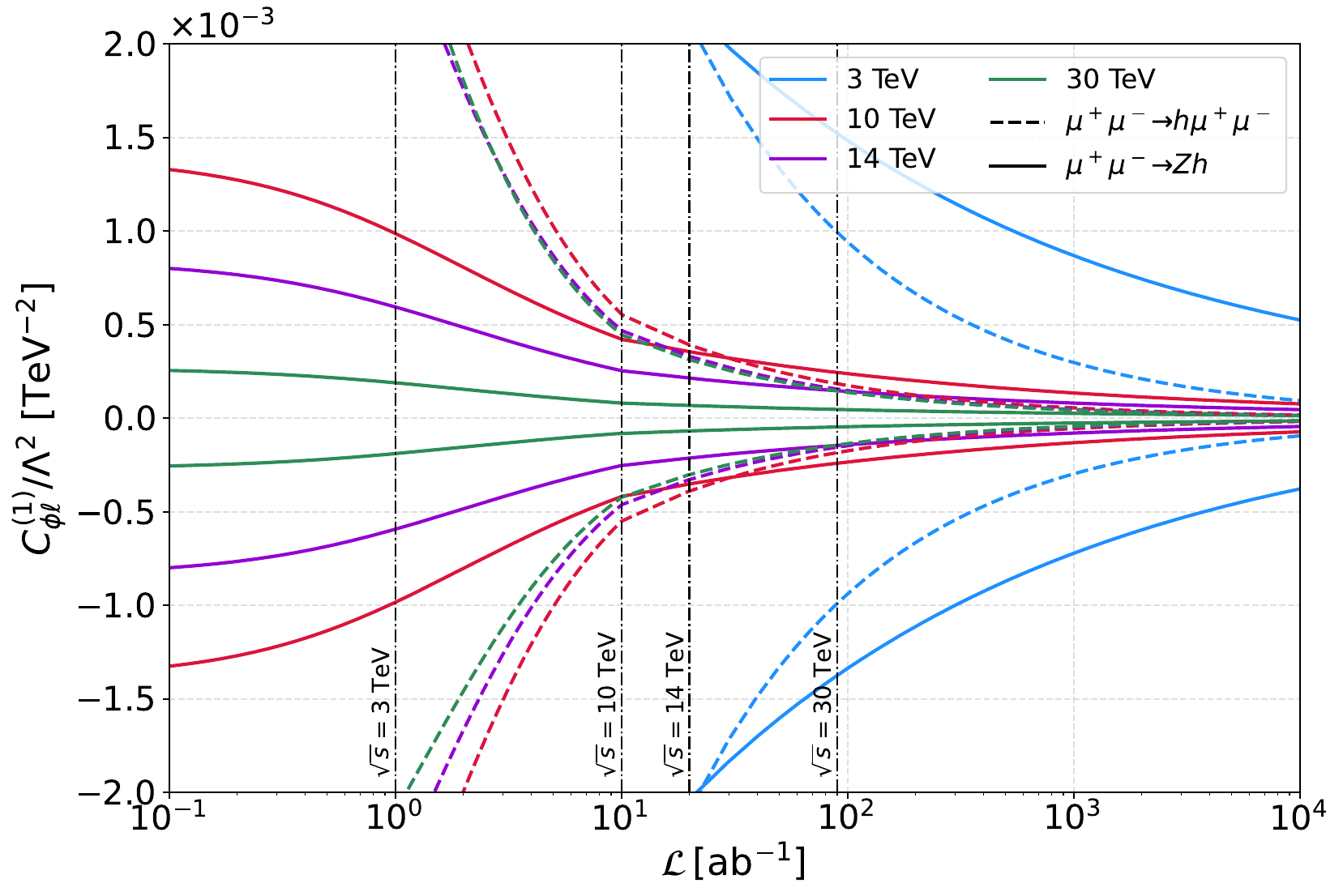}&
	\includegraphics[width=7.5cm,height=5.7cm]{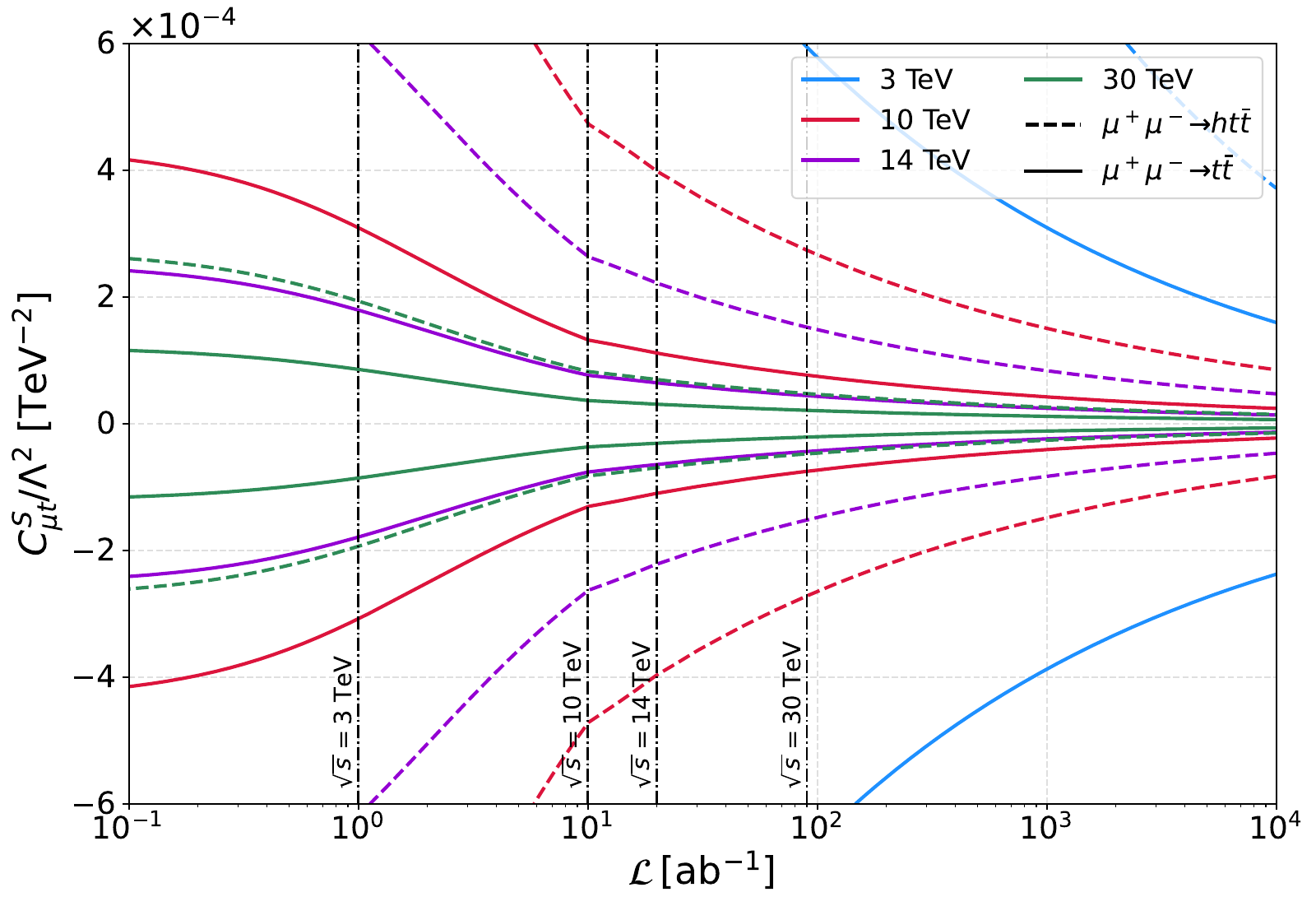}\\
    (a)&(b)
	  \end{tabular}
	\caption{Constraints on the Wilson coefficients (a) $\frac{C_{\varphi \ell}^{(1)}}{\Lambda^2}$ from $\mu^+\mu^- \to h Z$ (solid), $\mu^+\mu^- \to h \mu^+\mu^-$(dashed), (b) $\frac{C_{\mu t}^S}{\Lambda^2}$  from $\mu^+\mu^- \to t \bar{t}$ (solid) and  $\mu^+\mu^- \to t \bar{t} h $ (dashed) at $2 \sigma$ level, as a function of the integrated luminosity with $\sqrt{s}= [3,10,14,30]$ TeV.}
	\label{fig:reqd_lumi}
\end{figure}
Table~\ref{tab:future_limits_1} summarises the projected 95\% C.L. limits on various SMEFT coefficients obtained from our analysis at a $\sqrt{s}=10~\mathrm{TeV}$ muon collider. The limits are presented for both single-parameter and marginalised fits. The clean experimental environment with precisely known initial state and strong sensitivity to electroweak interactions allow the muon collider to probe both two-fermion and four-fermion operators with a precision that is improved by one to two orders of magnitude compared to those expected at HL-LHC and electron colliders (see Table~\ref{tab:future_limits}). This also highlights the ability of high-energy muon collisions to disentangle operator directions that are otherwise degenerate at the LHC and to probe chirality-flipping tensor and scalar structures through final states involving top quarks (see Fig.~\ref{ttbar_distributions} (right) and Fig.~\ref{fig:phimumu_asy_SM} (a)). For comparison, Table~\ref{tab:future_limits} presents the corresponding projected bounds from the HL-LHC and FCC\mbox{-}ee~\footnote{Throughout this work, ``FCC\mbox{-}ee'' refers to projections corresponding to the full running scenario with $\sqrt{s}=91$, $161$, $240$, and $365$ GeV.}, taken respectively from~\cite{Celada:2024mcf,Cornet-Gomez:2025jot}. At the HL-LHC, single-parameter fits yield meaningful bounds on most electroweak operators, but the marginalised limits for several four-fermion coefficients ($C_{\ell q}^{(-)}$, $C_{\ell u}$, $C_{eu}$, $C_{eq}$) remain unconstrained due to limited statistics and strong parameter correlations in processes such as $pp\to t\bar{t}\ell^+\ell^-$. The inclusion of precision electroweak and fermion-pair observables from FCC\mbox{-}ee breaks these degeneracies and provides independent sensitivity to distinct chiral and flavour structures, leading to well-defined limits on all operators~\cite{Bellafronte:2025ubi}.

Fig.~\ref{fig:significance_vs_C} presents the sensitivity reach for the aforementioned Higgs production processes from muon collisions  in terms of number of standard deviations for $\sqrt{s}: [3,10,14,30]$ TeV,  assuming the integrated luminosities of $[1,10,20,90]$ ab$^{-1}$, respectively. For the left-handed current $C_{\varphi \ell}^{(3)}/\Lambda^{2}$ (Fig.~\ref{fig:significance_vs_C} (a)), both $Zh$ and $\mu^+\mu^- h$ channels show a monotonic improvement with energy, reaching a $2\sigma$ sensitivity at the $\mathcal{O}(10^{-4})$ $\text{TeV}^{-2}$ level by 14 TeV, with further gains at 30 TeV where {\it{ZBF}} becomes competitive with $Zh$. 
For $C_{\mu Z}/\Lambda^2$-dependent processes, $Zh$ and {\it{ZBF}} (Fig.~\ref{fig:significance_vs_C} (b)), the significance increases steadily with energy, reaching $2\sigma$ for $C_{\mu Z}/\Lambda^2 \sim 10^{-5}~\text{TeV}^{-2}$ at $\sqrt{s}=30$ TeV. The purple band in Fig.~\ref{fig:significance_vs_C} (b) denotes the 95\% C.L. range of $C_{\mu Z}$ inferred from the anomalous magnetic moment of the muon, $\Delta a_\mu$. A 30 TeV muon collider in the $Zh$  channel can probe this coefficient independently at a precision comparable to that required to explain  $\Delta a_\mu$, whereas a 10 TeV collider would fall short of reaching this small value at the $2\sigma$ level.

Fig.~\ref{fig:reqd_lumi} presents the $2\sigma$ constraints on the Wilson coefficients $\frac{C_{\varphi \ell}^{(1)}}{\Lambda^2}$ and $\frac{C_{\mu t}^S}{\Lambda^2}$ as a function of the integrated luminosity, for the processes (a) $\mu^+\mu^- \to Zh$ (solid), $\mu^+\mu^- h$ (dashed), (b) $\mu^+\mu^- \to t\bar{t}$ (solid)  and $t\bar{t}h$ (dashed). The results are shown for various c.o.m energies in the range $\sqrt{s} = 3$--$30$ TeV. A 10 TeV collider with $\mathcal{L}_{\rm int}=10$ ab$^{-1}$ can constrain $|C_{\phi \ell}^{(1)}|$ down to the $10^{-4}$ TeV$^{-2}$ level (Fig.~\ref{fig:reqd_lumi} (a)), while higher-energy stages such as 14 TeV or 30 TeV could improve the precision by nearly an order of magnitude. Fig.~\ref{fig:reqd_lumi} (b) shows a similar dependence for the scalar four-fermion operator $C_{\mu t}^{S}$ extracted from the  $t \bar{t}$ and  $t \bar{t} h$ final states. As is expected, the sensitivity to the WCs improves gradually with increasing luminosity. Thus, high-energy measurements at muon colliders can match, and in some cases surpass, the reach of precision low-energy flavour observables for several classes of non-standard interactions.

We now discuss how our results depend on the assumed detector performance. The muon-collider Delphes card employed in this analysis represents, at present, a benchmark parametrisation rather than a finalised detector design. For the $Zh$ and $ZBF$ analyses with $h\to b\bar b$, the key detector-level inputs are the $b$-tagging efficiency and the jet momentum resolution, the latter controlling the reconstructed $b\bar b$ invariant-mass resolution. For the $ZBF$ topology, the acceptance for the final-state muons also plays an important role. For the $t\bar t$ and $t\bar t h$ channels, the relevant detector effects are expected to be associated with boosted-top reconstruction, jet-energy resolution, and flavour tagging. To assess the impact of detector performance on our projections, we vary two representative parameters. First, we study the $b$-tagging efficiency by comparing the baseline working point, with a tagging efficiency of $70\%$, to a degraded case with $50\%$ efficiency and an idealised case with $100\%$ efficiency. The corresponding effects on the signal and background event yields for the benchmark points considered in this study are summarised in Table~\ref{tab:detector_btag}. Second, we vary the reconstructed Higgs-mass resolution by scaling the jet momentum smearing $\Delta p_T/p_T$. We take $5\%$ as the baseline, while the degraded and improved cases correspond to $10\%$ and $2.5\%$, respectively. We also include the perfect-resolution limit in Table~\ref{tab:higgs_mass_resolution}.

\begin{table}[t]
\centering
\scriptsize
\renewcommand{\arraystretch}{1.25}
\resizebox{\textwidth}{!}{
\begin{tabular}{||c|c|ccc|ccc|ccc||}
\hline\hline
{Channel} 
&{Benchmark}
& \multicolumn{3}{c|}{Degraded $b$-tagging}
& \multicolumn{3}{c|}{Baseline $b$-tagging}
& \multicolumn{3}{c||}{Perfect $b$-tagging} \\
\cline{3-11}
& & $N_{\rm total}$ & $B$ & $\mathcal Z$
& $N_{\rm total}$ & $B$ & $\mathcal Z$
& $N_{\rm total}$ & $B$ & $\mathcal Z$ \\
\hline
{$Zh$} 
& BP1 &318  & 136 & 13.3 & 875 & 289 & 27.8& 1047 &381  & 28.2 \\
& BP2 &235  & 136 & 7.67 & 434 & 289 & 8.07 & 623 & 381 & 11.6 \\
& BP3 & 171 & 136 & 2.89 & 368 & 289 & 4.61 & 491 & 381 & 5.67 \\
\hline
{ZBF $h$} 
& BP1 & 252 & 217 & 3.45 & 555 & 490 & 4.22 & 697 & 551 & 7.45 \\
& BP2 & 225 & 217 & 1.66 & 600 & 490 & 6.14 & 733 & 551 & 8.33 \\
& BP3 & 200 & 217 & 0.09 & 537 & 490 & 3.49 & 622 & 551 & 4.50 \\
\hline
{$t\bar t$} 
& BP4 & 1289 & 1078 & 6.24 & 2965 & 2724 & 14.4 & 3600 & 3422 & 19.5 \\
& BP5 & 1793 & 1078 & 19.9 & 4613 & 2724  & 32.9 & 4735 & 3422 & 36.4 \\
& BP6 & 450 & 1078 & 5.28 & 1406 & 2724 & 6.49 & 1745 & 3422 & 8.70 \\
& BP7 & 1256 & 1078 & 5.82 & 2525 & 2724 & 6.83 & 2746 & 3422 & 9.68 \\
\hline
{$t\bar t h$} 
& BP4 & 48 & 14 & 7.73 & 129 & 49 & 10.7 & 164 & 68 & 10.9 \\
& BP5 & 58 & 14 & 9.36 & 186 & 49 & 16.0 & 241 & 68 & 17.8 \\
& BP6 & 16 & 14 & 1.51 & 72 & 49 & 4.53 & 107 & 68 & 5.06 \\
& BP7 & 19 & 14 & 2.20 & 105 & 49 & 8.10  & 129 & 68 & 8.19 \\
\hline\hline
\end{tabular}}
\caption{Dependence of the projected sensitivities on the assumed $b$-tagging performance at a $10~{\rm TeV}$ muon collider with ${\cal L}=10~{\rm ab}^{-1}$. We present the effects on the signal and background events for the various benchmark points in this study (Table~\ref{BM_points}) with three cases : degraded b-tagging efficiency of $50\%$, the baseline $b$-tagging efficiency of $70\%$ and perfect case of $100\%$  efficiency. For each benchmark point, $N_{\rm total}$ is the total event yield in the SMEFT, $B$ is the total SM background events after applying the selection efficiency cuts, the signal events correspond to $S = N_{\rm total}-N_{\rm SM}$ (Eq.~\ref{eq:S}), and $\mathcal{Z}$ is the signal significance from Eq.~\ref{signal_signi}.}
\label{tab:detector_btag}
\end{table}
\begin{table}[h]
\centering
\scriptsize
\renewcommand{\arraystretch}{1.25}
\resizebox{\textwidth}{!}{
\begin{tabular}{||c|c|ccc|ccc|ccc|ccc||}
\hline\hline
Channel 
& Benchmark
& \multicolumn{3}{c|}{Degraded}
& \multicolumn{3}{c|}{Baseline}
& \multicolumn{3}{c|}{Improved} 
& \multicolumn{3}{c||}{Perfect} \\
\cline{3-14}
& 
& $N_{\rm total}$ & $B$ & $\mathcal{Z}$
& $N_{\rm total}$ & $B$ & $\mathcal{Z}$
& $N_{\rm total}$ & $B$ & $\mathcal{Z}$ 
& $N_{\rm total}$ & $B$ & $\mathcal{Z}$ \\
\hline
$Zh$ 
& BP1 & 596 & 199 & 22.9 & 875 & 289 & 27.8 & 916 & 303 & 28.4 &916 & 302 & 28.4 \\
& BP2 & 295 & 199 & 6.59 & 434 & 289 & 8.07 & 465 & 303 & 8.77  &465 & 302 & 8.78 \\
& BP3 & 251 & 199 & 3.80 & 368 & 289 & 4.61 & 389 & 303 & 4.89  & 389 & 302 & 4.90 \\
\hline
ZBF $h$ 
& BP1 & 395 & 353 & 3.61 & 555 & 490 & 4.22 & 581 & 515 & 4.40  & 581 & 515 & 4.40 \\
& BP2 & 429 & 353 & 5.29 & 600 & 490 & 6.14 & 631 & 515 & 6.44  & 631 & 515 & 6.44 \\
& BP3 & 381 & 353 & 2.91 & 537 & 490 & 3.49 & 561 & 515 & 3.57  & 561 & 515 & 3.57 \\
\hline
$t\bar t h$ 
& BP4 & 93 & 42 & 8.26 & 129 & 49 & 10.7 & 133 & 51 & 10.8  & 133 & 50 & 10.9 \\
& BP5 & 132 & 42 & 12.4 & 186 & 49 & 16.0 & 191 & 51 & 16.1  &191 & 50 & 16.2 \\
& BP6 & 49 & 42 & 2.88 & 72 & 49 & 4.53 & 75 & 51 & 4.69  & 75 & 50 & 4.73\\
& BP7 & 75 & 42 & 5.79 & 105 & 49 & 8.10 & 106 & 51 & 8.12  & 106 & 50 & 8.14\\
\hline\hline
\end{tabular}}
\caption{Dependence of the signal sensitivity on the reconstructed Higgs mass resolution at a $10~{\rm TeV}$ muon collider with ${\cal L}=10~{\rm ab}^{-1}$. The Higgs candidate is reconstructed from the $b\bar b$ system, whose invariant-mass resolution is governed by the jet momentum resolution $\Delta p_T/p_T$. The baseline corresponds to a fractional Higgs mass resolution $\sigma_{m_{b\bar b}}/m_h \simeq 5\%$, while the degraded and improved cases correspond to $10\%$ and $2.5\%$, respectively, and the ``Perfect'' column to the unsmeared limit. The Higgs mass window $m_h \in [115,135]$~GeV and all other selections are fixed across the four cases, so that only the mass resolution is varied. The definitions of $N_{\rm total}$, $B$, and $\mathcal{Z}$ are the same as in the $b$-tagging case of Table~\ref{tab:detector_btag}.}
\label{tab:higgs_mass_resolution}
\end{table}

We find that the significance is strongly sensitive to the $b$-tagging performance.  Degrading the efficiency to $50\%$ reduces $\mathcal{Z}$ substantially for several benchmark points. The dependence on the Higgs-mass resolution is weaker. Improving the resolution beyond the baseline gives only a marginal gain, while degrading it by a factor of two changes the sensitivity by up to about $30\%$. For the $Zh$ and $ZBF$ channels, this behaviour can be understood from the fact that the dominant SM $Zh$ and $\mu^+\mu^-h$ contributions peak in the same $b\bar b$ invariant-mass region as the SMEFT signal. Thus, although the Higgs mass-window selection is important for suppressing non-Higgs backgrounds, it provides limited additional discrimination when the mass resolution is varied. The sensitivity is therefore driven mainly by the energy-enhanced high-$p_T$ tails.

\section{Implications on BSM physics}
\label{section7}
So far, we have analysed the effects of dimension-6 operators involving muons and derived projections on the values of WCs  at a muon collider setup with sub-per-mille level sensitivity. A key objective is to interpret these constraints in terms of UV complete scenarios and assess the resulting implications for new physics. Within a bottom-up framework, SMEFT serves as an interface connecting  the {\emph {low-energy}} theory to various UV models. Thus, the various BSM parameters of different UV models which need not be related at the UV scale, are on the same footing in the EFT framework. 
We focus on BSM scenarios that can be described as weakly coupled, renormalisable theories. In such frameworks, operators may be classified according to whether they appear at tree level or are necessarily loop-induced in the presence of decoupling dynamics~\cite{Arzt:1994gp}. 
In the following, we consider two explicit UV scenarios, namely the SM extended with vector-like leptons and scalar leptoquarks and map their parameters onto the SMEFT operator basis relevant for our analysis in the heavy-mass limit. The resulting constraints are then compared with current direct search bounds and projected sensitivities at a 10 TeV muon collider.

\subsection{Model with Vector-like Lepton}
We extend the SM with a new {\emph {vector-like lepton}} $E$,  an electroweak singlet with hypercharge $-1$. The relevant part of the Lagrangian involving the heavy lepton is given by~\cite{DasBakshi:2020ejz}:
\begin{equation}
  \mathcal{L}_E = 
   \bar{E}\,(i\slashed{D} - M_E)\,E
  - \Big[ \lambda_i\, \bar{\ell}_i\, \phi\, E_R + \text{h.c.} \Big],
  \label{Lag_VLL}
\end{equation}
where $\ell_i$ denotes the SM lepton doublets,  $\phi$ is the Higgs field, and $i$ corresponds to the lepton flavours.  The parameter $M_E$  is the mass of $E$ and  $\lambda_i^{'}s$ are complex Yukawa couplings, respectively. 
The dominantly contributing  SMEFT interactions appear in the muon sector through the operator coefficients $[C_{\varphi \ell}^{(1)}]_{\mu\mu}$ and $[C_{\varphi \ell}^{(3)}]_{\mu\mu}$. We present the matching relations for the model with these coefficients, after integrating out the heavy lepton up to one-loop in Appendix~\ref{VLL_matching}~\cite{Chakrabortty:2023yke,Fuentes-Martin:2022jrf,Carmona:2021xtq,Guedes:2023azv, Guedes:2024vuf}.  At tree level, these coefficients are equal, and their sum $(S_\mu)$ directly modifies the effective left-handed $Z\mu\mu$ coupling:
\begin{equation}
[C_{\varphi \ell}^{(1)}]_{\mu\mu} = [C_{\varphi \ell}^{(3)}]_{\mu\mu} 
= -\frac{|\lambda_\mu|^2}{4 M_E^2}, 
\qquad
S_\mu \equiv \big([C_{\varphi \ell}^{(1)}]+[C_{\varphi \ell}^{(3)}]\big)_{\mu\mu} 
= -\frac{|\lambda_\mu|^2}{2 M_E^2}.
\end{equation}
\begin{table}[b]
\centering
\renewcommand{\arraystretch}{1.25}
\resizebox{\textwidth}{!}{%
\begin{tabular}{|c|c|c|c|}
\hline
Decay channel
& \multicolumn{3}{c|}{$M_E$ reach at $95\%$ C.L.} \\
\cline{2-4}
at  [$\sqrt{s} ~/~\mathcal{L}~$]& LHC [$13$ TeV, $139~\mathrm{fb}^{-1}$] 
& HL--LHC [$14$ TeV, $3~\mathrm{ab}^{-1}$ ]
& HE--LHC [$27$ TeV, $15~\mathrm{ab}^{-1}$] \\
\hline\hline
SM-like BRs 
& $M_E \gtrsim 405~\mathrm{GeV}$ 
& $M_E \gtrsim 785~\mathrm{GeV}$ 
& $M_E \gtrsim 1295~\mathrm{GeV}$ \\
\hline
$\mathrm{BR}(E \to Z\ell)=1$ 
& $M_E \gtrsim 630~\mathrm{GeV}$ 
& $M_E \gtrsim 1090~\mathrm{GeV}$ 
& $M_E \gtrsim 1770~\mathrm{GeV}$ \\
\hline
$\mathrm{BR}(E \to \ell A_H)=1$ 
& $M_E \gtrsim 895~\mathrm{GeV}$ 
& $M_E \gtrsim 1450~\mathrm{GeV}$ 
& $M_E \gtrsim 1965~\mathrm{GeV}$ \\
\hline
\end{tabular}}
\caption{Direct search exclusions and projected sensitivities for a vector-like lepton singlet from pair production at hadron colliders~\cite{Guedes:2021oqx}.}
\label{tab:vll_direct_recast}
\end{table}
\begin{figure}[t]
 \centering
		\includegraphics[width=10.2cm,height=6.2cm]{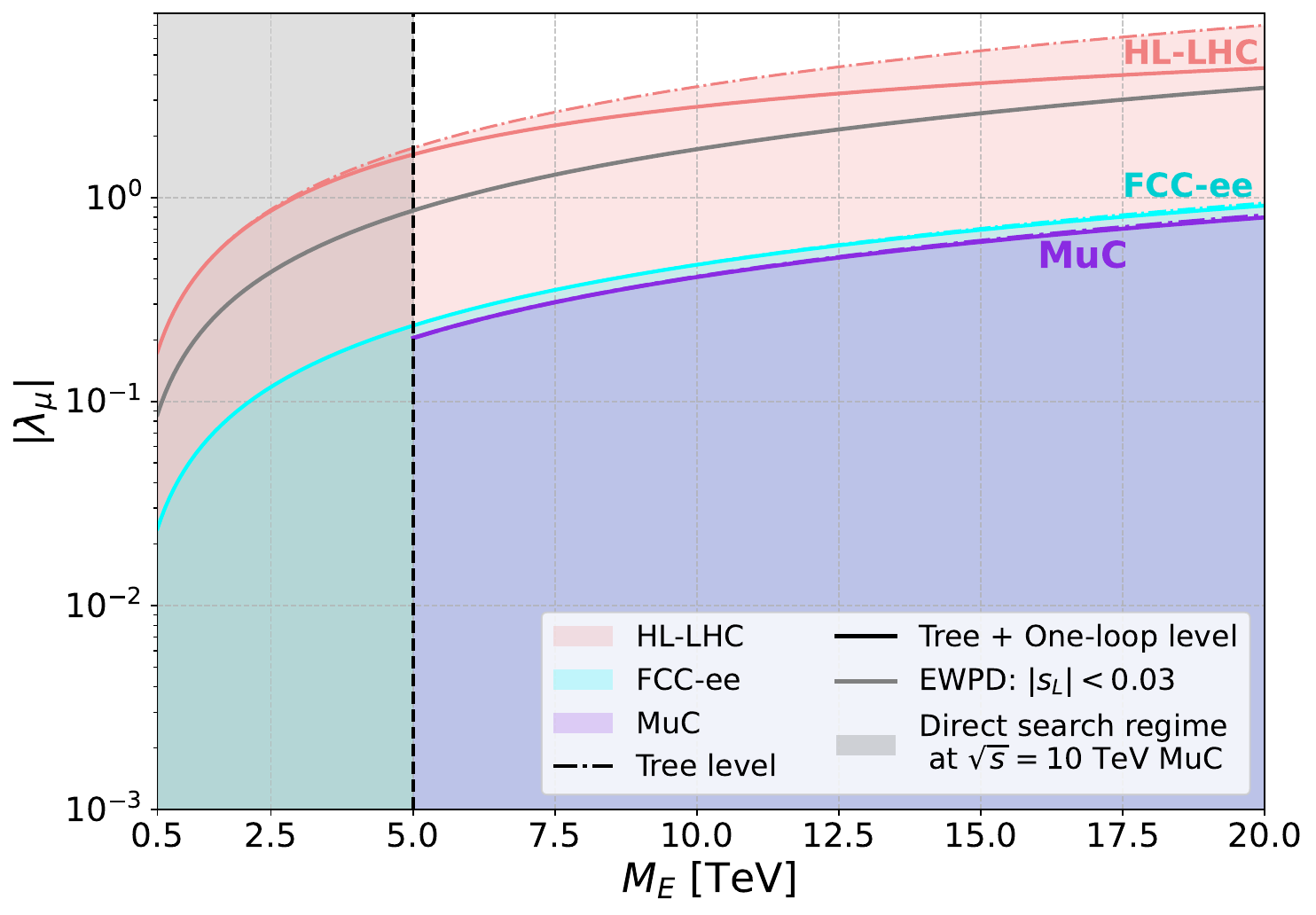}
\caption{Projected bounds on the vector-like lepton Yukawa coupling $|\lambda_\mu|$ as a function of the vector-like lepton mass $M_E$. Filled-bands with dot-dashed (solid) curves show the $95\%$ C.L. upper bound on $|\lambda_\mu|$ from $\delta g_L$ at HL-LHC (red), FCC\mbox{-}ee (blue), and a muon collider (violet) using tree level matching (up to one-loop corrections). The solid grey line shows the electroweak precision constraint from \(|sin~\theta_L| < 0.03\). The grey shaded region denotes the kinematic regime where the vector-like lepton can be produced on shell at a 10 TeV muon collider. In this region, a description in terms of the full UV theory is applicable, while for larger masses the SMEFT interpretation remains valid.} \label{fig:compamu}
\end{figure}
Coefficients such as $[C_{\varphi e}]_{\mu\mu}$ and $\{[C_{e W}]_{\mu\mu}, [C_{e B}]_{\mu\mu}\}$ are loop-suppressed and do not provide competitive constraints with respect to $[C_{\varphi \ell}^{(1,3)}]_{\mu \mu}$. Beyond the Higgs-current operators, four-fermion contact terms are generated at one-loop, suppressed by $1/(16\pi^{2}M_E^{2})$ and gauge factors. They include $\mathcal{O}_{\ell q}^{(1,3)}$, $\mathcal{O}_{eq}$, $\mathcal{O}_{eu}$, $\mathcal{O}_{\ell u}$, and $\mathcal{O}_{\ell equ}^{S}$. 

In Fig.~\ref{fig:compamu}, we present the indirect constraints from $\delta g_L$ (bounded by dot-dashed (at tree level) and solid (up to one-loop) curves, respectively) from the collider analysis in the previous section. The shaded region below each curve is allowed at $95\%$ C.L. for the corresponding collider scenarios. We also show a model-independent mixing bound (grey curve) where the muon–$E$ mixing angle satisfies ${\rm{sin}}~\theta_L\simeq \lambda_\mu v/(\sqrt{2} M_E)$, and electroweak precision data leads to $|\rm{sin}~\theta_L|\lesssim 0.03$ at $95\%$ C.L.~\cite{Chala:2020odv} for muon mixing. 
 We have considered $S_\mu$ both at tree level and including one-loop effects, with the renormalisation scale fixed to $\mu_M = m_Z$ in the logarithmic terms (see Appendix~\ref{VLL_matching}). We find that the one-loop corrections are numerically subdominant compared to the tree level contributions.
In Table~\ref{tab:vll_direct_recast}, we summarise the current LHC limits on $M_E$ from direct searches for pair production for three benchmark decay patterns: SM-like branching fractions, $\mathrm{BR}(E \to Z\ell)=1$, and $\mathrm{BR}(E \to \ell A_H)=1$. At $\sqrt{s}=13~\mathrm{TeV}$ and $\mathcal{L}_{\rm int}=139~\mathrm{fb}^{-1}$, the corresponding bounds are $M_E \gtrsim \{405,630,895\} ~\mathrm{GeV}$~\cite{Guedes:2021oqx}. Fig.~\ref{fig:compamu} and Table~\ref{tab:vll_direct_recast} illustrate the strong complementarity between direct searches and indirect probes. While direct pair production sets lower limits on $M_E$, indirect constraints from $\delta g_L$ probe much smaller Yukawa couplings and dominate the sensitivity for $M_E$ beyond the direct reach. In particular, FCC\mbox{-}ee and a multi-TeV muon collider significantly extend the accessible parameter space, providing sensitivity to multi-TeV vector-like leptons well within the regime where the SMEFT description is valid.

\subsection{Model with Scalar Leptoquarks : The $S_1 +S_3$ Scenario}
\label{SLQ}
 Leptoquarks are bosonic fields that carry QCD colour and weak isospin, allowing renormalisable couplings between leptons and quarks. We consider a UV completion that extends the SM by adding  two colour triplet {\emph{scalar leptoquarks}},
  $S_{1} \;(\bar{3},\,1,\tfrac{1}{3}),$
  and
  $S_{3} \;(\bar{3},\,3,\tfrac{1}{3})$,
with their  $(SU(3)_c,\,SU(2)_L,\, {U(1)_Y})$ representations. This pair has been studied as a simultaneous explanation of the charged and neutral current $B$ anomalies~\cite{Bernardi:2022hny,Buttazzo:2017ixm,Crivellin:2017zlb,Chakraverty:2001yg,Arnan:2019olv,Yan:2019hpm,Bigaran:2019bqv,Crivellin:2019dwb}. 
The part of the Lagrangian involving $S_{1}$ and $S_{3}$, relevant for our discussion below is given by~\cite{Buttazzo:2017ixm}
\begin{eqnarray}
	\mathcal{L}_{\text{LQ}}^{\rm Yukawa} &=& \left( (\lambda^{1L})_{i\alpha} \bar q^c _i \epsilon \ell _\alpha
			+ (\lambda^{1R})_{i\alpha} \bar u^c _i   e _\alpha  \right) S_1 
			+ (\lambda^{3L})_{i\alpha}\bar q^c _i \epsilon \sigma^I \ell _\alpha S_3^I + \text{h.c.}
\end{eqnarray}
where $\epsilon=i\sigma _2$ and  $(\lambda^{1L})_{i\alpha}, (\lambda^{1R})_{i\alpha}, (\lambda^{3L})_{i\alpha}  \in \mathbb{C}$.

Integrating out the heavy scalar leptoquarks $S_{1,3}$ generates at tree level, the semi-leptonic four-fermion operators
$O_{\ell q}^{(1,3)}$, $O_{eu}$, and $O_{\ell equ}^{S,T}$ with the corresponding matching relations~\cite{Gherardi:2020det}:
\begin{eqnarray}
\big[C_{\ell q}^{(1)}\big]_{ij \alpha\beta}
=\frac{\lambda^{1L*}_{\alpha i}\lambda^{1L}_{\beta j}}{4M_1^2}
+\frac{3\,\lambda^{3L*}_{\alpha i}\lambda^{3L}_{\beta j}}{4M_3^2},\quad
\big[C_{\ell q}^{(3)}\big]_{ij \alpha\beta}
=-\frac{\lambda^{1L*}_{\alpha i}\lambda^{1L}_{\beta j}}{4M_1^2}
+\frac{3\lambda^{3L*}_{\alpha i}\lambda^{3L}_{\beta j}}{4M_3^2}, \nonumber \\
\big[C_{eu}\big]_{ij \alpha\beta}
=\frac{\lambda^{1R*}_{\alpha i}\lambda^{1R}_{\beta j}}{2M_1^2},\qquad
\big[C_{\ell equ}^{S}\big]_{ij \alpha\beta}
=\frac{\lambda^{1L*}_{\alpha i}\lambda^{1R}_{\beta j}}{2M_1^2},\qquad
\big[C_{\ell equ}^{T}\big]_{ij \alpha\beta}
=-\frac{\lambda^{1L*}_{\alpha i}\lambda^{1R}_{\beta j}}{8M_1^2}.
\label{matchingrelation_treelevel}
\end{eqnarray}

\begin{figure}[b]
			\begin{center}
				\subfloat{
					\begin{tabular}{cc}
		\includegraphics[width=7.5cm,height=5.7cm]{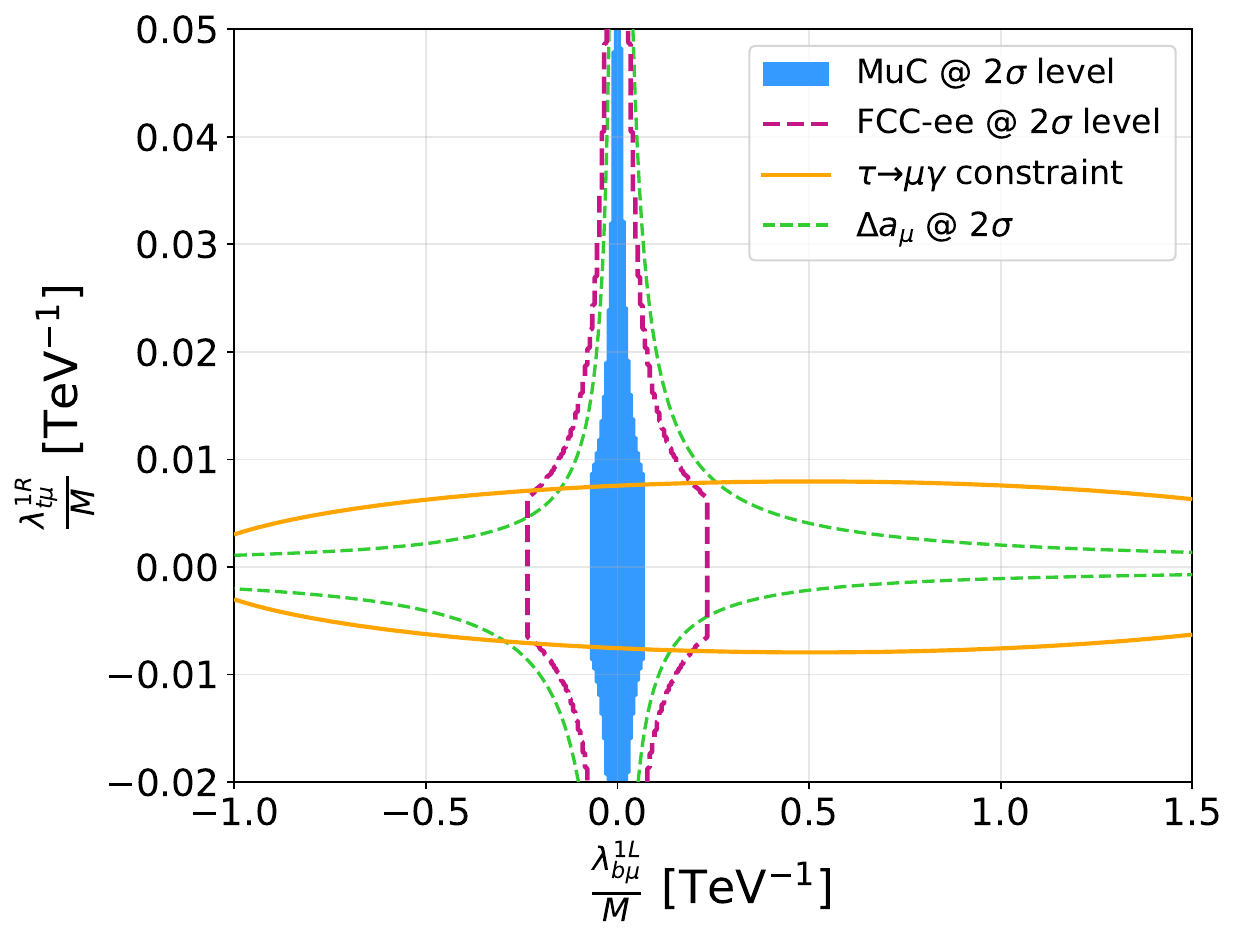}&
		\includegraphics[width=7.5cm,height=5.7cm]{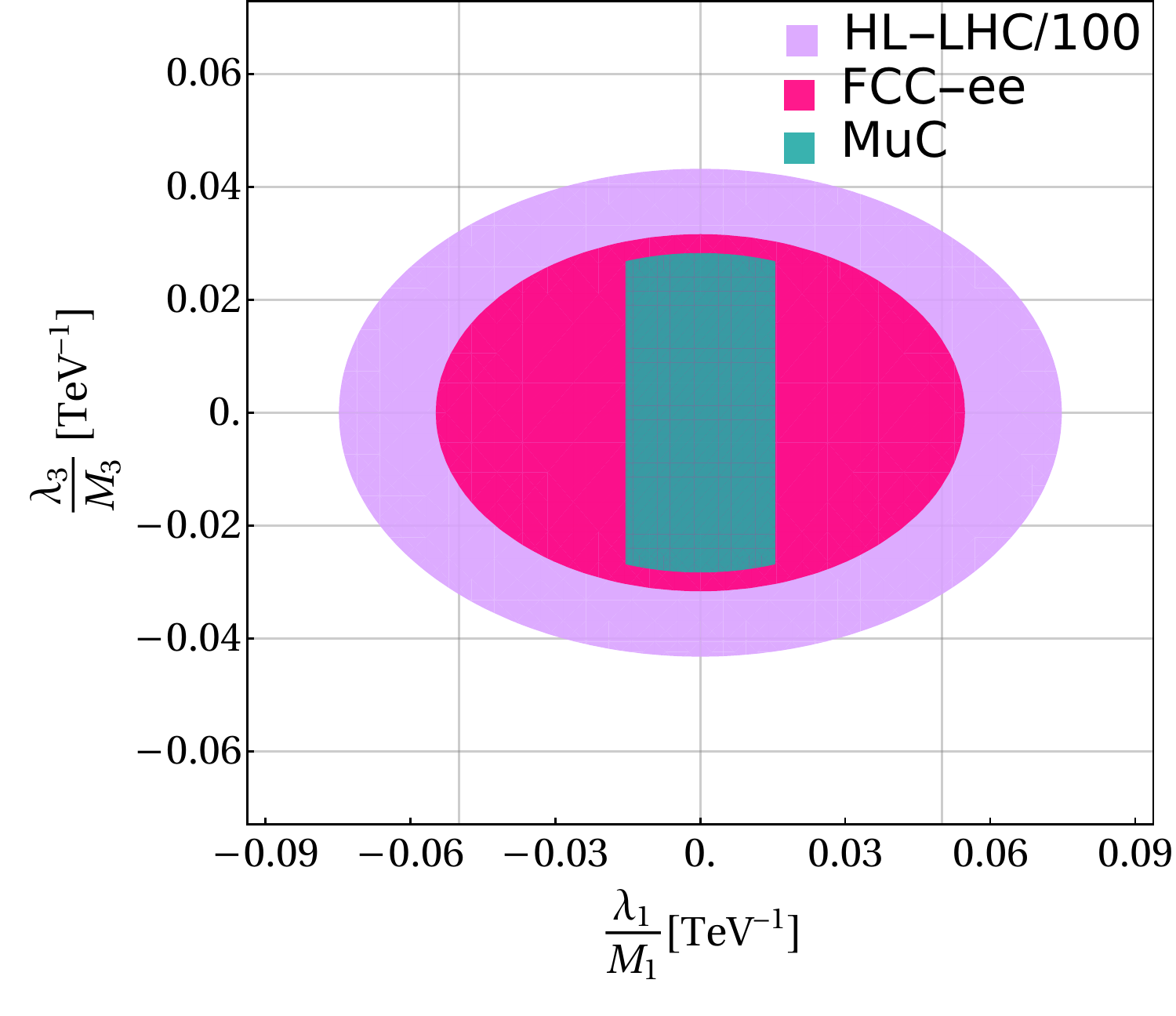}\\
						(a)&(b)
				\end{tabular}}
		\caption{(a) Contours in the $(\lambda^{1L}_{b\mu}/M,\,\lambda^{1R}_{t\mu}/M)$ plane for $M_1=M_3=M$. The blue region corresponds to the parameter space that can be constrained at $2\sigma$ level at a 10 TeV muon collider with 10 ab$^{-1}$. The solid orange contour indicates the current $95\%$ C.L. exclusion from $\text{Br}(\tau \to \mu \gamma) < 5.24 \times 10^{-8}$.  Green dashed lines correspond to the $2\sigma$ interval for $\Delta a_\mu$ and violet dashed line corresponds to constraint from FCC\mbox{-}ee at $2\sigma$ level. (b) Constraints on the scalar leptoquark model in $\lambda_1/M_1-\lambda_3/M_3$ plane from the four-fermion fit. The HL-LHC sensitivity is 100 times weaker along both axes than shown in the plot for illustrative comparison.}
        \label{scalarlqconstraints}
			\end{center}
\end{figure}
At one-loop level, the gauge, Yukawa and leptoquark-quartic interactions generate two-fermion operators $\mathcal{O}_{\varphi\ell}^{(1,3)}$, $\mathcal{O}_{\varphi e}$ and the dipole operators $\mathcal{O}_{\mu W}$, $\mathcal{O}_{\mu B}$ (along with purely bosonic terms), with explicit $\log(\mu_M^2/M_{1,3}^2)$ dependence~\cite{Gherardi:2020det}. These contributions are subject to strong limits from precision electroweak observables and dipole measurements, and together they constitute the low-energy operator basis that captures  the radiative imprints in presence of the leptoquark sector.
 For corroborating collider searches for $S_1,S_3$ see Ref.~\cite{Desai:2023jxh} and for global, low-energy fits including $(g-2)_\mu$ and $\tau\to\mu\gamma$, the benchmark values are given in Ref.~\cite{Calibbi:2020emz}. In Fig.~\ref{scalarlqconstraints} (a), we show the plane $(\lambda^{1L}_{b\mu}/M,\lambda^{1R}_{t\mu}/M)$ that is scanned at $M_1=M_3=M$ using the SMEFT mappings 
$C_{\ell equ}^{(S,T)} \propto \lambda^{1L}\lambda^{1R}/M^2$, $C_{eu} \propto (\lambda^{1R})^2/M^2$, $C_{\ell q}^{(-)} \equiv C_{\ell q}^{(1)}-C_{\ell q}^{(3)} \propto (\lambda^{1L})^2/M^2$ at tree level (Details are given in Appendix~\ref{S13LQ_matching}).
The blue filled region satisfies the $2\sigma$ bounds for the WCs expected at a $10$~TeV muon collider with $10~\mathrm{ab}^{-1}$ luminosity (given in Table~\ref{tab:future_limits_1}). The dashed magenta contour is the analogous FCC\mbox{-}ee bounds (given in Table~\ref{tab:future_limits}). The orange solid line shows the current $95\%$\,C.L. limit from $\mathrm{Br}(\tau\to\mu\gamma)$, obtained from the dipole structures. The green dashed lines mark the $2\sigma$ band around the central value for $\Delta a_\mu$. 
\begin{figure}[t]
			\begin{center}
				\subfloat{
					\begin{tabular}{cc}
						\includegraphics[width=7.5cm,height=5.7cm]{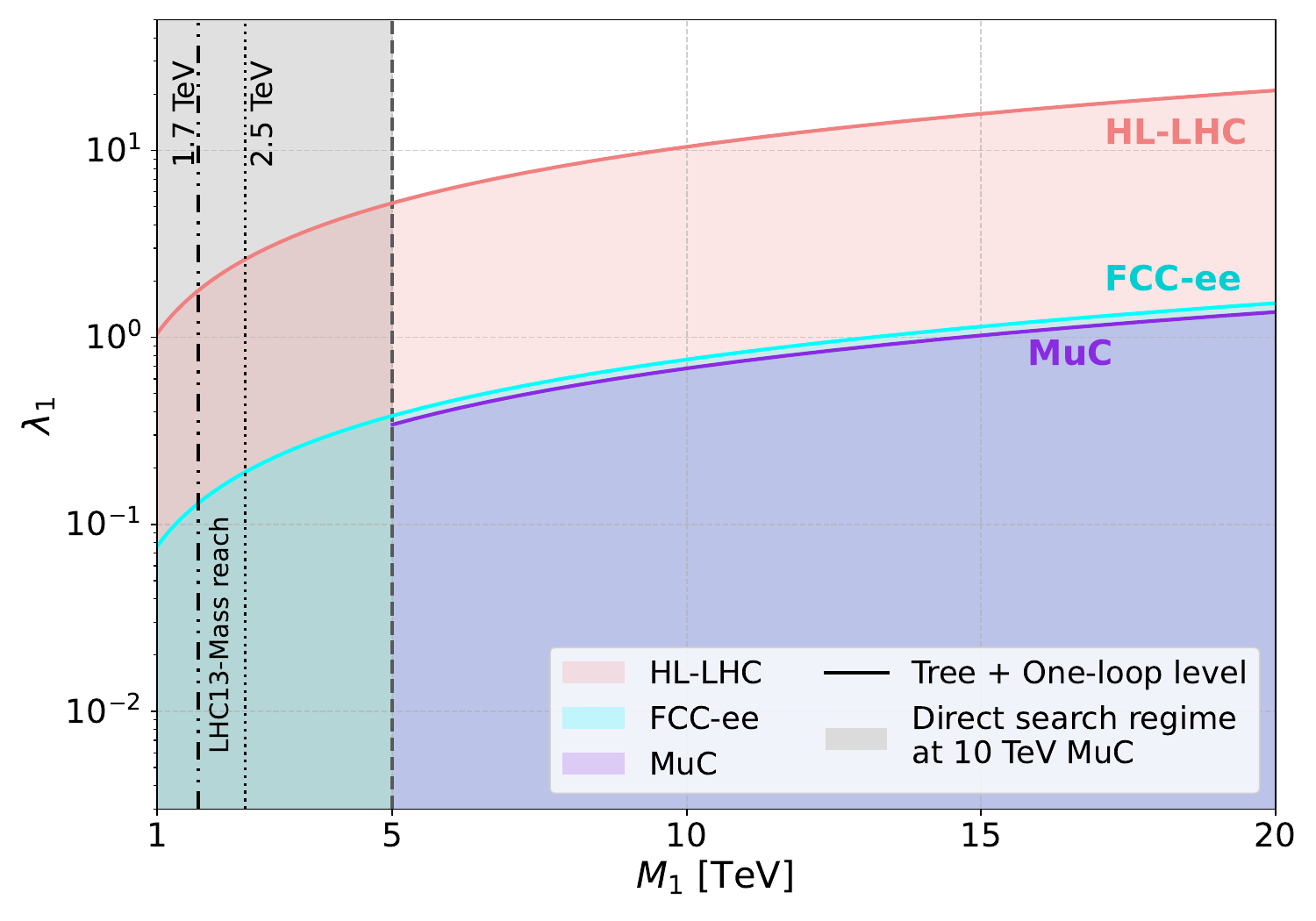}&
						\includegraphics[width=7.5cm,height=5.7cm]{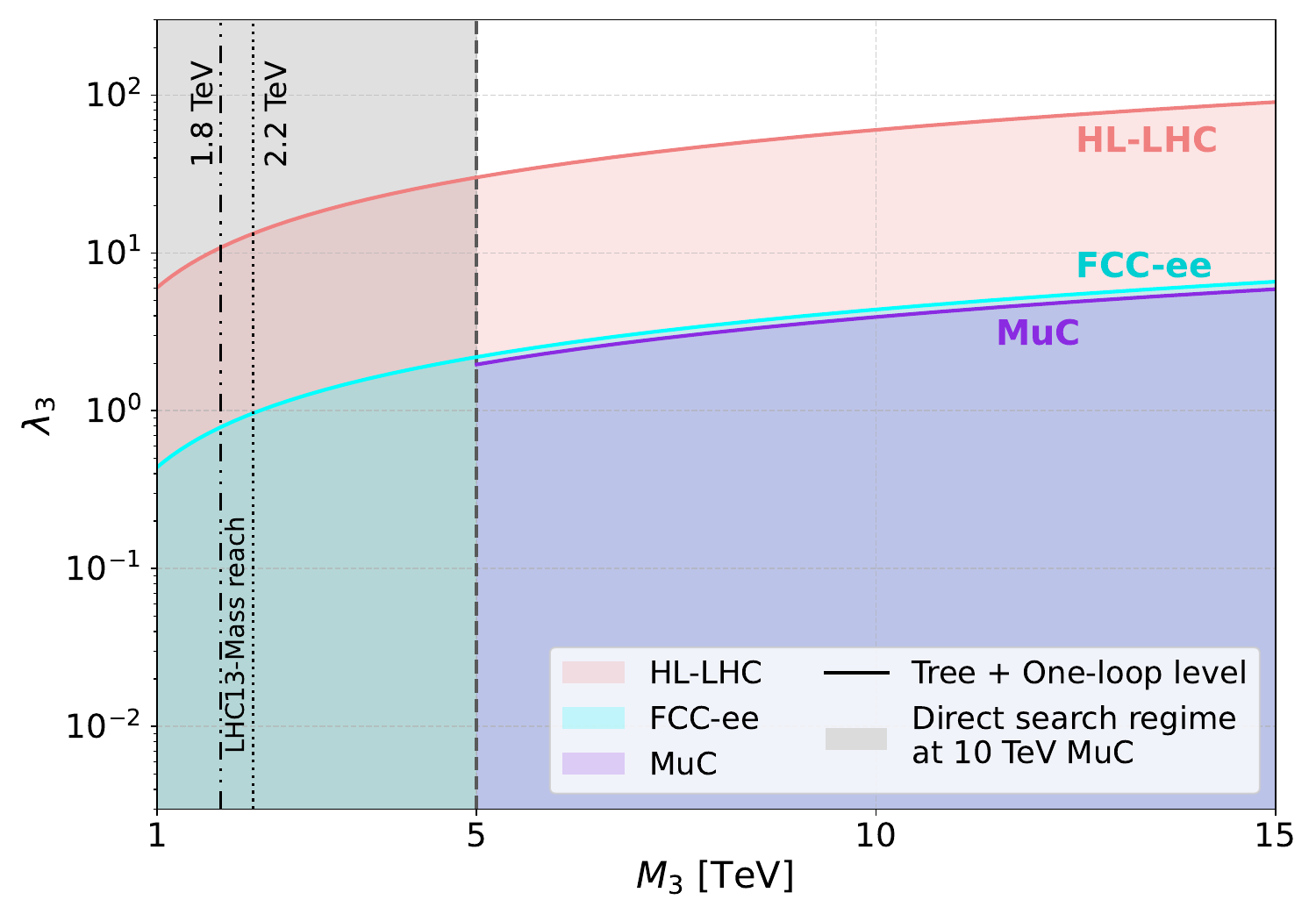}    \\
						(a)&(b)
				\end{tabular}}
				\caption{Projected bounds on the leptoquark couplings $\lambda_i$ as functions of the leptoquark mass $M_i$ for two scenarios: (a) the $S_1$ singlet including up to one-loop effects, for which the tree and one-loop results are nearly identical and (b) the $S_3$ triplet including up to one-loop effects, for which the tree level contributions are vanishing. The curves are derived from effective-operator constraints at an FCC\mbox{-}ee (blue) and a 10 TeV muon collider (purple). The dash-dot-dotted and dotted vertical lines indicate the mass bounds from pair production obtained at LHC, corresponding to the $139$ fb$^{-1}$ and $3$ ab$^{-1}$ luminosity, respectively. The grey shaded region denotes the kinematic regime where the leptoquark can be produced on shell at a 10 TeV muon collider. In this region, a description in terms of the full UV theory is applicable, while beyond it, the SMEFT interpretation is applicable.}
				\label{fig:collider_constraints}
			\end{center}
		\end{figure}
We emphasise that we consider in Fig.~\ref{scalarlqconstraints}  (a)  the dipole and flavour-violating structures, while in Fig.~\ref{scalarlqconstraints} (b) the constraints obtained from the four-fermion operator fit. Considering a simple case of the universal Yukawa scenario in the following discussion, the five WCs in Eq.~\ref{matchingrelation_treelevel} reduce to functions of only two ratios: $\lambda_1/M_1$ and $\lambda_3/M_3$. Constraints on this model from the bounds derived in Table~\ref{tab:future_limits_1} and ~\ref{tab:future_limits} are then shown in Fig.~\ref{scalarlqconstraints} (b). These results indicate that future lepton colliders will probe this parameter space well beyond the reach of the HL-LHC. Owing to its higher centre-of-mass energy and luminosity, the muon collider outperforms FCC\mbox{-}ee in this regard, providing stronger sensitivity to  $\lambda_1/M_1$. 

We show in Figs.~\ref{fig:collider_constraints} (a) and (b), coefficient bounds on SMEFT operators recast into limits on leptoquark–fermion couplings versus the leptoquark masses for two respective cases: (i) when only $S_1$ is present and $\lambda_3=0$; ii) when only $S_3$ is present and $\lambda_1=0$. For case (i) of $S_1$ we use $C_{\ell q}^{(-)} \sim (\lambda_{1L})^2/(2M_1^2)$ and $C_{eu} \sim (\lambda_{1R})^2/(2M_1^2)$ at tree level (we also include the one-loop corrections, depicted with the solid curve). For case (ii) of $S_3$ we obtain the constraints from these SMEFT operators at one-loop level. We find that for the scalar leptoquark-induced dimension-6 operator effects, the FCC\mbox{-}ee and the 10~TeV muon collider exhibit comparable sensitivities within the SMEFT valid regime. For the scalar leptoquark triplet $S_3$, no tree level contributions arise to $C_{eu}$ or to the combination $C_{\ell q}^{(-)} \equiv C_{\ell q}^{(1)} - C_{\ell q}^{(3)}$. The absence of $C_{eu}$ follows from the left-handed structure of the $S_3$ Yukawa interaction, while the vanishing of $C_{\ell q}^{(-)}$ is due to the $\mathrm{SU}(2)_L$ structure of the triplet, which yields equal contributions to $C_{\ell q}^{(1)}$ and $C_{\ell q}^{(3)}$ at tree level.
Consequently, the relevant semileptonic four-fermion interaction in the $S_3$ scenario is generated only at one loop and is therefore numerically suppressed. As a result, the corresponding bounds on the coupling $\lambda_3$ are significantly weaker compared to the $S_1$ case. Overall, Fig.~\ref{fig:collider_constraints} illustrates the distinct phenomenological implications of the $S_1$ and $S_3$ scenarios for indirect probes at future colliders.
\section{Conclusions}
\label{section8}
The SMEFT framework provides a systematic approach to parametrise indirect effects of heavy new physics at colliders~\cite{Giudice:2007fh,  Craig:2014una,  Englert:2015hrx, Degrande:2016dqg, deBlas:2016ojx,  Barklow:2017awn, Khanpour:2017cfq, Biekotter:2018ohn, Araz:2020zyh,Banerjee:2021huv, Ellis:2020unq,  Almeida:2021asy}.
While global SMEFT fits aim to constrain the full operator basis, the kinematic reach of future colliders allows one to isolate energy-enhanced contributions that probe the highest accessible new physics scales. Following this approach, we performed a high-energy analysis of electroweak processes at a 10 TeV muon collider, focusing on the leading dimension-6 operators relevant for Higgs and top interactions with \emph{muons}.

We first established the current constraints on the relevant Wilson coefficients by combining information from LHC Higgs and top measurements with electroweak precision observables and low-energy flavour data. This provided a benchmark against which the sensitivity of a high-energy muon collider was assessed.

Building on these results, we studied the reach of a 10 TeV muon collider for SMEFT effects in muon-Higgs–gauge and muon–top interactions. The analysis focused on the dominant electroweak operators relevant at high energies, with different production modes probing complementary structures. We examined the processes of $\mu^+\mu^- \to Zh$, $\mu^+\mu^- \to \mu^+\mu^-h$, $\mu^+\mu^- \to t\bar t$, and $\mu^+\mu^- \to t\bar t h$. Among these, $t\bar t$ production provides the strongest sensitivity on the new interactions, followed by $Zh$, $t\bar t h$, and the $Z$ boson fusion channel $\mu^+\mu^- \to \mu^+\mu^- h$. The sensitivity is driven by energy-growing contributions in the high-energy tails of kinematic distributions, emphasising the role of differential observables. Our results show that the electroweak dipole operators $C_{\mu\gamma}$ and $C_{\mu Z}$ can be probed down to $\mathcal{O}(10^{-4})~\text{TeV}^{-2}$, corresponding to scales $\Lambda \gtrsim 100$ TeV for $\mathcal{O}(1)$ couplings. Vector-like muon–top interactions can be tested up to the tens-of-TeV range. 
Although this study focused on dimension-6 operators, dimension-8 effects, expected to be relevant at multi-TeV energies, will be explored in future work. A more detailed treatment of systematic and detector effects will further improve the robustness of these projections.

We have also compared these sensitivities with projections from the HL-LHC and FCC\mbox{-}ee, where the latter correspond to the full FCC\mbox{-}ee running scenario at $\sqrt{s}=91$, $161$, $240$, and $365$ GeV within a global SMEFT fit. At the level of Wilson coefficients, a multi-TeV muon collider is particularly sensitive to operators with energy-growing effects and to four-fermion interactions involving the top quarks and muons. We further interpret these results in representative UV scenarios with vector-like leptons and scalar leptoquarks, finding strong constraints that push the characteristic mass scales well beyond current experimental reach. In the decoupling regime, where the SMEFT expansion remains valid, the corresponding bounds on UV parameters can be comparable to those from FCC\mbox{-}ee. Overall, a multi-TeV muon collider, especially at $\sqrt{s}\sim 10$ TeV, offers sensitivity to scales of several tens of TeV within a clean environment and broad operator coverage, making it a compelling probe of new physics in the next collider era.

\section*{Acknowledgements}
We thank Satyaki Bhattacharya and Subir Sarkar for helpful suggestions on studying the detector effects. TB is grateful to Joydeep Chakrabortty and Shankha Banerjee for useful discussions on EFT. TB also acknowledges the support of University of Calcutta where part of the work was done and the Indian Institute of Technology, Kanpur for support towards completing this work.
\appendix
\section{Matching Relations}
\label{matching}
\subsection{Model with Vector-like Lepton Singlet}	
\label{VLL_matching}
In this appendix, we summarise the matching relations for the Wilson coefficients defined in section~\ref{section2} within a UV completion that augments the SM by a heavy vector-like lepton singlet. We perform a one-loop matching onto the dimension-6 operator basis using {\tt{Matchmakereft}}~\cite{Carmona:2021xtq} and {\tt{Matchete}}~\cite{Fuentes-Martin:2022jrf}. For Warsaw basis operators with two fermion fields, the model yields the following non-vanishing contributions.\footnote{We use the notation $\ftr{\la{}\lab{}} \equiv \lambda_i \lambda_i^\ast$, with $i$ denoting the SM flavour index.} Clearly, $C_{\varphi \ell}^{(1)}$ and $C_{\varphi \ell}^{(3)}$ are induced at tree level whereas  remaining two-fermionic operators arise only at one-loop level. The four-fermion operators also receive contributions, arising first at one-loop.
\begin{align}
    [C_{\varphi \ell}^{(1)}]_{ij}=&
  -\frac{\la{i} \lab{j}}{4M_E^2}
  +\frac{1}{16\pi^2} \frac{1}{M_E^2}
  \left[
    \left(\frac{g'^{4} }{15}-
    \frac{13 g'^2 \ftr{\la{}\lab{}}}{72} \right)\delta_{ij}
    +\left(
    \frac{31 g'^2}{288} -\frac{33 g^2}{32}+\frac{13}{16} \ftr{\la{}\lab{}}
    \right) \la{i}\lab{j}
    \right.\nonumber\\&\left.
    -\frac{1}{2} \la{i} \lab{k} (y_{\ell})_{kl} (y_{\ell}^\dagger)_{lj}
	-\frac{1}{2} (y_{\ell})_{ik} (y_{\ell}^\dagger)_{kl}\la{l} \lab{j}
    \right]
  \nonumber  \\
  &
+\frac{1}{16\pi^2} \frac{1}{M_E^2}
  \left[
    -\frac{g'^2\ftr{\la{}\lab{}}}{12}\delta_{ij}
    +
    \left(
    \frac{25 g'^2}{48}-\frac{9g^2}{16}+\frac{3\ftr{\la{}\lab{}}}{8}
    \right)\la{i}\lab{j}
    \right.\nonumber\\&\left.    
-\frac{1}{2} \la{i} \lab{k} (y_{\ell})_{kl} (y_{\ell}^\dagger)_{lj}
	-\frac{1}{2} (y_{\ell})_{ik} (y_{\ell}^\dagger)_{kl}\la{l} \lab{j}    
\right] \log \tfrac{\mu_M^2}{M_E^2},
  \\
  [C_{\varphi \ell}^{(3)}]_{ij}=&
  -\frac{\la{i} \lab{j}}{4M_E^2}
  +\frac{1}{16\pi^2} \frac{1}{M_E^2}
  \left[
   - 
    \frac{5 g^2 \ftr{\la{}\lab{}}}{72}\delta_{ij} 
    +\left(
    \frac{9 g'^2}{32} +\frac{77 g^2}{288}+\frac{5}{16} \ftr{\la{}\lab{}}
    \right) \la{i}\lab{j}
    \right]  \\
  &
  +\frac{1}{16\pi^2} \frac{1}{M_E^2}
  \left[
    -\frac{g^2\ftr{\la{}\lab{}}}{12}\delta_{ij}
    +
    \left(
    \frac{9g'^2}{16}
    +\frac{7g^2}{48}+\frac{3\ftr{\la{}\lab{}}}{8}
    \right)\la{i}\lab{j}
\right] \log \tfrac{\mu_M^2}{M_E^2},
  \nonumber
  \\
  [C_{\varphi  e}]_{ij}=&
  \frac{1}{16\pi^2} \frac{1}{M_E^2}
  \left[
     g'^2 \left(
    \frac{2g'^2}{15}-\frac{13 \ftr{\la{}\lab{}}}{36}
    \right)\delta_{ij}
	+\frac{1}{24} (y_{\ell}^\dagger)_{ik} \la{k} \lab{l} (y_{\ell})_{lj}
    \right.\nonumber\\&\left.    
+\left(
    -\frac{g'^2}{6}  \ftr{\la{}\lab{}} \delta_{ij}
	+ \frac{1}{4} (y_{\ell}^\dagger)_{ik} \la{k} \lab{l} (y_{\ell})_{lj}
\right)    \log \tfrac{\mu_M^2}{M_E^2}
    \right]\\
  [C_{\mu W}]_{ij}=&-\frac{1}{16\pi^2} \frac{g}
  {24 M_E^2}
	\la{i} \lab{k}  (y_{\ell})_{kj},
  \quad \quad [C_{\mu B}]_{ij}=-\frac{1}{16\pi^2} \frac{g'}
  {12 M_E^2}
	\la{i} \lab{k}  (y_{\ell})_{kj}
\end{align}
\subsection{Model with Scalar Leptoquarks : The $S_1 +S_3$ Scenario}	
\label{S13LQ_matching}
After integrating out the heavy scalar leptoquarks up to one-loop and matching the model parameters with the generated dimension-6 operators, only semi-leptonic structures are generated at tree level. Gauge or Yukawa loops and the quartic interactions in the scalar potential  induce the Higgs current operators $\mathcal{O}_{\varphi\ell}^{(1,3)}$, $\mathcal{O}_{\varphi e}$ and the dipole operators $\mathcal{O}_{\mu W}$ and $\mathcal{O}_{\mu B}$. The corresponding Wilson coefficients contribute to observables such as $\Delta a_\mu$ and $Z \to \mu^+ \mu^-$, thereby leading to constraints from precision electroweak data. The effects of quartic leptoquark interactions are suppressed both by loop factors and by the heavy mass scales $M_{1,3}^{-2}$ and are therefore mainly constrained through electroweak precision observables, such as oblique parameters measured at LEP and analogous QCD effects, cf.\ the discussion in \cite{Calibbi:2020emz}. We summarise the notation entering the one-loop matching expressions in Table~\ref{tab:notation_S1S3} and present the matching relations for the generated four-fermion operators in the following.
\begin{table}[b]
\centering
\renewcommand{\arraystretch}{1.1}
\setlength{\tabcolsep}{7pt}
\begin{tabular}{|c|c|l|}
\hline
Notation & Type & Description \\
\hline
$S_{1}$, $S_{3}$ & Scalar LQs & $(\bar{3},\,1,\,1/3)$,\; $(\bar{3},\,3,\,1/3)$ \\[2pt]
$M_{n}$ & Mass & $M_{1}$, $M_{3}$: masses of $S_{1}$, $S_{3}$ \\[2pt]
$L_{n}$ & Logarithmic term & $\ln(\mu^{2}/M_{n}^{2})$ \\[2pt]
$G$ & Loop factor & $1/(16\pi^{2})$ \\[2pt]
$c_1$&LQ self-coupling& $(S_1^\dagger S_1)^2$\\[2pt]
$\kappa_3^{(1)}$&LQ self-coupling& $(S_3^\dagger S_3)(S_3^\dagger S_3)$\\[2pt]
$\kappa_3^{(3)}$&LQ self-coupling& $(S_3^{I \dagger} \epsilon^{IJK} S_3^J)(S_3^{L \dagger} \epsilon^{LMK}S_3^M)$\\[2pt]
$\kappa_3^{(5)}$&LQ self-coupling& $\frac{1}{4}(S_3^{I\dagger} S_3^J)(S_3^{I\dagger} S_3^J)-\frac{1}{6}(S_3^\dagger S_3)(S_3^\dagger S_3)$ \\[2pt]
$\kappa_{13}^{(1)}$&LQ self-coupling& $(S_1^\dagger S_1)(S_3^\dagger S_3)$\\[2pt]
\hline
\end{tabular}
\caption{Notations used in the $S_{1}+S_{3}$ scalar leptoquark (LQ) model. Indices $n=1,3$ denote the singlet ($S_{1}$) and triplet ($S_{3}$) leptoquark fields, respectively.}
\label{tab:notation_S1S3}
\end{table}
\hspace{-20mm}
\begin{align}
[C_{\mu W}]_{ij}= & G\Bigg[ -\frac{N_{c}}{8}g\left(\frac{{3}}{2}+ L_1 \right)\frac{(\lambda^{1L \dagger} y_u^\ast \lambda^{1R})_{ij}}{M_{1}^{2}}
-\frac{N_{c}}{24}g\left(3\frac{(\lambda^{3L \dagger} \lambda^{3L}y_{e})_{ij}}{M_{3}^{2}} -\frac{(\lambda^{1L \dagger} \lambda^{1L}y_{e})_{ij}}{M_{1}^{2}}\right) \Bigg], \nonumber\\{}
[C_{\mu B}]_{ij}= & G\Bigg[\frac{N_{c}}{4}g^{\prime}\left[(Y_{q}+Y_{u}) L_1  + \frac{1}{2}Y_{q}+\frac{3}{2}Y_{u}-Y_{e} { - Y_H} \right]\frac{(\lambda^{1L \dagger} y_u^\ast \lambda^{1R})_{ij}}{M_{1}^{2}} - \frac{N_{c}}{24}g^{\prime}(Y_{e}{+}3Y_{u})\frac{(y_{e} \lambda^{1R \dagger} \lambda^{1R})_{ij}}{M_{1}^{2}} \Bigg]. \nonumber
\end{align}
Matching relations for the four-fermionic operators, generated up to one-loop are given by: 
\begin{align}
[C_{\ell q}^{(1)}]_{ij \alpha\beta} &=
\frac{\lambda_{\alpha i}^{1L*}\lambda_{\beta j}^{1L}}{4M_{1}^{2}}
+\frac{3\lambda_{\alpha i}^{3L*}\lambda_{\beta j}^{3L}}{4M_{3}^{2}} + G \Bigg[
\frac{1}{4}\left(\frac{3}{2}\right)\left[g_{s}^{2}\frac{N_c^2-1}{2N_c}+g^{\prime2}(Y_{q}-Y_{\ell})^{2}\right]
\left(\frac{3\lambda_{\alpha i}^{3L*}\lambda_{\beta j}^{3L}}{M_{3}^{2}}+\frac{\lambda_{\alpha i}^{1L*}\lambda_{\beta j}^{1L}}{M_{1}^{2}}\right) \nonumber\\
\hspace{-20mm}&+\frac{1}{4}\left(\frac{3}{2}\right) g^2 \cdot 3
\left(\frac{\lambda_{\alpha i}^{3L*}\lambda_{\beta j}^{3L}}{M_{3}^{2}}+\frac{\lambda_{\alpha i}^{1L*}\lambda_{\beta j}^{1L}}{M_{1}^{2}}\right) -\frac{N_{c}}{30}g^{\prime4}Y_{\ell}Y_{q}\delta_{ij} \delta_{\alpha\beta}
\left(\frac{3Y_{S_{3}}^{2}}{M_{3}^{2}}+\frac{Y_{S_{1}}^{2}}{M_{1}^{2}}\right)  \nonumber\\
&-\Bigg\{ \left(
\frac{1}{2}\,\frac{N_c}{2}\Big(\tfrac{1}{2}+ L_1 \Big)(\lambda^{1L\dagger}\lambda^{1L})_{ik}
+\frac{3N_c}{2}\Big(\tfrac{1}{2}+ L_3 \Big)(\lambda^{3L\dagger}\lambda^{3L})_{ik}
\right)\delta_{jl}\delta_{\alpha \gamma}\delta_{\beta\delta} \nonumber\\
\hspace{-20mm}&+ \left(
\frac{1}{2}\,\delta_{ik}\,\frac{N_c}{2}\Big(\tfrac{1}{2}+L_1 \Big)(\lambda^{1L\dagger}\lambda^{1L})_{jl}
+\frac{3N_c}{2}\Big(\tfrac{1}{2}+ L_3 \Big)(\lambda^{3L\dagger}\lambda^{3L})_{jl}
\right) \delta_{\alpha \gamma}\delta_{\beta\delta} \nonumber\\
\hspace{-20mm}&+ \frac{1}{4}\,\delta_{ik}\delta_{jl}
\left( \Big(\tfrac{1}{2}+L_1 \Big)(\lambda^{1L\ast}\lambda^{1LT})_{\alpha \gamma}
+3\Big(\tfrac{1}{2}+ L_3 \Big)(\lambda^{3L\ast}\lambda^{3LT})_{\alpha \gamma}\right)\delta_{\beta\delta} \nonumber\\
&+\frac{1}{4}\,\delta_{ik}\delta_{jl}\delta_{\alpha \gamma}
\left(\Big(\tfrac{1}{2}+L_1 \Big)(\lambda^{1L\ast}\lambda^{1LT})_{\beta\delta}
+3\Big(\tfrac{1}{2}+ L_3 \Big)(\lambda^{3L\ast}\lambda^{3LT})_{\beta\delta}\right)
\Bigg\}\left(\frac{\lambda_{k\gamma}^{1L*}\lambda_{l\delta}^{1L}}{4M_{1}^{2}}
+\frac{3\lambda_{k\gamma}^{3L\ast}\lambda_{l\delta}^{3L}}{4M_{3}^{2}}\right) \nonumber\\
& +\frac{g^{\prime2}}{3}Y_{\ell}\delta_{ij}\Bigg\{
3\left(\frac{8Y_\ell-Y_{S_3}}{6}+Y_{\ell}L_3 \right)
\frac{(\lambda^{3 L \ast} \lambda^{3LT})_{\alpha \beta}}{M_{3}^{2}}
+\left(\frac{8Y_\ell-Y_{S_1}}{6}+Y_{\ell} L_1 \right)
\frac{(\lambda^{1 L \ast} \lambda^{1LT})_{\alpha \beta}}{M_{1}^{2}}\Bigg\} \nonumber\\
&+\frac{N_{c} g^{\prime2}}{3}Y_{q}\delta_{\alpha\beta}\Bigg[
3\left(\frac{8Y_q-Y_{S_3}}{6}+Y_{q} L_3 \right)
\frac{(\lambda^{3 L \dagger} \lambda^{3L})_{ij}}{M_{3}^{2}}
+\left(\frac{8Y_q-Y_{S_1}}{6}+Y_{q} L_1 \right)
\frac{(\lambda^{1 L \dagger} \lambda^{1L})_{ij}}{M_{1}^{2}}\Bigg] \nonumber\\
& -\frac{1}{4}\left(
3\frac{(\lambda^{3 L \dagger} \lambda^{3L})_{ij}(\lambda^{3 L \ast} \lambda^{3LT})_{\alpha \beta}}{M_{3}^{2}}
+\frac{(\lambda^{1 L \dagger} \lambda^{1L})_{ij}(\lambda^{1 L \ast} \lambda^{1LT})_{\alpha\beta}}{M_{1}^{2}}
\right) \\
&  +\left(c_{1}\Big(1+ L_1 \Big)
+\frac{9}{4}\Big(1+L_3 \Big)\kappa_{13}^{(1)}\frac{M_{3}^{2}}{M_{1}^{2}}\right)
\frac{\lambda_{\alpha i}^{1L\dagger}\lambda_{\beta j}^{1L}}{M_{1}^{2}} \nonumber\\
&  +\left(\frac{9}{4}\Big(1+ L_1 \Big)\kappa_{13}^{(1)}\frac{M_{1}^{2}}{M_{3}^{2}}
+\frac{3}{2}\Big(1+ L_3 \Big)\left[5 \kappa_{3}^{(1)}-\kappa_{3}^{(3)}+\tfrac{5}{6}\kappa_{3}^{(5)}\right]\right)
\frac{\lambda_{\alpha i}^{3L\dagger}\lambda_{\beta j}^{3L}}{M_{3}^{2}}
\Bigg], \nonumber \\
[C_{\ell q}^{(3)}]_{ij\alpha\beta} &= 
-\frac{\lambda_{\alpha i}^{1L*}\lambda_{\beta j}^{1L}}{4 M_{1}^{2}}
+\frac{3\lambda_{\alpha i}^{3L*}\lambda_{\beta j}^{3L}}{4M_{3}^{2}}
+ G\Bigg[
\frac{1}{4}\!\left(\frac{3}{2}\right)\!\left[g_s^{2}\frac{N_c^2-1}{2N_c}+g^{\prime2}(Y_q-Y_{\ell})^{2}\right]
\left(\frac{\lambda_{\alpha i}^{3L\ast}\lambda_{\beta j}^{3L}}{M_{3}^{2}}
      -\frac{\lambda_{\alpha i}^{1L\ast}\lambda_{\beta j}^{1L}}{M_{1}^{2}}\right) \nonumber\\
& + \frac{3}{8}\! g^2 
\left(-\frac{3\,\lambda_{\alpha i}^{3L\ast}\lambda_{\beta j}^{3L}}{M_3^{2}}
      +\frac{\lambda_{\alpha i}^{1L\ast}\lambda_{\beta j}^{1L}}{M_1^{2}}\right) + \frac{N_c}{12}g^{2}\,\delta_{\alpha\beta}\!\left[
\left(2+ L_3 \right)\frac{(\lambda^{3L\dagger}\lambda^{3L})_{ij}}{M_3^{2}}
-\left(\tfrac{4}{3}+L_1 \right)\frac{(\lambda^{1L\dagger}\lambda^{1L})_{ij}}{M_1^{2}}
\right] \nonumber \\
& - \Bigg[\frac{N_c}{4} \left( \left(\tfrac{1}{2}+L_1 \right) 
(\lambda^{1L\dagger} \lambda^{1L})_{ik} 
+ 3 \left(\tfrac{1}{2}+ L_3 \right) 
(\lambda^{3L\dagger} \lambda^{3L})_{ik}\right)\delta_{jl}\delta_{\alpha \gamma}\delta_{\beta\delta} \nonumber \\
& +\frac{N_c}{4} \left( \left(\tfrac{1}{2}+L_1 \right) 
(\lambda^{1L\dagger} \lambda^{1L})_{jl} 
+ 3 \left(\tfrac{1}{2}+ L_3 \right) 
(\lambda^{3L\dagger} \lambda^{3L})_{jl}\right)\delta_{ik} \delta_{\alpha \gamma}\delta_{\beta\delta} \nonumber\\
&  +\frac{1}{4} \left( \left(\tfrac{1}{2}+L_1 \right) 
(\lambda^{1L\ast} \lambda^{1LT})_{\alpha \gamma} 
+ 3 \left(\tfrac{1}{2}+L_3 \right) 
(\lambda^{3L\ast} \lambda^{3LT})_{\alpha \gamma}\right) \delta_{ik}\delta_{jl} \delta_{\beta\delta} \nonumber\\
&  +\frac{1}{4} \delta_{ik}\delta_{jl}\delta_{\alpha \gamma} 
\left( \left(\tfrac{1}{2}+L_1 \right) (\lambda^{1L\ast} \lambda^{1LT})_{\beta\delta} 
+ 3 \left(\tfrac{1}{2}+ L_3 \right) 
(\lambda^{3L\ast} \lambda^{3LT})_{\beta\delta}\right)\Bigg] . \left( -\frac{\lambda_{\gamma k}^{1L*}\lambda_{\delta l }^{1L}}{4 M_{1}^{2}} 
+\frac{3\lambda_{\gamma k}^{3L*}\lambda_{\delta l}^{3L}}{4M_{3}^{2}}\right) \nonumber\\
&  - \frac{N_c}{60}g^{4}\,\delta_{ij}\delta_{\alpha\beta}\,\frac{1}{M_3^{2}} 
+ \frac{1}{12}g^{2}\,\delta_{ij}\!\left[
\left(2+ L_3 \right)\frac{(\lambda^{3L\ast}\lambda^{3L\,T})_{\alpha\beta}}{M_3^{2}}
-\left(\tfrac{4}{3}+L_1 \right)\frac{(\lambda^{1L\ast}\lambda^{1L\,T})_{\alpha\beta}}{M_1^{2}}
\right]- \frac{1}{2} \times \nonumber\\
&  \times \!\left[
\frac{(\lambda^{3L\dagger}\lambda^{3L})_{ij}(\lambda^{3L\ast}\lambda^{3L\,T})_{\alpha\beta}}{M_3^{2}}
+ \frac{\log\!\tfrac{M_3^{2}}{M_1^{2}}\left[(\lambda^{3 L \dagger} \lambda^{1L})_{ij}
(\lambda^{1 L \ast} \lambda^{3LT})_{\alpha\beta}
+(\lambda^{1 L \dagger} \lambda^{3L})_{ij}
(\lambda^{3 L \ast} \lambda^{1LT})_{\alpha\beta}\right]}{2(M_3^{2}-M_1^{2})}
\right] \nonumber\\
&  - (\kappa_{1}\!\left(1+L_1 \right)
+\frac{9}{4}\!\left(1+ L_3 \right) \kappa_{13}^{(1)}\frac{M_3^{2}}{M_1^{2}})
\tfrac{\lambda_{\alpha i}^{1L\dagger}\lambda_{\beta j}^{1L}}{M_1^{2}} \\
& + \left(\frac{3}{4}\!\left(1+L_1 \right)\kappa_{13}^{(1)}\frac{M_1^{2}}{M_3^{2}}
+\frac{1}{2}\!\left(1+ L_3 \right)
\!\left[5 \kappa_{3}^{(1)}-\kappa_{3}^{(3)}+\tfrac{5}{6}\kappa_{3}^{(5)}\right]\right)
\frac{\lambda_{\alpha i}^{3L\dagger}\lambda_{\beta j}^{3L}}{M_3^{2}} 
\Bigg] . \nonumber \\
\hspace{-20mm}[C_{eu}]_{ij \alpha\beta} &=  
\frac{\lambda_{\alpha i}^{1R\ast} \lambda_{\beta j}^{1R}}{2 M_1^2} 
+G\Bigg[
\frac{1}{2}\!\left(\frac{3}{2}\right)\!
\left[g_{s}^{2}\frac{N_c^2-1}{2N_c}+g^{\prime2}(Y_{u}-Y_{e})^{2}\right]
\frac{\lambda_{\alpha i}^{1R*}\lambda_{\beta j}^{1R}}{M_{1}^{2}} \nonumber\\
\hspace{-20mm}&  -\Bigg \{
\frac{1}{2}\,\frac{N_c}{2}\!\left(\tfrac{1}{2}+L_1 \right)(\lambda^{1L\dagger} \lambda^{1L})_{ik} \delta_{jl}\delta_{\alpha \gamma}\delta_{\beta\delta}+ 
\frac{1}{2}\,\delta_{ik}\,\frac{N_c}{2}\!\left(\tfrac{1}{2}+L_1 \right)(\lambda^{1L\dagger} \lambda^{1L})_{\delta\beta}
\delta_{\alpha \gamma}\delta_{\beta\delta} \nonumber\\
\hspace{-20mm}&  +\frac{1}{4}\,\delta_{ik}\delta_{jl}
\left(\tfrac{1}{2}+L_1 \right)(\lambda^{1R\ast} \lambda^{1RT})_{\alpha \gamma}\,\delta_{\beta\delta}
+\frac{1}{4}\,\delta_{ik}\delta_{jl}\delta_{\alpha \gamma}
\left(\tfrac{1}{2}+L_1 \right)(\lambda^{1R\ast} \lambda^{1RT})_{\beta\delta}
\Bigg \} \frac{\lambda_{\gamma k}^{1R\ast} \lambda_{\delta l}^{1R}}{2 M_1^2} \nonumber\\
\hspace{-20mm}&  -\frac{N_{c}}{30}g^{\prime4}Y_{e}Y_{u}\delta_{ij}\delta_{\alpha\beta}
\left(\frac{3Y_{S_{3}}^{2}}{M_{3}^{2}}+\frac{Y_{S_{1}}^{2}}{M_{1}^{2}}\right)+\frac{1}{3}g^{\prime2}Y_{e}
\left(\frac{-Y_{e}-Y_{S_{1}}}{6}+Y_{e}L_1 \right)
\frac{\delta_{ij}(\lambda^{1R \ast} \lambda^{1RT})_{\alpha\beta}}{M_{1}^{2}} \nonumber\\
\hspace{-20mm}&+\frac{N_{c}}{3}g^{\prime2}Y_{u}
\left(\frac{-Y_{u}-Y_{S_{1}}}{6}+Y_{u} L_1 \right)
\frac{\delta_{\alpha\beta}(\lambda^{R \dagger} \lambda^R)_{ij}}{M_{1}^{2}}\\
\hspace{-20mm}& -\frac{1}{4}\frac{(\lambda^{R \dagger} \lambda^R)_{ij}(\lambda^{1R \ast} \lambda^{1RT})_{\alpha\beta}}{M_{1}^{2}}
+\left(2 \kappa_{1}(1+L_1 )
+\frac{9}{2}(1+L_3 )\kappa_{13}^{(1)}\frac{M_{3}^{2}}{M_{1}^{2}}\right)
\frac{\lambda_{\alpha i}^{1R\dagger}\lambda_{\beta j}^{1R}}{M_{1}^{2}}
\Bigg] ,  \nonumber\\
[C_{\ell equ}^{S}]_{ij \alpha\beta} &= 
\frac{\lambda_{\alpha i}^{1L\ast} \lambda_{\beta j}^{1R}}{2 M_1^2} 
- G\Bigg[ \Bigg(
\Bigg[\frac{N_c}{4}\!\left(\tfrac{1}{2}+L_1\right)(\lambda^{1L\ast}\lambda^{1LT})_{ik}
+\frac{3N_c}{4}\!\left(\tfrac{1}{2}+L_3\right)(\lambda^{3L\ast}\lambda^{3LT})_{ik}
\Bigg]\delta_{jl}\delta_{\alpha \gamma}\delta_{\beta\delta} \nonumber \\
& +\tfrac{1}{2}\,\delta_{ik}\delta_{jl}
[\left(\tfrac{1}{2}+L_1 \right)(\lambda^{1L\dagger}\lambda^{1L})_{\alpha \gamma} 
+ 3\left(\tfrac{1}{2}+L_3 \right)(\lambda^{3L\dagger}\lambda^{3L})_{\alpha \gamma}]\delta_{\beta\delta} \\
& +
\tfrac{1}{2}\,\delta_{ik}\,\tfrac{N_c}{2}\!\left(\tfrac{1}{2}+L_1 \right)(\lambda^{1R\ast}\lambda^{1RT})_{\delta\beta}
\delta_{\alpha \gamma}\delta_{\beta\delta}  +\tfrac{1}{2}\,\delta_{ik}\delta_{jl}\delta_{\alpha \gamma}
\left(\tfrac{1}{2}+L_1 \right)(\lambda^{1R\dagger}\lambda^{1R})_{\beta\delta}
\Bigg) \tfrac{\lambda_{k \gamma}^{1L\ast} \lambda_{l \delta}^{1R}}{2 M_1^2} \nonumber\\
& +\frac{1}{2}\left(\tfrac{3}{2} + L_1 \right)
\frac{(y_{e})_{ij}\,\big[(\lambda^{1R } y_e^{\dagger} \lambda^{1L\dagger})^T\big]_{\alpha\beta}}{M_{1}^{2}} +\frac{N_{c}}{2}\left(\tfrac{3}{2}+L_1 \right)
\frac{(y_{u})_{\alpha\beta}\,(\lambda^{1L \dagger} y_u^\ast \lambda^{1R})_{ij}}{M_{1}^{2}} + \frac{\lambda_{\alpha i}^{1L\dagger}\lambda_{\beta j}^{1R}}{M_{1}^{2}} \times \nonumber\\
&  \times \left\{ 2\left(1+ L_1 \right)\kappa_{1}
+\frac{9}{2}\left(1+L_3 \right)\kappa_{13}^{(1)}\frac{M_{3}^{2}}{M_{1}^{2}} \right. \left.-\frac{3}{2}\left(\tfrac{3}{2}+L_1 \right)
\left[(Y_{q}-Y_{\ell})(Y_{u}-Y_{e})g^{\prime2}
+\frac{N_c^2-1}{2 N_c}g_{s}^{2}\right]\right\} 
\Bigg], \nonumber\\
[C_{\ell equ}^{T}]_{ij \alpha\beta} &=
-\frac{\lambda_{\alpha i}^{1L\ast} \lambda_{\beta j}^{1R}}{8 M_1^2}
- G\Bigg[ \Bigg\{
\left( \frac{N_c}{4}\!\left(\tfrac{1}{2}+L_1 \right)(\lambda^{1L\ast}\lambda^{1LT})_{ik}
+\frac{3N_c}{4}\!\left(\tfrac{1}{2}+L_3 \right)(\lambda^{3L\ast}\lambda^{3LT})_{ik}
\right)\delta_{jl}\delta_{\alpha \gamma}\delta_{\beta\delta} \nonumber\\
& 
+\frac{1}{2}\,\delta_{ik}\delta_{jl}
\Big[
\left(\tfrac{1}{2}+ L_1 \right)(\lambda^{1L\dagger}\lambda^{1L})_{\alpha \gamma}
+3\left(\tfrac{1}{2}+ L_3 \right)(\lambda^{3L\dagger}\lambda^{3L})_{\alpha \gamma}
\Big]\delta_{\beta\delta} \\
& +\delta_{ik}\,\frac{N_c}{4}\!\left(\tfrac{1}{2}+  L_1 \right)
(\lambda^{1R\ast}\lambda^{1RT})_{\delta\beta}
\delta_{\alpha \gamma}\delta_{\beta\delta} 
+\frac{1}{2}\,\delta_{ik}\delta_{jl}\delta_{\alpha \gamma}
\left(\tfrac{1}{2}+ L_1 \right)(\lambda^{1R\dagger}\lambda^{1R})_{\beta\delta}
\Bigg \}\!
\left(\frac{-\lambda_{\gamma k}^{1L\ast} \lambda_{\delta l}^{1R}}{8 M_1^2}\right) \nonumber\\
& 
+\frac{\lambda_{\alpha i}^{1L\dagger}\lambda_{\beta j}^{1R}}{4 M_{1}^{2}}\!\Bigg(
\!\left(1+ L_1 \right)\kappa_{1}
+\frac{9}{2}\!\left(1+ L_3 \right)\kappa_{13}^{(1)}\frac{M_{3}^{2}}{M_{1}^{2}} +\frac{1}{2}\!\left(\tfrac{3}{2}+ L_1 \right)
\!\left[(Y_{q}-Y_{\ell})(Y_{u}-Y_{e})g^{\prime2}+\frac{N_c^2-1}{2 N_c}g_{s}^{2}\right]
\Bigg)
\Bigg] . \nonumber
\end{align}
\bibliographystyle{JHEP}
\bibliography{references}
\end{document}